# The Spatiotemporal Organization of Motor Cortex Activity Supporting Manual Dexterity

by

**Nicholas G. Chehade**

Neuroscience and Computer & Information Science, The Ohio State University, 2017

Submitted to the Graduate Faculty of the

School of Medicine in partial fulfillment

of the requirements for the degree of

Doctor of Philosophy

University of Pittsburgh

2024

UNIVERSITY OF PITTSBURGH

SCHOOL OF MEDICINE

This dissertation was presented

by

**Nicholas G. Chehade**

It was defended on

November 20, 2024

and approved by

Committee Chair: Robert S. Turner, PhD, Neurobiology, University of Pittsburgh

Aaron P. Batista, PhD, Bioengineering, University of Pittsburgh

Jennifer L. Collinger, PhD, Physical Medicine and Rehabilitation, University of Pittsburgh

Nicholas G. Hatsopoulos, PhD, Neuroscience, University of Chicago

Dissertation Advisor: Omar A. Gharbawie, PhD, Neurobiology, University of Pittsburgh

**The Spatiotemporal Organization of Motor Cortex Activity Supporting Manual Dexterity**

Nicholas G. Chehade

University of Pittsburgh, 2024

Motor cortex (M1) is a crucial brain area for controlling voluntary movements, such as reaching and grasping for a cup of coffee. M1 is organized in a somatotopic manner, such that M1 output driving movement to different parts of the body is organized along the cortical surface. In primates, the arm and hand are represented in M1 as separate but overlapping territories. Unit activity recorded from the M1 forelimb representation comodulates with parameters related to reaching and/or grasping. The overall aim of this dissertation is to understand the spatiotemporal dynamics of M1 activity that produces reach-to-grasp movements. To address this goal, intracortical microstimulation (ICMS) is delivered along the precentral gyrus of two macaque monkeys to define the M1 motor map. Subsequently, cortical activity is recorded from the M1 forelimb representation using intrinsic signal optical imaging (ISOI) while macaques execute an instructed reach-to-grasp task. Results from imaging experiments produce spatial maps that define cortical territories with increased activity during reach-to-grasp movements. Next, unit activity was recorded from the M1 forelimb representation with a laminar multielectrode while macaques completed the same reach-to-grasp task. Recording site locations differed between sessions to comprehensively sample unit responses throughout the M1 forelimb representation. Imaging experiments reveal that activity supporting reach-to-grasp movements was concentrated in patches that comprise less than half of the M1 forelimb representation. Electrophysiology recordings reveal that activity related to reaching is spatially organized within M1 distinctly from activity related to grasping. The results support the idea that spatial organizing principles are inherent in M1 activity that supports reach-to-grasp movements.

**Table of Contents**

**List of Abbreviations**

| | |
|---|---|
| CM | Corticomotoneuronal |
| CMAd | Dorsal Cingulate Motor Area |
| CMAr | Rostral Cingulate Motor Area |
| CMAv | Ventral Cingulate Motor Area |
| CST | Corticospinal Tract |
| ECRB | Extensor Carpi Radialis Brevis |
| ED,4-5 | Extensor Digitorum 4-5 |
| EMG | Electromyography |
| FCR | Flexor Carpi Radialis |
| FDS | Flexor Digitorum Superficialis |
| FOV | Field of View |
| ICMS | Intracortical Microstimulation |
| ISOI | Intrinsic Signal Optical Imaging |
| M1 | Primary Motor Cortex |
| MAP | Maximum a Posteriori Probability |
| NHP | Non-Human Primate |
| PMd | Dorsal Premotor Cortex |
| PMv | Ventral Premotor Cortex |
| ROI | Region of Interest |
| SMA | Supplementary Motor Area |

**List of Equations**

**List of Figures**

## Acknowledgements

Thank you to my advisor, Omar Gharbawie, for his committed dedication to teaching me. Thank you to my mother and father for encouraging and supporting me. Thank you to ToniAnn, who I am completely indebted to, because none of the work in this dissertation would have been completed without her efforts and constant emotional support. Thank you to the amazing female doctorates in my cohort, Kim and Kati, who kept me from giving up countless of times and have given me lifelong friendships. Thank you to my brother and sister and all my friends for keeping me sane and enabling me to complete my graduate training. I dedicate this dissertation to my grandfather, George Chehade, who spent half a century serving as a doctor and giving back to his community.

# 1.0 Introduction

Movement is an indispensable facility for organisms to interact and react in their environment. For example, all organisms need to find and obtain sustenance to survive. And whether an organism forages for food or retrieves it from the front door, these movements are planned and executed seamlessly hundreds of times a day. Despite the perceived ease of actuating, the neural processes underlying even the simplest movements are hardly uncomplicated. The neural basis of movement has remained a complex and significant subject of investigation in neuroscience throughout the past century.

Early studies of neocortex were largely committed to mapping cognitive and behavioral functions to brain regions. Localization of motor function in cortex was first theorized by Jackson, who observed the convulsions of patients suffering from epileptic seizures (Jackson, 1870). He hypothesized that various “centers” in the brain were responsible for controlling the movements of various body parts (M. S. A. Graziano, 2009). Jackson’s conjectures were fundamental to studying the cortical control of movement as he was the first to introduce the idea that parts of the body were represented in certain parts of the brain.

Jackson’s work was later validated from experiments that injected electrical current into various locations of the brain. Fritsch and Hitzig found that electrically stimulating cortex in frontal brain areas evoked movements in the contralateral limbs of dogs (Fritsch, 1870). Subsequently, Broadmann’s meticulous characterization of the cortical cytoarchitecture revealed that these frontal motor areas were distinguishable by an absent cortical layer IV (Palomero-Gallagher & Zilles, 2018; Zilles, 2018). These findings demonstrate one of the many examples in which cortical structure and function demonstrate interdependence. Namely, layer IV is prominent in cortical sensory areas from ascending thalamic projections, whereas layer IV is compacted in agranular cortex. Moreover, layer V in agranular cortex is larger than in granular (sensory) cortex due to the inflated number of descending projections (Barbas & García-Cabezas, 2015; Shipp, 2005). The discovery of functional motor areas in cortex illustrates an example of the ubiquitous principle that is consistently observed in the life sciences—structure gives rise to function.

The discovery of localized motor function in the cortex of dogs motivated subsequent investigations of cortex from other mammals. From rabbits and cats (Ferrier, 1873, 1875) to primates and humans (Cushing, 1909; Horsley & Schafer, 1888; Penfield & Boldrey, 1937; Woolsey, 1958), electrical stimulation of frontal cortex consistently evoked movements in contralateral limbs. Specifically, Hitzig identified cortex anterior to the central sulcus as an important landmark for localizing these "excitable" areas in cortex (Hitzig, 1874; Huber, 1934). It was clear that a cortical homology existed in mammals that contributed to movement execution and control.

Indeed, histological staining and pathway tracing via lesions confirmed that mammals, ranging from rodents to anthropoids, sustained projections emanating from excitable cortical motor areas that descended the contralateral side of the spinal cord (Ariëns Kappers & Fortuyn, 1920; Huber, 1934). M1 lesion studies indicated that these projections, which comprised the corticospinal tract (CST), were crucial for controlling voluntary movements, especially dexterous movement (Lawrence & Kuypers, 1968; Leyton & Sherrington, 1917). M1 neurons comprise approximately half of all descending projections in the CST (Crevel & Verhaart, 1963; Kuypers, 1981; Russell & DeMyer, 1961) . However, additional lesion and tracer studies indicated that these projections originated from various cortical locations. Additionally, the ratio of descending corticospinal projections among these cortical motor areas varied between species (Holmes & May, 1909; Levin & Beadford, 1938). Furthermore, CM cells making direct connections with spinal motoneurons from M1 were identified in primates, but not cats (Bernhard, 1954). Bernhard & Bohm hypothesized in 1954 that these projections were found in non-human primates (NHPs) to facilitate control of dexterous movements involving the extremities (e.g., digits). This conjecture launched decades of investigation concerning the neural control of dexterous movements.

For this thesis, I will focus on NHPs. NHPs are suitable models for studying the motor control of movement due to their (1) dexterous abilities and (2) cortical homology. This investigation modeled primate dexterity using macaque monkeys (Macaca mulatta) trained on a reach-to-grasp task, that involves a precision grip. The precision grip is defined as the stabilization or manipulation of an object using the pads of the thumb and one or more digits that develops uniquely in primates (Kivell, 2015). Increased descending fibers in the CST in tandem with direct corticomotoneuronal (CM) connections in primates is posited to facilitate the neural control of fine motor abilities, such as the precision grip (Courtine et al., 2007). In fact, adult mice that were genetically modified to maintain postnatal CM connections demonstrated superior prehensile dexterity (albeit with a < 60% success rate) compared to controls that lacked CM connections (Gu et al., 2017). Thus, the comparable dexterity of NHPs and cortical homology to humans render NHPs valuable models for studying the neural control of dexterous movement.

The objective of this dissertation is to understand the relationship between the neural activity regulating coordinated movement and the structural organization in M1. In the following sections, I will discuss relevant structural properties found in M1. Next, I will review functional properties that have been identified from the neural activity in M1. Finally, I will discuss the shortcomings of investigating the relationship between function and structure in M1and delineate what the subsequent experiments revealed about the structural and functional relationship in M1.

## 1.1 M1 Structural Properties

### 1.1.1 Descending M1 projections to the spinal cord

Early studies of electrical stimulation of cortex led to further detailing cortical locations with evoked movements of specific body parts. Meticulous mapping between cortical stimulation locations to evoked movements of the body provided a somatotopic organization in M1 that has influenced the investigation of M1 for decades (Barnard & Woolsey, 1956; Penfield & Boldrey, 1937). These works delineated a somatotopic organization in M1 along the precentral gyrus with a similar sensory somatotopic organization along the postcentral gyrus. The overall motor somatotopy, from medial to lateral, is as follows: leg, trunk, arm, hand and face. This localization of motor function in the precentral gyrus was supported by the striking cytoarchitectural distinction marked by the presence of giant Betz cells (Campbell, 1905; Guillery, 2005; Leyton & Sherrington, 1917). Due to the exclusive existence of these cells in the precentral gyrus of primate cortex, it was hypothesized that these cells comprised the descending projections in the CST. However, subsequent experiments made it evident that Betz cells were not the sole source of descending projections in the CST (Ralston & Ralston, 1985; Wise et al., 1979).

As investigators analyzed the fibers that formed the CST, it became clear that fibers comprising the CST originated from various origins in cortex. There was early doubt that M1 was the sole source of descending projections that formed the CST (Russell & DeMyer, 1961). Yet, this was not corroborated until Strick traced the origins of monosynaptic connections in the spinal cord and specific forelimb muscles using retrograde transport tracers (e.g., rabies virus) (Bortoff & Strick, 1993; R. P. Dum & Strick, 1991; R. Dum & Strick, 2002; S. He et al., 1993; S. Q. He et al., 1995; Rathelot & Strick, 2009). Retrograde tracer injections into cervical sections of the spinal cord revealed that fibers in the CST also emanated from supplementary motor area, anterior cingulate, parietal and insular cortex (Darian-Smith et al., 1996). These descending projections from SMA, PMd, PMv, CMAd, CMAv and CMAr each maintained a distinct somatotopic organization analogous to the one from M1. Though, of all the projections descending in the CST, it is estimated that one-third originate from M1 (He et al., 1993).

These tracer studies also provided additional support to the emergence of dexterous motor ability in primates. In primates, injections into spinal segments of the arm and leg yield non-overlapping cortical origins for these projections. However, in rodents, the forelimb and hindlimb representations indicated overlapping territories from tracer injections into forelimb and hindlimb spinal segments (Strick et al., 2021). Overlapping somatotopic representations may facilitate coordination between the forelimbs and hindlimbs of rodents, activation of a small cluster in M1 with overlapping somatotopic representations could simultaneously evoke movement in both limbs. In contrast, primates developed a need for independent control of the forelimbs and hindlimbs, which is reflected in the somatotopic organization of cortical motor areas (Strick et al., 2021).

### 1.1.2 New and old M1

A subset of descending projections in the CST makes direct monosynaptic connections to spinal motoneurons (CM cells). CM cells are uniquely found in primates (Bortoff & Strick, 1993), notably absent in cortex of the cat, rat, racoon or mouse (Lemon, 2008; Yang & Lemon, 2003). Though descending projections in other pathways (e.g., rubrospinal, reticulospinal) may form monosynaptic connections, the pattern and timing of CM cell activity appropriately corresponds to the muscle responses supporting voluntary, skilled movements (Lemon et al., 2004). The dexterous abilities of NHPs coupled with the homology of CM cells in primates further corroborates the importance of CM cells for dexterous movement.

Retrograde transneuronal tracer injections into single muscles localized the majority of CM cells in the caudal portion of M1 on the anterior bank of the central sulcus (Rathelot & Strick, 2009a). Whereas other CST projections terminate on spinal interneurons, which then synapse to motoneurons to control muscles, CM cells bypass circuitry inherent in the spinal cord by synapsing directly onto motoneurons. Though CM cells are entirely absent in less dexterous carnivores, approximately half of M1 CST projections constitute CM cells in macaque monkeys (Xu et al., 2024). The localized presence of CM cells in caudal M1 in certain primate species has been identified as an evolutionarily "new M1", which contrasts the "old M1" located rostrally on the cortical surface. The large volume of these cells with direct termination onto motoneurons of forelimb muscles emphasize the importance of new M1 in highly dexterous movements.

Comparable to old M1, new M1 also exhibits a somatotopic organization of the forelimb. The organization of the forelimb in new M1 roughly matches that of old M1; the proximal forelimb is most medial and the distal forelimb most lateral. However, there appears to be even greater spatial intermingling of the proximal and distal forelimb in new M1 compared to old M1 (discussed further in next section). In fact, CM cells innervating digit motoneurons were found in cortical zones that evoked shoulder movements when stimulated using ICMS (Rathelot & Strick, 2006a). Additionally, CM cells usually facilitate multiples muscles though the pattern of CM activity to the facilitated muscles usually deviates from simply acting as an agonist to these muscles (Griffin et al., 2015). The direct and complex facilitation of CM cells with motoneurons has recognized new M1, separately from old M1, as crucial to understanding the neural control of dexterous movements.

### 1.1.3 Intermingling of somatotopic representations

Early M1 studies that established the original somatotopic motor maps used direct electrical stimulation to evoke movements. Though the prevailing somatotopic M1 map conveys body representations as discrete zones with defined borders along the precentral gyrus, it is now well established that the somatotopic organization in M1 is not nearly so orderly. Additionally, the motor maps hypothesized from studies in the 1950s (body representations in discrete M1 locations) were not simply a product of archaic methodology, there are established anatomical features in M1 that facilitate coactivation of muscles. This is most apparent in the M1 forelimb representation where spatial intermingling of the arm and hand representation enables movement of both the arm and hand to be evoked from a single site in cortex.

Intracortical microstimulation (ICMS) enabled cortical stimulation with greater spatial resolution than via direct electrical stimulation. Nonetheless, multiple investigators noted overlapping cortical representations of different body parts (e.g., arm and hand) (Kwan, MacKay, et al., 1978a; Lemon, 1981; Sessle & Wiesendanger, 1982). Additionally, stimulation administered at a single site in M1 often activated more than one forelimb muscle  (Donoghue et al., 1992a). In other words, M1 projections may diverge and terminate on various motoroneurons, which may even span different somatotopic representations in M1. Conversely, any given muscle could be activated via cortical stimulation from multiple M1 locations—demonstrating convergence from wide-spread cortical locations in M1 onto a muscle (Schneider et al., 2001). The intermixing of somatotopic representations, namely the arm and hand, was hypothesized to facilitate coordinated movement by activating common “muscle synergies”. However, to corroborate these principles, methods aside from ICMS would have to be employed to verify that forelimb intermingling in M1 was a genuine principle and not a result of stimulation spread.

A series of experiments from Fetz & Cheney that evaluated the response of CM cells in the EMG activity of forelimb muscles established the principles of convergence and divergence of M1 projections to muscles (Fetz et al., 1976; Fetz & Cheney, 1980a). EMG activity of forelimb muscles was summed over a time interval centered on a spike recorded from a M1 neuron. Neurons projecting to any of the recorded muscles would demonstrate post-spike facilitation or inhibition in the average EMG activity, which lagged the cortical spike by 5-10 ms. The speed at which the facilitation occurred suggested that these projections were monosynaptic (later verified using antidromic stimulation). Apart from introducing this clever utility to study CM cells, Fetz & Cheney verified that CM cells often facilitated the activity of multiple muscles simultaneously (muscle fields) (Cheney & Fetz, 1985). This work was paramount in corroborating the divergence of CM cell projections and introducing a methodology to identify and investigate muscle fields.

A redefining study in regard to the somatotopic organization of the forelimb representation utilized stimulus-triggered averaging of EMG activity from 24 forelimb muscles at various sites within the M1 forelimb representation (Hudson et al., 2017; Park, Belhaj-Saïf, et al., 2001a). The strength of facilitation of the given muscles were mapped to the stimulation site in M1 and, ultimately, established a new standard mapping of the forelimb representation in M1. Namely, a core representation of the distal forelimb is incased along the central sulcus by a horseshoe shaped representation of the proximal forelimb. Moreover, between the proximal and distal forelimb representation, there was a separate representation defined as a proximal + distal representation. Spatial overlap in M1 is also noted between CM populations projecting to proximal and distal forelimb muscles (Rathelot & Strick, 2009a). The noted spatial intermingling between the distal and proximal forelimb representations further supports the concept of cortical synergies serving coordinated movements.

An additional way that M1 motor outputs are intermingled was elucidated from work by Huntley & Jones (1991). After using ICMS to map the M1 somatotopic organization, tracers were injected into sites that evoked thumb movements and generated maps showing discontinuous, patchy terminations that formed bidirectional networks throughout the M1 forelimb representation. Explicitly, horizontal interconnections bridged communication between representations of the proximal and distal forelimb. These labeled horizontal connections were specific to the forelimb representation; no intrinsic connections were found from sites representing the thumb to the face representation. Aside from demonstrating clear cross-talk between somatotopic representations of the forelimb, this study highlights the importance of considering the activity of intrinsic connections and other circuits apart from the descending projections in M1 to study the cortical control of movement.

As a final example, descending projections (including but not limited to CM cells) emanating from the semi-organized cortical somatotopy demonstrate intermingling of function as branching collaterals can diverge onto motoneurons comprising separate somatotopic representations, further scrambling the cortical somatotopic organization (Dum & Strick, 1991; Nathan & Smith, 1955). Diverging descending projections are another example of how cortical structure may serve to facilitate coordinated movements (Lemon & Morecraft, 2023; Morecraft et al., 2022). The convergence and divergence of cortical projections onto motoneurons, the spatial correspondence between somatotopic representations and the intrinsic and recurrent cortical connections in the M1 forelimb representation provide anatomical architecture in which complex, coordinated movements can be simplified. Overall, though the cortical somatotopy did not maintain the clean-cut organization originally depicted from earlier maps, core elements of the somatotopic organization have persisted through the century. Thus, while advancing technology and methodologies enabled further refinement and alteration of it, the somatotopic organization remains a crucial characteristic of M1.

## 1.2 M1 Functional Properties

### 1.2.1 The neural coding of M1 unit activity

As the knowledge and capability of recording neural activity improved from the 1930s-50s, Jasper, Ricci and Doane developed a technique to record single unit activity in behaving monkeys (Jasper, 1991). This marked a pivotal moment in the field of neuroscience as it introduced the ability to record neural activity concurrent with natural behavior. Following the paradigm of localization of function in cortex, investigators could now validate previous conjectures regarding cortical function—hypothesized from lesion and stimulation studies— by correlating neural activity to behavioral events. Quantifying the amount of information in a unit response regarding certain behavioral events or stimuli provide insight about a units' "neural code", which is formulated using information theory (Borst & Theunissen, 1999). Perhaps one of the most recognizable implementations of utilizing the neural code to characterize a cortical structure and function relationship came from Hubel and Weisel's 1960s investigations.

Upon discovering that units in the striate cortex of cats selectively fired to dots of light at specific locations of the cat's visual field, Hubel and Weisel demonstrated how cortical circuitry and organization could induce visual processes (Hubel & Wiesel, 1959). Briefly, they proposed that neurons recorded from various depths in a given cortical location coded the same orientation of light bars, whereas neurons recorded from adjacent sites across the cortical surface coded different orientations (Hubel & Wiesel, 1962). Hubel and Weisel proposed that these cortical columns coded one value (e.g., 90°) of a given variable (e.g., orientation) to form a hypercolumn. The hypercolumn represented the functional building block of cortex; interconnected modules throughout cortex that process and relay information. Though the validity of cortical columns and hypercolumns remain the subject of much debate (Horton & Adams, 2005; Ts'o et al., 2009), these studies mark a shift in neuroscience from speculating on the function of certain cortical processes based on anatomical principles to investigating cortical organization by analyzing cortical activity.

Around the time that Hubel and Weisel analyzed unit activity of visual cortex, the successful recording of unit activity in behaving monkeys (Jasper, 1991) introduced the possibility to record and analyze single unit activity concurrent with behavior. Subsequently, the spiking activity of individual neurons in M1 could be correlated with muscle activity or kinematic variables to investigate the neural code of movements. Naturally, as CM cells had been identified to project directly onto spinal motoneurons, investigators found M1 neurons with spiking rates that correlated to the EMG activity of certain muscles (Fetz & Finocchio, 1971). And though this finding offers an intuitive and valid basis for the neural encoding of movement, subsequent reports demonstrated the complication in defining the neural coding of movements as the equivalent of muscle activity.

It is straightforward to imagine that M1 neurons simply modulate activity in accordance with the use of the muscle it terminates onto, such that as the required use of a certain muscle increases, the neural activity from that neuron increases (Evarts, 1966). However, it became clear that M1 unit coding was much more complicated than a direct mapping of muscle responses (Humphrey, 1972). For example, whereas some M1 neurons fired in a phasic manner that correlated with muscle activity, other neurons fired tonically even in the absence of any movement (Evarts, 1968). Additionally, CM cells indicated activity modulation that was correlated both positively and negatively to EMG activity (Maier et al., 1993). Ultimately, this led researchers to consider alternative properties that M1 units could encode.

The ability to record M1 unit activity while subjects executed movements introduced alternative perspectives regarding the encoding M1 activity. A principal investigation demonstrated that M1 neural activity correlated with exerted force and others correlated with the direction of movement (Evarts, 1968). Resulting from this work, M1 movement encoding could be considered in relation to behavior that varied qualitatively (e.g., displacement direction) or quantitatively (e.g., exerted force), as opposed to muscle activity. The appeal to this hypothesis is that it simplifies the encoding from muscles, which is burdened by interactions between end effectors resulting in a limitless combination of responses to achieve the same movement, to movements, which ultimately *is* the end effector. Therefore, although this paradigm generally ignored time-variant information inherent in the muscles that implement voluntary movements, considering the encoding of M1 activity as a variable related to movement, led to revolutionary work that maintains a strong influence in M1 research today.

One of the most renowned experiments regarding M1 activity came from Georgopoulos in the 1980s. In his famous reports, Georgopoulos found that M1 units preferentially fired to arm movements in a certain direction and the population response of these directionally tuned units could accurately predict the resulting movement direction of the arm (Georgopoulos et al., 1986). Explicitly, NHPs were instructed to move their arm from a central fixation point to a radial target. When the preferred direction of neurons was modeled by cosine tuning functions, the summed population response of M1 neurons accurately reported the implemented movement direction of the arm. Not only was this work influential in shaping subsequent studies that decode neural activity using brain computer interfaces (Lebedev & Nicolelis, 2006), but it was also instrumental in considering M1 encoding at the level of population activity, rather than of individual units. Though M1 unit "encoding" is not definitively defined, it undoubtedly demonstrates strong correlations to both muscle activity and movement parameters.

### 1.3 Investigating the Relationship Between M1 Structure and Function

As previously discussed, the relationship between cortical structure and function has been leveraged to help elucidate how the brain encodes certain processes. Logically, we can assume that the same investigation in M1 would help to understand the neural coding of dexterous movements. However, the structural (i.e., spatial) and functional (i.e., temporal) relationship in M1 is often overlooked. Yet, the combined knowledge of separate investigations that study M1 structure and function indicate that there is a clear relationship between the two. For example, tracer studies in M1 indicated strong horizontal M1 connections within ~1 mm of an injection site (Capaday et al., 2009). These local, intrinsic connections give a basis for similar preferred directions of neurons within close proximity to each other (Merchant et al., 2012). Very few studies have considered both the characteristics of M1 activity according to its spatial location and temporal response.

Investigations into the spatiotemporal M1 activity of dexterous movements are limited. However, two studies recorded M1 unit activity in the forelimb representation and concluded that the coding to reach and grasp was intermixed throughout the M1 forelimb representation (Rouse & Schieber, 2016a; Saleh et al., 2012a). These investigations recorded unit activity from Utah arrays, which limited the spatial sampling from M1. Additionally, another study investigating the unit activity in zones of the arm and hand representation found intermingling of reach and grasp functional properties that was distinct between the arm and hand representation (Friedman, Chehade, et al., 2020a).

The objective of this dissertation is to better understand the spatiotemporal organization of M1 unit activity that underlies dexterous movements. Specifically, I will quantify the ratio of M1 that is active during reach-to-grasp movements by imaging the intrinsic signal of cortex while NHPs implement reach-to-grasp movements. The intrinsic signal measures hemodynamic changes in cortex to indirectly denote locations with increased neural activity. In contrast to physiology recordings, cortical imaging enables contiguous surveil of cortex, which provides a comprehensive perspective of the spatial spread of cortical activity in M1. Additionally, using ISOI in M1 enables evaluation of signal changes at a spatial resolution of approximately 100 microns, which is spatially acute enough to distinguish between and within somatotopic representations in M1. The following chapter discusses the experiments and results that I acquired to address the spatial organization of M1 activity during reach-to-grasp movement.

The succeeding section addresses the functional organization of M1 activity. Briefly, I recorded from 100+ sites in M1 of two NHPs to acquire a sample of ~1500 single units while NHPs conducted a reach-to-grasp task. By recording the spiking activity of single neurons, I can evaluate the instantaneous firing rate (FR) of neurons during certain phases of reach-to-grasp movement (e.g., reach or grasp phase). I assess the organization of units that selectively modulate activity, or not, to the reach or grasp phase. The high temporal resolution from single unit recordings enables distinguishing between neural activity related to reach and grasp, which is indecipherable during imaging due to the slow temporal dynamics of the intrinsic signal. Whereas the spatial resolution is superior (in terms of continuity) in imaging experiments. This multimethod approach offers powerful interpretations when each approach yields the same result pertaining to the spatiotemporal organization of M1 activity.

# 2.0 The Spatial Organization of Neural Activity in Motor and Premotor Cortex[1]

## 2.1 Introduction

Motor (M1) and dorsal premotor (PMd) cortex in monkeys are widely used as models for studying cortical control of movement. Both cortical areas contain complete motor representations of the forelimb (Boudrias et al., 2010a; Gould et al., 1986; Kwan, MacKay, et al., 1978b; Raos et al., 2003). Somatotopy of the forelimb representations varies across studies, but there is consensus that muscles and cortical columns do not have a one-to-one mapping. Instead, an arm or a hand muscle receives corticospinal inputs from a population of cortical columns (Andersen et al., 1975; Rathelot & Strick, 2006b). Similarly, a cortical column can influence activity in several muscles (Donoghue et al., 1992b; Fetz & Cheney, 1980b; Lemon et al., 1987). The convergence and divergence of corticospinal projections means that the same motor output is present in many locations within a forelimb representation. For example, intracortical microstimulation (ICMS) evokes shoulder flexion from many M1 sites where the affiliate corticospinal projections target the same group of arm muscles (Park, Belhaj-Saïf, et al., 2001b). But how can a seemingly simple motor map control a rich repertoire of arm and hand actions?

To address this question, we must first understand the spatial relationship between neural activity that supports arm and hand actions (i.e., function), and the forelimb motor maps (i.e., structure). We consider two different perspectives. The first perspective is that neural codes for arm and hand actions are not spatially organized within the forelimb motor maps. Instead, neural activity affiliated with a particular function (e.g., reach) is present throughout the arm zones, or even throughout entire forelimb representations. This scheme has support in electrophysiological studies that examined relationships between neural coding, recording site location, and forelimb behavior (Rouse & Schieber, 2016b; Saleh et al., 2012b; Vargas-Irwin et al., 2010a). Nevertheless, most electrophysiological investigations pool results across recording sites, offering only limited insight into the spatial organization of functions. An alternative perspective is that neural codes for arm and hand actions are more spatially organized within the forelimb motor maps. Here, neural activity affiliated with a particular action (e.g., reach) is concentrated in subzones within the forelimb representations. This spatial organization of function is consistent with maps obtained using long train ICMS (500 ms), which approximates the duration of natural actions (M. S. A. Graziano et al., 2002a). The same motor mapping parameters revealed functional subzones in both frontal and parietal cortical areas (Cooke & Graziano, 2004; Gharbawie et al., 2011a; M. S. A. Graziano et al., 2002a; Stepniewska et al., 2005).

---

[1] *The contents of this chapter were previously published* (Chehade & Gharbawie, 2023).

Adjudicating between the two perspectives can be facilitated with neuroimaging because it provides uninterrupted spatial sampling and retains the spatial dimension of the recorded activity. The necessary implementation here is to image entire forelimb representations on the cortical surface during arm and hand actions then quantify the spatial organization of movement-related cortical activity. Intrinsic signal optical imaging (ISOI), 2-photon imaging, and fMRI have been successfully used in measuring cortical activity related to arm and hand actions in monkeys (Ebina et al., 2018a; Friedman, Chehade, et al., 2020b; Kondo et al., 2018a; Nelissen & Vanduffel, 2011a). The mesoscopic field-of-view in ISOI is particularly well-suited for the size of the forelimb representations in M1 and PMd. Moreover, ISOI affords high spatial resolution and contrast without extrinsic agents (e.g., GCaMP, dyes, MION). These reasons motivated us to use ISOI to measure M1 and PMd activity in head-fixed macaques engaged in an instructed reach-to-grasp task (Figure 1A-D). In the same field-of-view, we used ICMS for high density mapping of the forelimb representations and surrounding territories. We then quantified the spatial overlap between cortical activity determined from ISOI and the forelimb motor map. We reasoned that if the neural code for reaching and grasping is not spatially organized within the forelimb representation, then ISOI should report diffuse task-related activity that overlaps most of the M1 and PMd forelimb representations (Figure 1B top). Alternatively, if the neural code is more spatially organized, then ISOI should report focal task-related activity in subzones of the M1 and PMd forelimb representations (Figure 1B bottom).

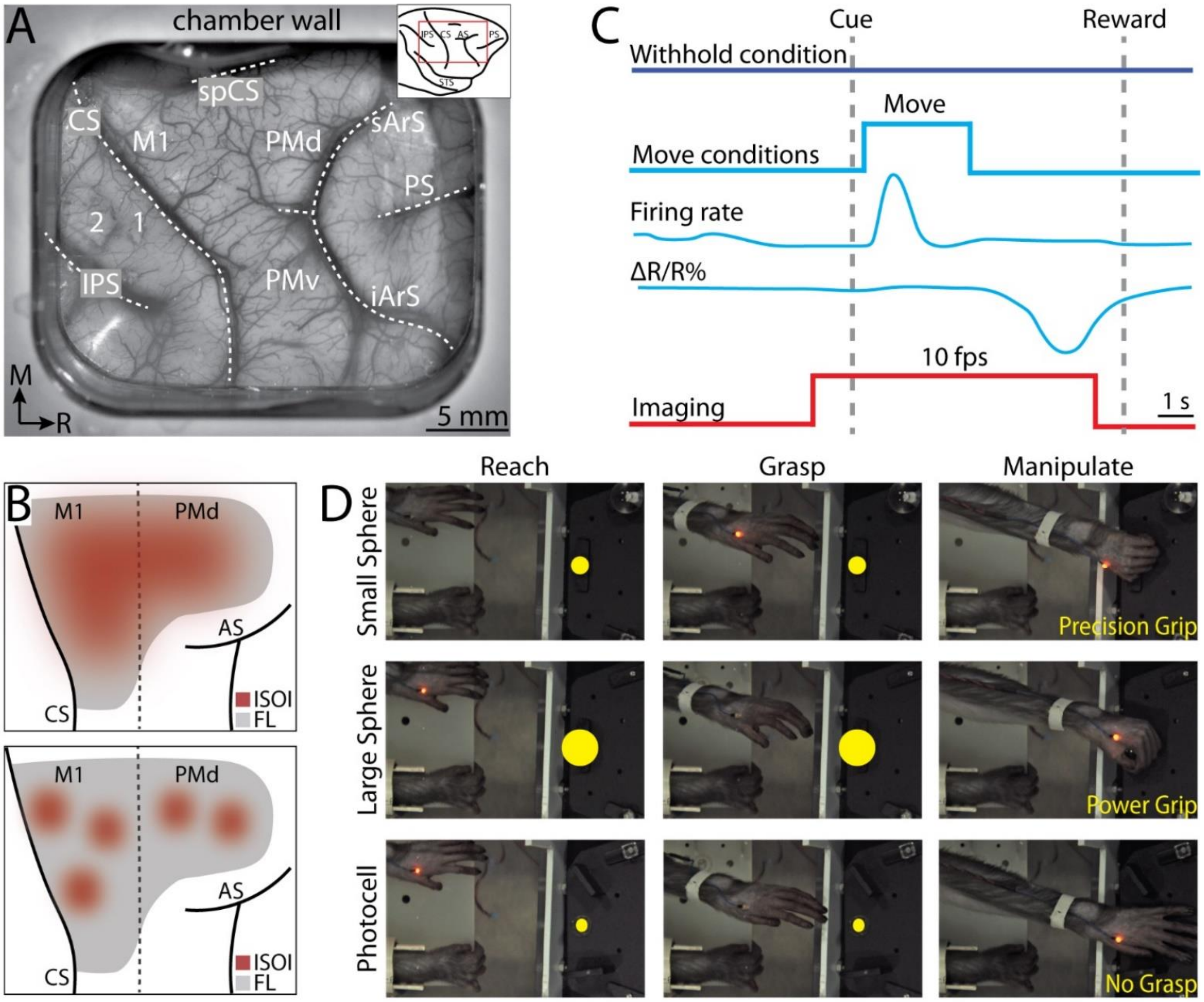

**Figure 1. Intrinsic signal optical imaging during forelimb arm and hand actions (A) Chronic recording chamber provides access to motor and somatosensory areas in the right hemisphere. Native dura was replaced with transparent membrane. Dashed lines mark central sulcus (CS), intraparietal sulcus (IPS), and arcuate sulcus (AS). Inset shows approximate location of chamber (red rectangle). (B) Schematic of potential results. Dotted line separates M1 from premotor areas. Forelimb representation (FL) is gray. Red patches are clusters of pixels that darkened (i.e., negative reflectance) after task-related increase in neural activity. (Top) Pixels reporting activity are in a large patch that overlaps most of FL. (Bottom) Pixels reporting activity are in several small patches that collectively overlap a smaller portion of FL than the patch in A. (C) Relative timing in task conditions. Blue square pulse indicates movement period, whereas there was no movement in the withhold condition. Increase in neural firing coincides with movement and precedes reflectance change (ΔR/R%) measured with ISOI. Red pulse depicts ISOI acquisition in all conditions (10 frames/s). (D) Still**

**frames from 3 phases (columns) in the 3 movement conditions (rows). Task was performed with the left forelimb and the right forelimb was restrained. Yellow circles were not visible to the monkey but were digitally added here to facilitate visualization of the three targets.**

## 2.2 Methods

### 2.2.1 Animals

The right hemisphere was studied in two male macaque monkeys (Macaca mulatta). Monkeys were 6-7 years old and weighed 9-11 kg. All procedures were approved by the University of Pittsburgh Animal Care and Use Committees and followed the guidelines of the National Institutes of Health guide for the care and use of laboratory animals.

### 2.2.2 Head post and recording chamber

After an animal acclimated to the primate chair and training environment, a head-fixation device was secured to the occipital bone and caudal parts of the parietal bone. Task training with head-fixation started after ~1 month (monkey G) or ~9 months (monkey S) and lasted for ~22 months. At the end of this training period, the monkey was considered ready for optical imaging. A craniotomy was performed for implanting a chronic recording chamber (30.5 x 25.5 mm internal dimensions) over motor and somatosensory cortical areas. The chamber was secured to the skull with ceramic screws and dental cement. Within the recording chamber, native dura was resected and replaced with a transparent silicone membrane (500 µm thickness) that we fabricated from a mold. Protocols for the artificial dura have been previously described in detail (Arieli & Grinvald, 2002; Ruiz et al., 2013). The walls of the artificial dura lined the walls of the recording chamber. The floor of the artificial dura (i.e., the optical window) was flush with the surface of cortex and facilitated visualization of cortical blood vessels and landmarks (Figure 1A). The walls and floor of the artificial dura delayed regrowing tissues from encroaching underneath the optical window. Single electrodes and linear electrode arrays were readily driven through the optical window without permanent deformation.

### 2.2.3 Reach-to-grasp task

Monkeys performed a reach-to-grasp task while head fixed in a primate chair. The left forelimb was used, and the right forelimb was secured to the waist plate. The task apparatus was positioned in front of the animal. A stepper motor rotated the carousel in between trials to present a target ~200 mm from the start position of the left hand. The relative position of the target made it directly visible and was also supposed to encourage consistent reach trajectories across conditions and trials. Task instruction was provided with LEDs mounted above the target. Photocells were embedded in multiple locations within the apparatus to monitor hand location and target manipulation. An Arduino board (Arduino Mega 2560, www.arduino.cc) running a custom script (1 kHz) controlled task parameters, timing, and logged the monkey’s performance on each trial. The task involved four conditions.

#### 2.2.3.1 Two reach-to-grasp conditions.

In a successful trial, the animal had to reach, grasp, lift, and hold a sphere. The small sphere condition (12.7 mm diameter) and the large sphere condition (31.8 mm diameter) were used to motivate precision and power grips, respectively (Figure 1D). Spheres were attached to rods that moved in a vertical axis only. Task rules were identical for both conditions and are therefore described once. To initiate a trial, an animal placed its left hand over a photocell embedded in the waist plate. Covering the photocell for 300 ms turned on an LED, which signaled the start of the trial. Holding this start position for 5000 ms triggered the Go Cue, which was a blinking LED. The animal had 2400 ms (monkey G) or 2550 ms (monkey S) to reach, grasp, and lift the sphere; time limits were also set for each phase. Lifting the sphere by 15 mm turned the blinking LED solid, which signaled the end of the lift phase. Maintaining the lifted position for 1000 ms turned off the LED, which instructed the animal to release the object and return its hand to the start position within 900 ms. Maintaining the start position for 5000 ms triggered a tone and LED blinking. After an additional 2000 ms in the start position the trial was considered successful; tone and LEDs turned off and water reward was delivered. The animal could not initiate a new trial for another 3000 ms. Failure to complete any step within the allotted time window resulted in an incorrect trial signaled by a 1500 ms tone and a 5000 ms timeout in which the apparatus was unresponsive to the monkey's actions. After the timeout, a new trial could be initiated with hand placement in the start position. Across both monkeys, the median failure rate per session was 18% (IQR = 13-39%) in the precision grip condition and 16% (IQR = 9-28%) in the power grip condition.

**2.2.3.2 Reach-only condition.**

The target was a photocell embedded into the surface of the carousel. The photocell was visible to the monkey but was not graspable. The Go Cue was the same as the one used in the reach-to-grasp conditions, but here it prompted the monkey to reach and place its hand over the photocell. The hand had to cover the photocell for >220 ms (monkey G) or >320 ms (monkey S).

All other task rules and steps were identical to the reach-to-grasp condition. Across both monkeys the median failure rate per session was 12% (IQR = 4-32%).

#### 2.2.3.3 Withhold condition.

In a successful trial, the monkey had to maintain its hand in the start position for ~10 s. Trial initiation was identical to the other conditions. Holding the start position for 5000 ms triggered the Withhold Cue, which was distinctly different from the Go Cue in the movement conditions. Maintaining the start position for another 2800 ms triggered a tone and LED blinking. After an additional 2000 ms in the start position the trial was considered successful and rewarded. Removing the hand from the start position at any time resulted in an incorrect trial and the same consequences described in a failed reach-to-grasp trial. Across both monkeys the median failure rate was 12% (IQR = 7-19%).

Conditions were presented in an event-related design (1 successful trial/condition/block). Condition order was randomized across blocks. We structured the trials and the inter-trial interval so that the hand would remain in the start position for ~13 s in between trials. Thus, in a successful trial from a movement condition, the arm and hand remained still in the start position for ~13 s and then moved for ~1-2 s. This relatively long period without movement was useful for relating the changes in intrinsic signal (slow) to movement onset (rapid). In the withhold condition, the arm and hand were in the start position for ~10 s and no movement was allowed.

### 2.2.4 Muscle activity

Electromyography (EMG) was conducted in 7 forelimb muscles on sessions that did not involve ISOI or other neural recordings. After head fixation, the monkey was lightly sedated with a single dose of ketamine (2-3 mg/kg, IM). Sedation was confirmed from a reduction in voluntary movements with the working forelimb. At that point, pairs of stainless-steel wires (27 gauge, AM Systems) were inserted percutaneously into each muscle (~15 mm below skin). Three arm muscles were targeted (1) deltoideus, (2) triceps brachii, and (3) biceps brachii. Four extrinsic hand muscles were targeted (1) extensor carpi radialis brevis, (2) flexor carpi radialis, (3) extensor digitorum 4-5, and (4) flexor digitorum superficialis. The task started 45-60 min after sedation. The monkey was fully alert by that point and showed no lingering effects of sedation. The non-working forelimb was restrained and therefore could not tamper with the EMG wires.

EMG signals were filtered (bandpass 15-350 Hz) and digitized (2 kHz) using a dedicated processor (Scout model, Ripple Neuro, Salt Lake City, UT). Recorded signals were segmented into trials and their power spectral density was estimated with a discrete Fourier transform (MATLAB *fft* function, Natick, MA). Trials with power >7 μV2 in the 1-14 Hz range were presumed to have artifact and were excluded from further analysis. EMG signals were rectified and smoothed with a 100 ms sliding window (MATLAB *filtfilt* function). Finally, for each muscle in each movement condition, the average signal was computed across trials and sessions (308-595 trials/muscle).

### 2.2.5 Joint kinematics

Forelimb joints were tracked in 3D on sessions that did not involve ISOI or other neural recordings. Before the task started, 6 LEDs were secured to sites on the left forelimb to track up to two joints/session. A motion capture system (Impulse X2, Phasespace) outfitted with six cameras recorded LED positions and logged x, y, z coordinates (480 Hz). For each tracked joint,

LEDs were configured to form imaginary vectors or planes. For example, to track the elbow, 3 LEDs were placed in a triangular formation on the upper arm (i.e., parallel to humerus) and another 3-LED formation was placed on the forearm (i.e., parallel to ulna). Elbow flexion/extension was calculated as the angle between the vector aligned with the upper arm and the vector aligned with the forearm. Pronation/supination was calculated as the angle between the normal of the plane of the upper arm and the normal of the plane of the forearm. A similar approach was adopted for (1) shoulder flexion/extension, (2) shoulder abduction/adduction, (3) wrist flexion/extension, and (4) digits flexion/extension. Time series of LED coordinates were segmented into trials. Trials were excluded from analyses if LED positions were not logged for >2% of trial duration. Such dropouts were typically due to LED occlusion by the forelimb, primate chair, or grasp apparatus. Kinematic profiles were calculated trial-by-trial and then averaged for each condition (143-405 trials/degree of freedom).

#### 2.2.6 Movement quantification

EMG and kinematics were quantified trial-by-trial. Time-resolved traces were generated for each recorded muscle and degree of freedom. The area under the curve (AUC) was calculated with trapezoidal approximation (MATLAB *trapz* function) applied to the time-resolved traces. We focused on the period from Cue onset until the end of forelimb withdrawal when the hand returned to the start position. For each muscle and degree of freedom, AUC values were standardized across task condition. Standardization was done separately for each session to account for recording variations across sessions. The average AUC for a muscle, or a joint, was calculated as the median of all trials acquired for a task condition. The average AUC for a group of muscles, or a group of joints, was calculated as a mean weighted by the number of trials.

### 2.2.7 Motor mapping

We used intracortical microstimulation (ICMS) to map the somatotopic organization of frontal motor areas. In monkey S, all sites (n=158) were investigated with a microelectrode in dedicated motor mapping sessions. We used the same approach in >50% of the sites (n=118) in monkey G. The remaining sites (n=99) were mapped with a linear electrode array at the end of electrophysiological recordings that will be presented in a separate report. In the dedicated motor mapping sessions, the monkey was head-fixed in the primate chair and sedated (ketamine, 2-3 mg/kg, IM, every 60-90 minutes). This mild sedation reduced voluntary movements but did not suppress reflexes or muscle tone. A hydraulic microdrive (Narishige MO-10) connected to a customized 3-axis micromanipulator was attached to the recording chamber for positioning a tungsten microelectrode [250 μm shaft diameter, impedance = 850 ± 97 kΩ (mean + SD)] or a platinum/iridium microelectrode [250 μm shaft diameter, impedance = 660 ± 153 kΩ (mean + SD)]. A surgical microscope aided with microelectrode placement in relation to cortical microvessels. The microelectrode was in recording mode at the start of every penetration. Voltage differential was amplified (10,000×) and filtered (bandpass 300–5000 Hz) using an AC Amplifier (Model 2800, AM Systems, Sequim, WA). The signal was passed through a 50/60 Hz noise eliminator (HumBug, Quest Scientific Instruments Inc.) and monitored with an oscilloscope and a loudspeaker. As the electrode was lowered, the first evidence of neural activity was considered 500 μm below the pial surface. The microelectrode was then switched to stimulation mode and the effects of ICMS were tested at >4 depths (500, 1000, 1500, 2000 μm). Microstimulation trains (18 monophasic, cathodal pulses, 0.2 ms pulse width, 300 Hz) were delivered from an 8-Channel Stimulator (model 3800, AM Systems). Current amplitude, controlled with a stimulus isolation unit (model BSI-2A, BAK Electronics), was increased until a movement was evoked (max 300 μA). The stimulation threshold for each depth was the current amplitude that evoked movement on 50% of stimulation trains.

One experimenter controlled the location and depth of the microelectrode. A second experimenter, blind to microelectrode location, controlled the microstimulation. Both experimenters inspected the evoked response and discussed their observations to reach consensus about the active joints (i.e., digits, elbow, etc.) and movement type (flexion, extension, etc.). Movement classification was not cross-checked against EMG recording or motion tracking. The overall classification for a given penetration included all movements evoked within 30% of the lowest threshold across depths. The location of each penetration (500-1000 μm apart) was recorded in relation to cortical microvessels. Color-coded maps were generated from this data using a voronoi diagram (MATLAB *voronoi* function) with a maximum tile radius of 750 μm (Figure 2B & 2F). The rostral border of M1 was marked to separate sites with thresholds <30 μA from higher threshold sites (Figure 2C & 2G).

The mapping approach was similar for penetration sites stimulated with a linear electrode array (32 or 24 channels, 15 μm contact diameter, 100 μm inter-contact distance, 210-260 μm probe diameter; V-Probe, Plexon). Each penetration was mapped ~2.5 hours after the linear array was inserted into cortex, which was also the end of electrophysiological recordings during task performance. Only 1 penetration was mapped per session. Microstimulation parameters were identical to the ones used with the microelectrode but were controlled here using the Trellis Software Suite (Ripple Neuro). Channels were stimulated one at a time and every other channel was used. One experimenter controlled the microstimulation and classified the evoked movements.

### 2.2.8 Intrinsic signal optical imaging

The field-of-view (FOV) was illuminated with three independently controlled red LEDs (630 nm wavelength). Each LED was outfitted with a lens to diffuse the light emitted. The experimenters optimized LED positions and brightness with the aid of a real-time heatmap of the field-of-view. Camera frames 768x768 pixels (monkey G) or 1080x1310 pixels (monkey S) were captured with a 12-bit CMOS sensor (Photon Focus, Lachen, Switzerland). The tandem lens combination achieved a FOV of 15x15 mm or 22x26 mm, both at 20 $\mu m^2$/pixel. Frames were temporally averaged from 100 frames/s to 10 frames/s then saved. Image acquisition and parameters were controlled with an Imager 3001 system (Optical Imaging Ltd, Rehovot, Israel). In every trial, imaging started 1 s before Cue onset and lasted for 7 s unless otherwise stated. For spatial reference, a high contrast image of the cortical surface was captured with green illumination (528 nm wavelength) at the start of every session.

### 2.2.9 Image processing

Image processing was conducted on individual imaging sessions. Data frames were rigid-aligned in x, y coordinates (MATLAB *estimateGeometricTransform* function) to a reference frame from the middle of the session. A trial was excluded if any frame was out of register by >10 pixels (i.e., >200 μm). In the remaining trials, the first 10 frames (-1.0 to 0 s from Cue) of a trial were averaged then subtracted from all frames in the same trial. This subtraction converted pixel values to reflectance change with respect to baseline, which effectively normalized every trial to itself. To correct uneven illumination and residual motion artifact, frames were processed with a high-pass median filter (kernel = 250 pixels). Justification pertaining to the high-pass filter kernel size is discussed in Appendix 3. A low-pass Gaussian filter (kernel = 5 pixels) was used for smoothing. To accelerate the spatial filtering computations, frames were temporarily down sampled by a factor of 4.

Frames from different sessions were aligned to a common reference, which was a high contrast image of the cortical surface. For every imaging session, we marked 25-60 points that were apparent in the microvessel patterns from that session and in the common reference. These points were used to construct a mesh grid for the reference and session images. Non-rigid transformation (multilevel B-spline approximation) was then applied to fit the session mesh grid to the reference mesh grid (Koon, 2008; Lee et al., 1997; Rueckert et al., 1999). The transformation matrix was then used to co-register all frames from the session to the common reference. Frames could then be averaged within a session and across sessions. To enhance visualization of average frames, the distribution of pixel values was clipped in relation to the median pixel value. Clipping was excluded, however, from time courses and thresholded maps.

### 2.2.10 Reflectance change time course

Time courses were generated from two types of regions-of-interest (ROIs). (1) Small circles (radius = 20 pixels [400 μm]) that were placed in arm zones and hand zones in M1 and PMd. (2) A larger ROI that included the M1 and PMd forelimb representations. In both cases, pixel values within an ROI were averaged to obtain 1 value per frame. Pixels that overlapped blood vessels were excluded. Time courses were based on trial-averaged time series.

In the present illumination (630 nm, i.e., red), pixel darkening is accepted as a lagging indicator of neural activity (Grinvald et al., 1986). The increased consumption of oxygen by local neural activity is believed to increase deoxyhemoglobin concentrations, which absorbs red light and is therefore detected as pixel darkening (Malonek & Grinvald, 1996; Shtoyerman et al., 2000). Some have likened the early pixel darkening in ISOI to the initial dip in BOLD fMRI (Ances, 2004; Kim et al., 2000; Menon et al., 1995). Nevertheless, there is evidence from multi-spectral imaging to suggest that pixel darkening in ISOI is a more complicated response driven by changes in total hemoglobin, blood volume, and blood flow (Sirotin & Das, 2009). In contrast, pixel brightening reports increase in the concentration of oxygenated hemoglobin and blood flow/volume and may therefore resemble the increase in BOLD fMRI (Chen-Bee et al., 2007).

### 2.2.11 Thresholded activity maps

To identify pixels that darkened in response to task-related movements, we compared frames from the end of movement (i.e., movement frames) to frames acquired before movement onset (i.e., baseline frames). Every trial contributed 1 movement frame and 1 baseline frame (8 sessions, 236-432 trials/condition). A baseline frame was an average of the ten data frames acquired from -1.0 to Cue. Depending on the analysis, a movement frame was an average of 39 data frames, or an average of 5 data frames, acquired after movement completion.

A two-sample t-test was then conducted pixel-by-pixel to compare values between movement frames and baseline frames. Right tail values ($p<1e^{-4}$) indicated that the affiliate pixels brightened significantly in movement frames as compared to baseline frames. Left tail values ($p<1e^{-4}$) indicated that pixels darkened significantly in movement frames as compared to baseline frames. Pixels were also thresholded after Bonferroni correction for multiple comparisons ($p<1e^{-7}$); here was based on the number of pixels (excluding blood vessels) in the M1 and PMd forelimb representations. For both threshold levels, pixels that darkened significantly were considered the constituents of the thresholded activity map. To denoise those maps, significant pixels were removed if they did not belong in a group with >10 connected significant pixels.

### 2.2.12 Thresholded map quantification

Thresholded activity maps and the motor maps were both registered to the common reference. Pixels within the M1 and PMd forelimb representations were counted. Those pixel counts were then expressed as a percentage of the total number of pixels that make up the forelimb representations. The analysis excluded pixels that overlapped major blood vessels.

### 2.2.13 Paired comparisons of condition maps

Cortical activity was directly compared between condition pairs in two ways. (1) Two sample t-test as described earlier but deployed here for comparing movement frames between conditions. (2) Cross-correlation between condition pairs. For each condition, an average time series (n=70 frames) was generated from all trials. Each average frame was then converted into a 1-dimensional array. Pairs of arrays from different conditions but matched time points, were cross-correlated at zero lag (e.g., power grip condition Frame 1 X reach-only condition Frame 1). Thus, the full comparison between two conditions returned 70 correlation coefficients. We used the motor map to constrain the analysis to the M1 and PMd forelimb representations.

### 2.2.14 Statistical analyses

All analyses were done in MATLAB (*function name*). We used parametric tests after verifying that our data met parametric assumptions. Specifically, we confirmed that the variance was comparable between groups (*vartestn*) and that the data was normally distributed, which we established from skewness (*skewness*) and kurtosis (*kurtosis*) values between -1 and +1. We used paired t-tests (*ttest2*) for the pixel-by-pixel comparisons used to generate thresholded activity maps. We used one-way ANOVA (*anova1*) for condition comparisons of imaging time series and metrics of forelimb use. Post hoc comparisons (*multcompare*) were follow-up t-tests corrected by the number of comparisons.

## 2.3 Results

Two monkeys performed an instructed forelimb task that consisted of 4 conditions: (1) reach-to-grasp with precision grip, (2) reach-to-grasp with power grip, (3) reach-only, and (4) withhold (Figure 1D). Both monkeys completed 49 successful trials/condition/session (median; IQR = 37-51 trials [monkey G], IQR = 45-51 trials [monkey S]). We used intrinsic signal optical imaging (ISOI) to measure cortical activity during task performance (Figure 1). The spatial extent of activity in motor (M1) and dorsal premotor (PMd) cortex was quantified in relation to motor maps derived with intracortical microstimulation (ICMS) from the same recording chambers. Measurements of joint angles and EMG activity during the task provided context for condition differences in cortical activity.

### 2.3.1 Consistent somatotopy in M1 and PMd

To find the forelimb representations in M1 and PMd, we used ICMS to map frontal cortex from central sulcus to arcuate sulcus (Figure 2A & 2E). The general organization of the motor map was consistent between monkeys (Figure 2B-D & 2F-H). Most ICMS sites evoked forelimb movements (Figure 2B & 2F). We marked the rostral border of M1 such that (1) it was 3-5 mm from the central sulcus, and (2) separated high threshold sites (>30 µA) from low threshold sites (Figure 2C & 2G). To simplify the motor map, site classifications were consolidated into broader categories (e.g., elbow and shoulder became arm). The simplified maps (Figure 2D & 2H) showed that the forelimb representation was flanked medially by trunk zones and laterally by trunk and face zones. Within the M1 forelimb representation, the main hand zone (i.e., digits and wrist) was surrounded by an arm zone, or an arm and trunk zone. This nested organization is consistent with previous maps from macaque monkeys (Kwan, MacKay, et al., 1978a; Sessle & Wiesendanger, 1982) including maps obtained with stimulus triggered averaging of EMG activity (Park, Belhaj-Saïf, et al., 2001). The organization of the forelimb representations was less clear in PMd and PMv, which could have been related to higher thresholds, less extensive mapping, as well as more overlap between arm and hand zones than in M1 (Boudrias et al., 2010b).

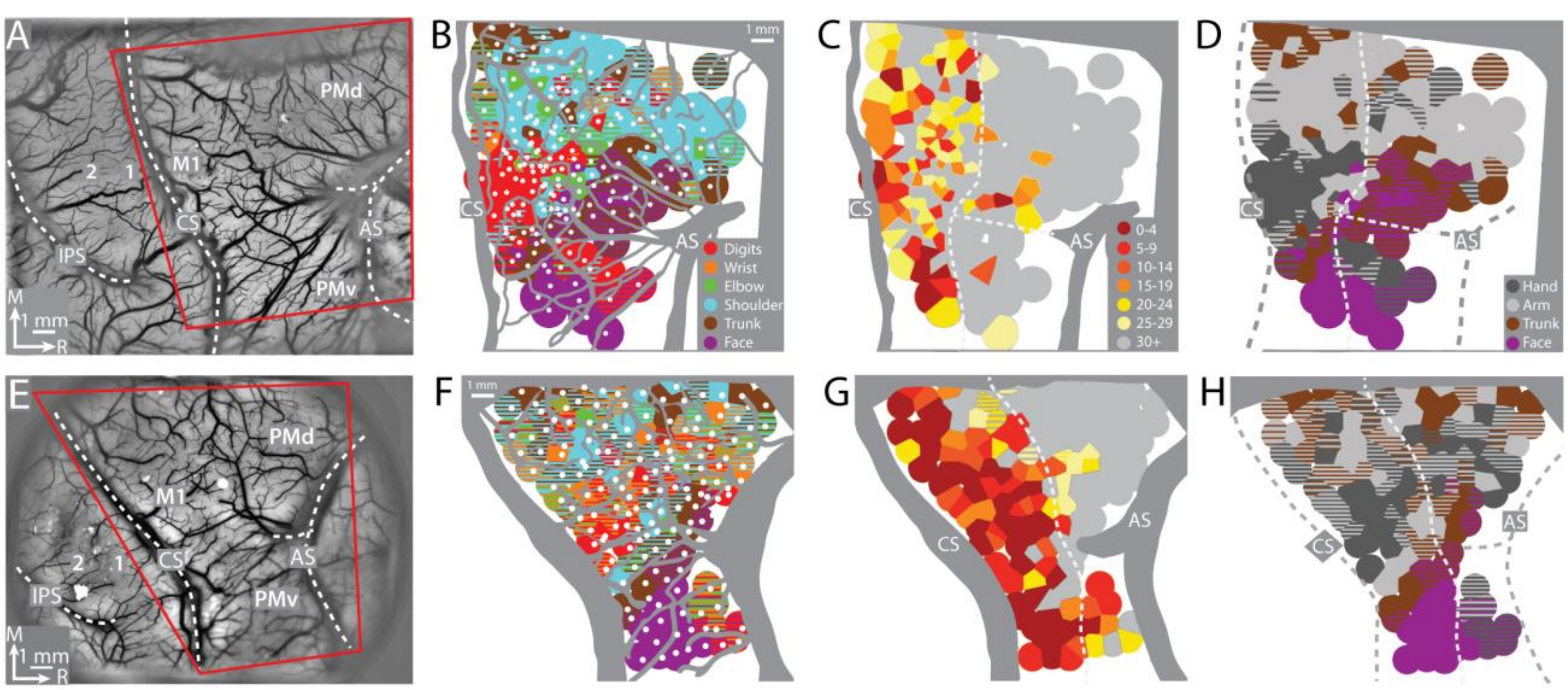

**Figure 2 Motor map organization in M1 and PMd (A-D) Right hemisphere of monkey G (A) Cropped image of a chronically implanted recorded chamber. Blood vessels and cortical landmarks are visible through the transparent membrane. Red outline is the field-of-view in subsequent panels and figures. (B) Major blood vessels and chamber walls are masked in gray. White dots are intracortical microstimulation (ICMS) sites (n = 218). Voronoi tiles (0.75 mm radius) are color-coded according to ICMS-evoked movement. Striped tiles represent dual movements. (C) Same motor map from (B) colored according to current amplitude (μA) for evoking movements. Border between M1 and premotor cortex is drawn at the transition from low (<30 μA) to high (≥30 μA) current thresholds. (D) Same motor map as (A), but here wrist and digit sites are classified as hand, and shoulder and elbow sites are classified as arm. (E-H) Same as top row, but for right hemisphere of monkey S. Motor map has 158 ICMS sites.**

### 2.3.2 Movement-related activity in M1 and PMd

Neural activity drives a hemodynamic response that is detectable as reflectance change in ISOI. Under red illumination (630 nm wavelength), which was used here, negative reflectance (i.e., pixel darkening) is a lagging indicator of increased neural activity (Figure 1C). Thus, in the movement conditions, we assumed that pixels would darken in locations where neural activity increased for movement execution, movement planning, or both. We refer to a spatial cluster of darkened pixels as an active "patch", which is conceptually similar to "domain" used in other studies (e.g., Bonhoeffer & Grinvald, 1991; Lu & Roe, 2007). Domain, however, is typically identified from the cortical response differential to orthogonal conditions (e.g., orientation domains in V1). If task-related neural activity increases throughout the forelimb representations, then we would expect a large patch to fill most of the field-of-view (Figure 1B top). If neural activity is more spatially confined, however, then we would expect multiple, smaller, patches (Figure 1B bottom).

First, we examined reflectance change in M1 and PMd in an average time series from a representative session (36 trials/condition). In the power grip condition (Figure 3A), there was no reflectance change from baseline to movement onset (-1.0 to +0.5 s from Cue). During reach, grasp, and hold (+0.5 to +2.0 s from Cue), pixels in the center of the FOV brightened (cool colors). In the same period, pixel darkening mostly overlapped the central sulcus and superior arcuate sulcus (landmarks on first panel). At first glance, these observations were counterintuitive as we expected pixel darkening (warm colors) in the center of FOV where neural activity presumably increased in support of arm and hand actions. We will explore the time course of reflectance change in subsequent panels and figures. For now, however, we interpret pixel darkening from this period as indication of activity in major vessels. Once the hand returned to the start position after movement completion (i.e., hold start), we observed the expected reflectance change: pixels in the center of the FOV started to darken and form patches in M1 and PMd (+3.5 s from Cue). Patches continued to darken and expand in the remaining frames. Nevertheless, even at peak intensity and size (+5.0 to +6.0 s from Cue), patches remained spatially separable within M1 and between M1 and PMd. For contrast, Figure 3B shows an average time series from the withhold condition where the Cue instructed the animal to hold the start position for the duration of the trial. Here, pixel darkening was limited to major blood vessels as most of the FOV was unchanged or brightened (Figure 3B, cyan pixels). Thus, the patches in Figure 3A were likely lagging indicators of movement-related neural activity.

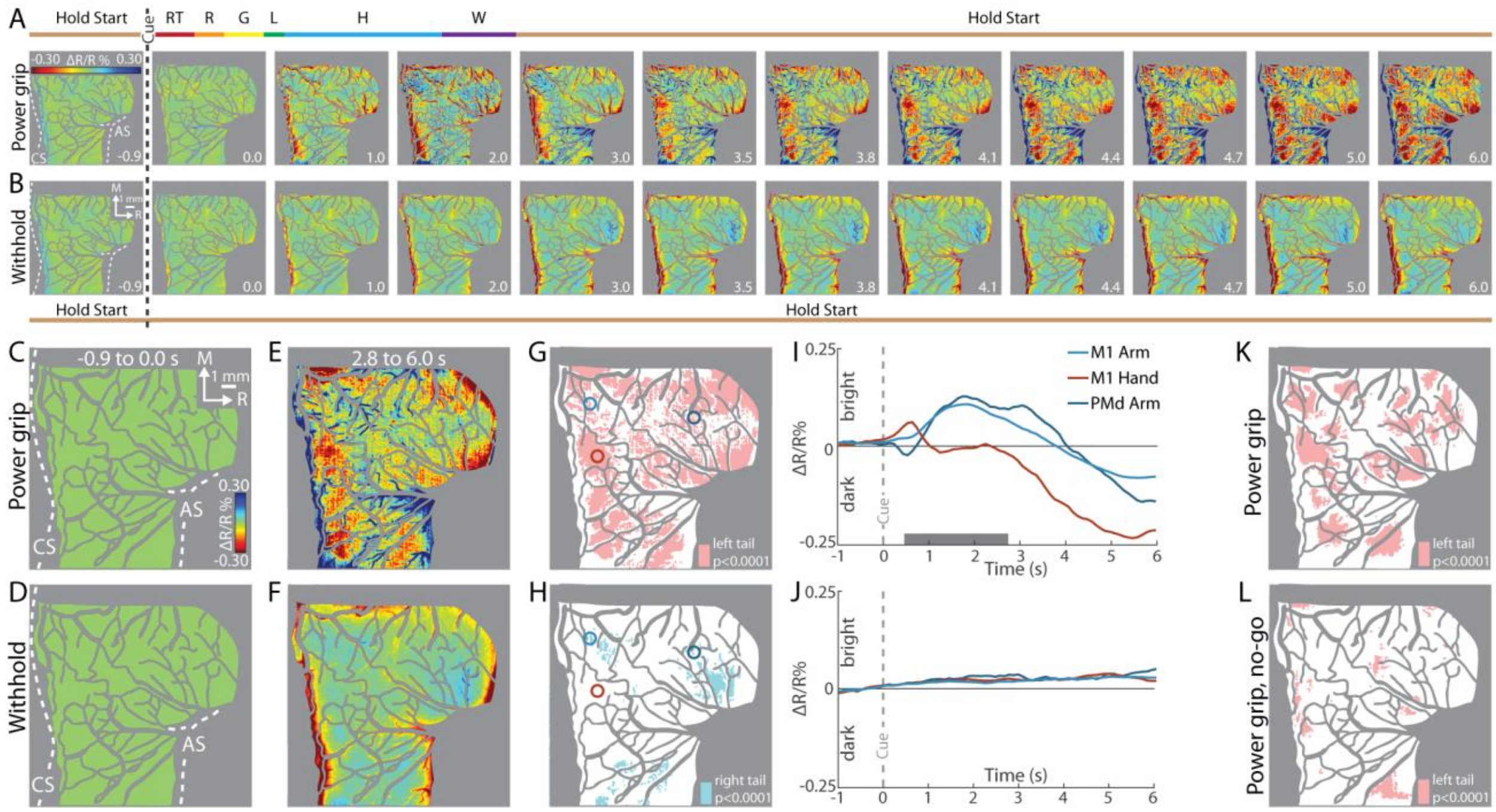

**Figure 3. ISOI detects movement-related activity in M1 and PMd (A-B) Average time series from a representative session (36 trials/condition, monkey G). (A) Timeline of average trial from the power grip condition. Hold start = hand in start position. Other phases include reaction time (RT), reach (R), grasp (G), lift (L), hold (H), and withdraw (W). Select frames with time (seconds from cue onset) in bottom right. Reflectance change was clipped to the median pixel value $\pm$ 0.3 SD. Clusters of pixels started to darken (hot colors) at 3.5 s and gradually increased in size and intensity. (B) Matching frames from the withhold condition. Color scale is same as (A). (C-D) Baseline frames in the power grip and withhold conditions. (E) Movement frame in the power grip condition: mean of 39 frames captured from movement completion until end of trial. (F) Temporally matched mean frame from the withhold condition. (G) Thresholded map from the power grip condition. Red pixels were significantly darker (t-test, left tail, $p < 0.0001$) in (E) than in (C). Colored circles (0.40 mm radius) are regions-of-interest (ROIs) placed in M1 hand, M1 arm, and PMd arm. (H) Thresholded map from the withhold condition. Cyan pixels were brighter (t-test, right tail, $p < 0.0001$) in (F) than in (D). (I) Time courses of reflectance change in the power grip condition. Line colors match the ROIs. Negative values indicate pixel darkening. Gray horizontal bar depicts mean movement duration. (J) Same as (I), but for the withhold condition. (K) Thresholded map from a reach-to-grasp condition (2 sessions,**

**159 trials). Red pixels darkened after movement as compared to baseline (t-test, left tail, p < 0.0001). (L) Thresholded map from the no-go condition (2 sessions, 159 trials). Only a few pixels darkened in M1 and PMd.**

To summarize the activity patterns in the time series, we averaged frames from two phases. For the baseline phase, we averaged 10 frames acquired -1.0 to 0 s from Cue. For the movement phase, we averaged 39 frame captured post-movement (+2.2 to +6.0 s from Cue). Average baseline frames showed no reflectance change and were indistinguishable between the movement and withhold conditions (Figure 3C-D). In contrast, reflectance change was apparent in the movement frame (Figure 3E), but not in the matching period from the withhold condition (Figure 3F). To threshold Figure 3E, we flagged pixels that darkened significantly in that frame as compared to the baseline frame (Figure 3G, t-test, left tail, p<0.0001). Thus, red pixels in Figure 3G show locations of movement-related neural activity from a single session. We thresholded Figure 3F for contrast, but to capture the predominant reflectance change here, we focused on pixels that brightened in this frame as compared to the baseline frame (Figure 3H, t-test, right tail, p<0.0001). The small number of zones and their small size indicates that reflectance change was limited in the absence of movement.

Next, we examined the time course of reflectance change in both conditions (Figure 3I-J). We placed ROIs with two guiding objectives (Figure 3G, colored circles). First, to target M1 hand, M1 arm, and PMd arm, which we achieved by consulting the motor maps and blood vessel patterns. Second, to target locations that showed activity in time series averaged across trials from all imaging sessions (e.g., Figure 5A). Time courses from two of the ROIs showed an abrupt, yet small, reflectance change with movement onset (+0.5 s from Cue). This brief change was likely due to motion artifact and will be more apparent in the next figures. After the artifact, the arm ROIs and the hand ROI had distinct time course profiles (Figure 3I). For the arm ROIs, positive reflectance started to increase ~0.3 s after movement onset (+0.8 s from Cue) and peaked during the hold phase (+1.7 s from Cue) when the object was grasped and maintained in the lifted position. Then, negative reflectance increased and peaked 2-3 s after movement completion (+5.0 to +6.0 s from Cue). The positive–negative sequence is discussed later in relation to triphasic time courses (negative-positive-negative) established for sensory cortex (Chen-Bee et al., 2007; Sirotin & Das, 2009). In contrast, the time course of the hand ROI did not show the positive reflectance that lasted for several seconds in the arm ROIs. Instead, the time course was flat until negative reflectance began to increase at +2.5 s from Cue. The timing of the negative peak was consistent, however, across the hand and arm ROIs. The distinctiveness of the time courses from the arm and hand may reflect functional differences between those zones. In contrast to the profiles in Figure 3I, the same ROIs showed no reflectance change in the withhold condition (Figure 3J).

### 2.3.3 Pixel darkening in M1 and PMd is locked to movement

Our next objective was to determine whether movement execution was necessary for the pixel darkening observed in the movement condition. We addressed this point in three separate experiments. In *Experiment 1,* one monkey performed reach-to-grasp trials interleaved with no-go trials. In a no-go trial, the go Cue and preceding steps were no different from a trial in a movement condition. In a no-go trial, however, 250 ms after the go Cue (i.e., during reaction time), another Cue (LED plus tone) instructed the monkey to hold the start position instead of move. Correct trials were typically achieved with the monkey releasing the start position to reach and then immediately returning its hand to the start position. Movements in no-go trials were therefore truncated but not inhibited. If the monkey reached past the midpoint between the start position and the target, then the trial was considered incorrect. We reasoned that no-go trials would minimize movement-related cortical activity without interfering with internal processes that precede movement (e.g., movement preparation). Results were obtained from two imaging sessions. In the movement condition, the thresholded map showed patches of activity in movement frames as compared to baseline frames (Figure 3K, t-test, $p<0.0001$). In contrast, the thresholded map from the no-go condition had fewer and smaller patches (Figure 3L). *Experiment 1* therefore shows that significant pixel darkening was contingent on movement execution.

In *Experiment 2*, we added conditions in which the monkey did not move but observed an experimenter perform the task instead. First, the monkey completed blocks of trials with regular movement conditions (performed trials). Next, the monkey observed the experimenter perform blocks of trials of the same movement conditions (observed trials). The primate chair was closed off in the observed trials to prevent the monkey from reaching. Cues, timing, and reward schedule were consistent between performed and observed trials. Time courses were measured from ROIs in M1 and PMd (Figure 4A). In M1, time courses had a clear negative peak in the performed trials (Figure 4B) but remained near baseline in the observed trials. In PMd however, the performed and observed trials had overlapping time courses. All negative peaks in PMd were smaller than the M1 peaks in the performed trials. Results from *Experiment 2* suggest that M1 activity was driven by movement execution, whereas PMd activity was likely related to movement execution, movement preparation, and motor cognition.

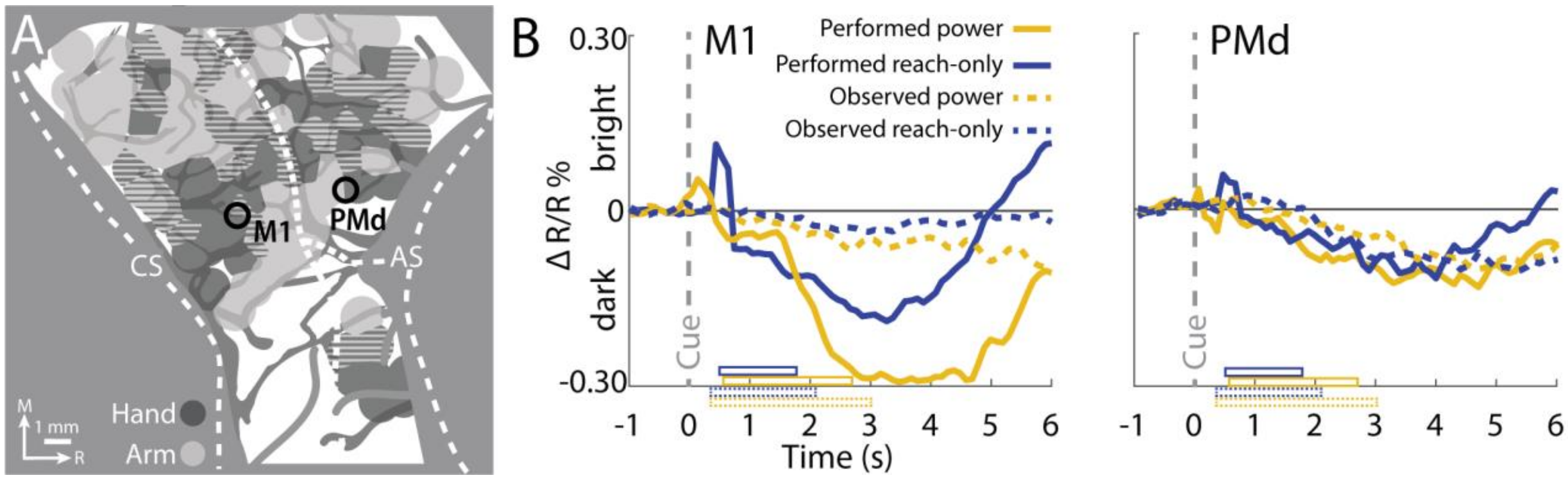

**Figure 4. Movement observation drives reflectance change in PMd but not in M1 (A) ROIs in M1 and PMd (0.4 mm radius, monkey S). (B) Time courses from 2 movement execution conditions (35 trials/condition) and 2 movement observation conditions (81 trials/condition). Horizontal bars show average movement duration for each condition.**

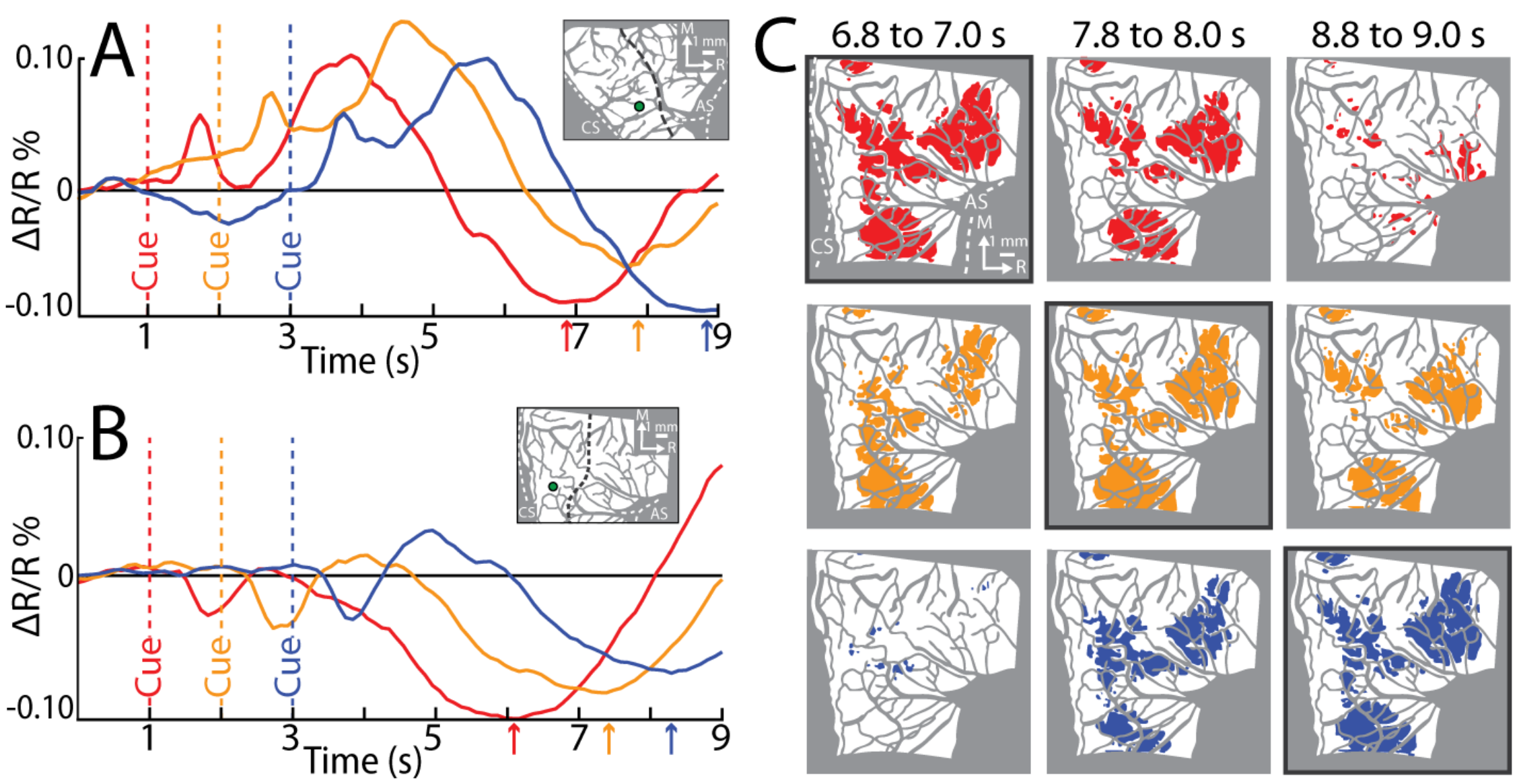

**Figure 5. Reflectance change is time locked to movement (A) Time course of reflectance change in M1. Average of 2 sessions (95 trials/condition, monkey S). Inset shows ROI (0.40 mm radius). Line plot colors match cue condition (dashed line). Arrows point to negative peaks. (B) Same as (A), but based for monkey G (1 session, 44 trials/condition). (C) Thresholded maps from same data in (B). Movement frames were an average of 3 frames, and the movement frames were taken from different time points in relation to Cue.**

**Colored pixels darkened in the movement frame as compared to baseline (t-test, p<0.0001). Pixel colors match cue condition (1 condition/row). A black border is drawn around the frame that is +5.8 to +6.0 s from Cue.**

In *Experiment 3*, we systematically varied the timing of the go Cue so that it was 1, 2, or 3 s from trial initiation. We reasoned that temporal shifts in movement onset would lead to predictable shifts in pixel darkening. Only one movement condition was tested in this experiment. We placed ROIs in the M1 hand zone (insets in Figure 5A-B). Time courses from both monkeys confirmed that the negative peaks were locked to Cue onset, and by extension movement onset. This temporal relationship was also evident in the spatial development of cortical activity (Figure 5C). For each Cue condition, we generated thresholded maps from three time windows near the negative peak. Each thresholded map therefore reported on pixels that darkened in a specific time window as compared to the baseline frame (Figure 5C, t-test, $p<0.0001$, $n=44$ trials). Across Cue conditions, the largest maps were in the time window +5.8 to +6.0 s from Cue (Figure 5C, outlined panels). Thus, *Experiment 3* shows that peak map size, and the timing of the negative peak, were both locked to movement onset. Collectively, *Experiments 1-3* confirm that the M1 activity reported here was linked to movement execution, whereas PMd activity was likely more complex.

### 2.3.4 Average time series reveal consistent features of M1 and PMd activity

To identify the most consistent activity patterns, we generated an average time series for each condition. Figure 6A summarizes the average time series for the precision grip condition (monkey G, 8 sessions, 326 trials). Animating the time series side-by-side with a representative behavioral trial (https://tinyurl.com/4n59sk9k) qualitatively revealed spatiotemporal features of the cortical activity. (1) During movement, pixel darkening was limited to major vessels; the rest of the field-of-view brightened or was unchanged. (2) Pixel darkening in cortex (i.e., beyond major vessels) occurred after movement completion. (3) The intensity of pixel darkening and the number of pixels that darkened, both progressed gradually from movement completion until the end of image acquisition. The same features were present in the other movement conditions (not shown).

From the average time series, we condensed the spatiotemporal organization of cortical activity into a single frame. To that end, we used the time course of every pixel to find the time of the negative peak (i.e., maximal darkening). Figure 6B color codes peak times from the time series in Figure 6A but omits information related to magnitude of reflectance. Peak times fit into three broad categories (Figure 6B). (1) *Early peaks*. Green pixels peaked ~2 s from Cue when the monkey was still engaged in task-related movements. Green pixels generally corresponded with the hot color pixels in Figure 6A panels 1.6 to 2.5, and therefore may have been affiliated with the early response in major vessels. (2) *Intermediate peaks*. Orange/Yellow pixels peaked ~3.5 s from Cue, which was within 1 s of movement completion. Orange/Yellow pixels were in the approximate location of yellow pixels in Figure 6A last panel. (3) *Late peaks*. Magenta/red pixels peaked ~5.5 s from Cue, which was ~3.0 s from movement completion. These pixels corresponded with the hot color pixels in Figure 6A last 3 panels. Late peak pixels were in clusters surrounded by intermediate peak pixels, but there was no regional divide between the two.

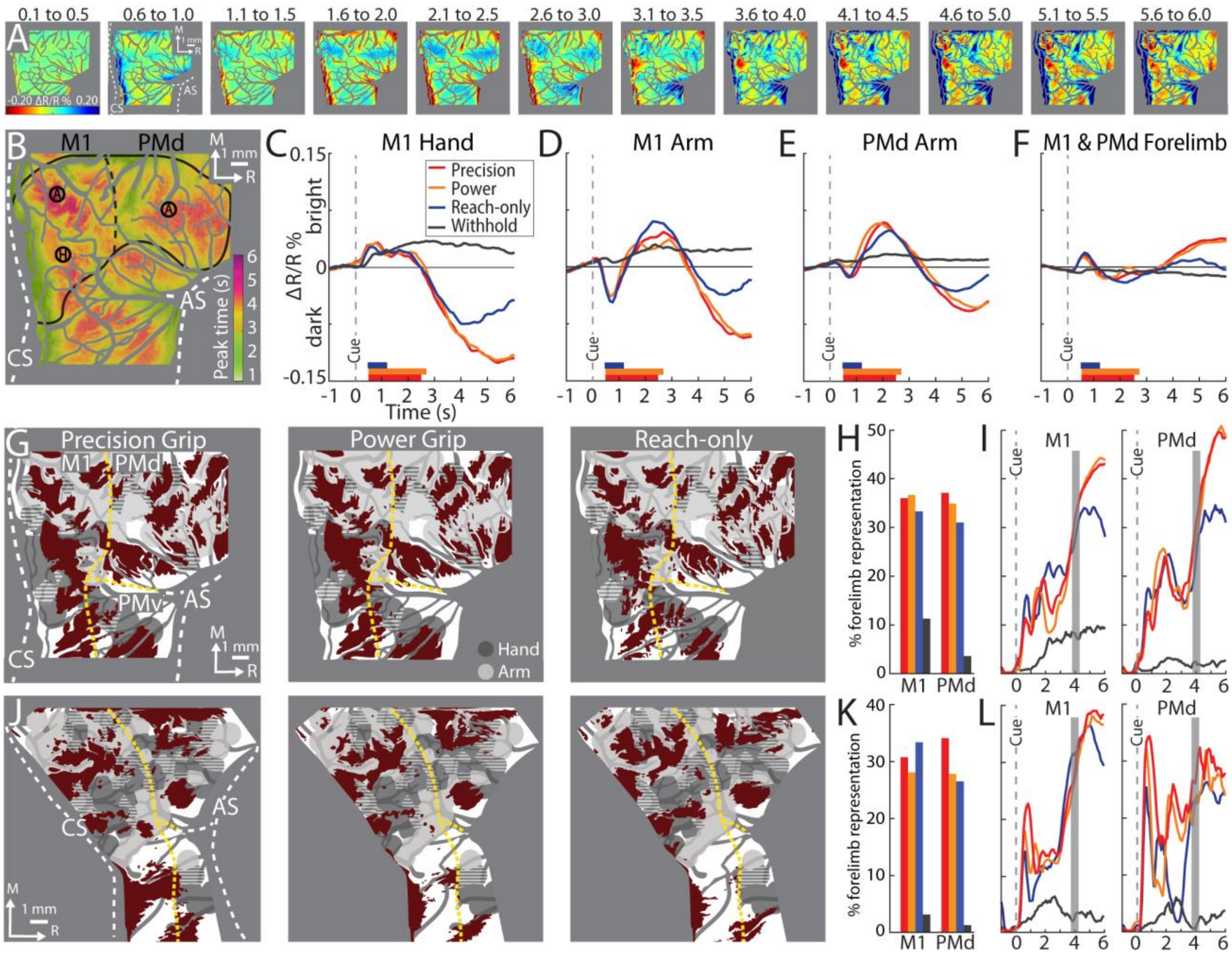

**Figure 6. Average time series reveal consistent features of cortical activity (A) Average time series for precision grip (8 sessions, 404 trials, monkey G). Each frame is the average of five successive frames, averaged across trials. Time above each frame is seconds from cue onset. Reflectance change was clipped to median ± SD pixel value across frames. Patches started to form at 3.1-3.5 s. (B) Full FOV colored according to the time (seconds from Cue) that pixels reached maximal negative reflectance. Hot colored pixels peaked late in the trial. Black circles (0.40 mm radius) in M1 and PMd are ROIs placed in arm (a) and hand (h) zones. The forelimb representations are outlined in black with the dotted line separating M1 from PMd. The entire outline is also used an ROI. (C-F) Each plot is based on one of the ROIs in (B). Horizontal bars on x-axis show the average movement duration in each condition. (G) Thresholded maps for each condition. Red flags pixels that darkened in the movement frame as compared to the baseline frame (t-test, p<0.0001). Dashed yellow lines mark cortical borders. (H) Percentage of forelimb representations with red pixels from (G). Bar colors match legend in (C). (I) Same quantification in (H) expressed as a function of trial duration.**

**Thresholded maps were generated at every time point (0.1 s) with a t-test (p<0.0001) comparison of the frame at the time point and the baseline frame. (J-L) Same as (G-I), but for monkey S.**

For another perspective on the spatiotemporal patterns of cortical activity, we examined time courses from ROIs that we placed in M1 and PMd (Figure 6B). The three circular ROIs matched those in Figure 3G and returned time courses consistent with Figure 3I & 3J. In the movement conditions, time courses from the three ROIs had motion artifact that started with reach onset (Figure 6C-E; ~0.5 s from Cue) and lasted for ~0.2 s. For the M1 hand ROI (Figure 6C), after the artifact, reflectance change was flat for ~1.5 s. Negative reflectance started to increase ~2.5 s from Cue, which coincided with the end of movement in the precision and power conditions. In those conditions, the negative peak was larger, and occurred later, than in the reach-only condition. This temporal difference could have been due to the larger size of the negative peaks, or longer movement durations, or both. ROIs in M1 and PMd arm zones returned time courses that were consistent with one another (Figure 6D-E). In both ROIs, the motion artifact was followed by an increase in positive reflectance, which is quite different from the time courses from the hand ROI (Figure 6C). Nevertheless, in the arm ROIs, negative reflectance increased from +3.5 from Cue and peaked at approximately the same time as the M1 hand ROI.

Two controls provided context for the time courses of the movement conditions (Figure 6C-E). First, in all ROIs, the withhold condition had time courses that remained close to baseline (Figure 6C-E). This control confirms that the reflectance change in the movement conditions was movement driven. Second, an ROI that spanned the M1 and PMd forelimb representations (Figure 6B) returned time courses that differed entirely from the other three ROIs. Most importantly, the forelimb ROI lacked the characteristic negative peaks in Figure 6C-E. This control confirms that time courses from movement conditions were spatially specific responses that did not generalize to the entire forelimb representation.

### 2.3.5 Thresholded activity maps overlap limited portions of the forelimb representations

Next, we generated thresholded maps to summarize the spatial patterns of movement-related cortical activity. For each condition, trials were pooled across sessions and every trial contributed a baseline frame and a movement frame. Pixels that darkened significantly (t-test) in the movement frame as compared to the baseline frame were included in the thresholded map. A baseline frame was the average of the first 10 frames in a trial. The movement frame was defined in two different ways. First, as an average of all frames (n=39) captured from end of movement (+2.2 to +6.0 s from Cue). This strategy was adopted for the thresholded maps in Figure 3. Second, as an average of 5 consecutive frames captured in the post movement period. A similar strategy was adopted in Figure 5, but here the 5-frame range was guided by the timing of the negative peaks. For example, we averaged the values in Figure 6B to determine the mean timing of negative peaks in the precision condition. We repeated the procedure for the other conditions and then averaged the times across conditions. From those averages, the 5 frames for calculating the movement frame were set to +4.1 to +4.4 s from Cue (monkey G) and +3.9 to +4.3 s from Cue (monkey S). Additional support for the chosen frame range is discussed in Appendix 3.

First, we examine the 5-frame thresholded maps (t-test, $p<0.0001$). We co-registered those maps with the motor maps for reference (Figure 6G & 6J). The general organization of the thresholded maps was more similar across conditions within an animal than for the same condition across animals (Figure 6G & 6J), which is consistent with human fMRI maps of M1 finger zones (Ejaz et al., 2015). In all thresholded maps, significant pixels were organized in patches that overlapped subzones of the M1 and PMd forelimb representations. The most lateral patches were in M1 face and PMv forelimb and were therefore excluded from subsequent analyses. In each map the patches collectively overlapped only 31% (median, IQR = 29-35%) of the forelimb representations (Figure 6H & 6K). This measurement takes into consideration both monkeys, both M1 and PMd, and the three movement conditions. For reference, thresholded maps from the withhold condition overlapped 3% (median, IQR = 1-7%) of the M1 and PMd forelimb representations.

We repeated the overlap measurements after re-thresholding the maps to correct for multiple comparisons based on the number of pixels in the forelimb representations (t-test, p<1.0e-7). This more conservative threshold did not impact the spatial organization of the maps, but it shrank their sizes to 14% (median, IQR = 16-20%) overlap with the forelimb representations (Figure 7). In contrast, expanding the frame range of thresholded maps to 39 (t-test, p<0.0001) led to only a small reduction in overlap with the forelimb representations (Figure 8; median= 27%, IQR = 25-34%). Thus, generating thresholded maps using a less conservative threshold, or using different frame ranges from the post movement period, returned the same overall result: <40% overlap with the M1 and PMd forelimb representations.

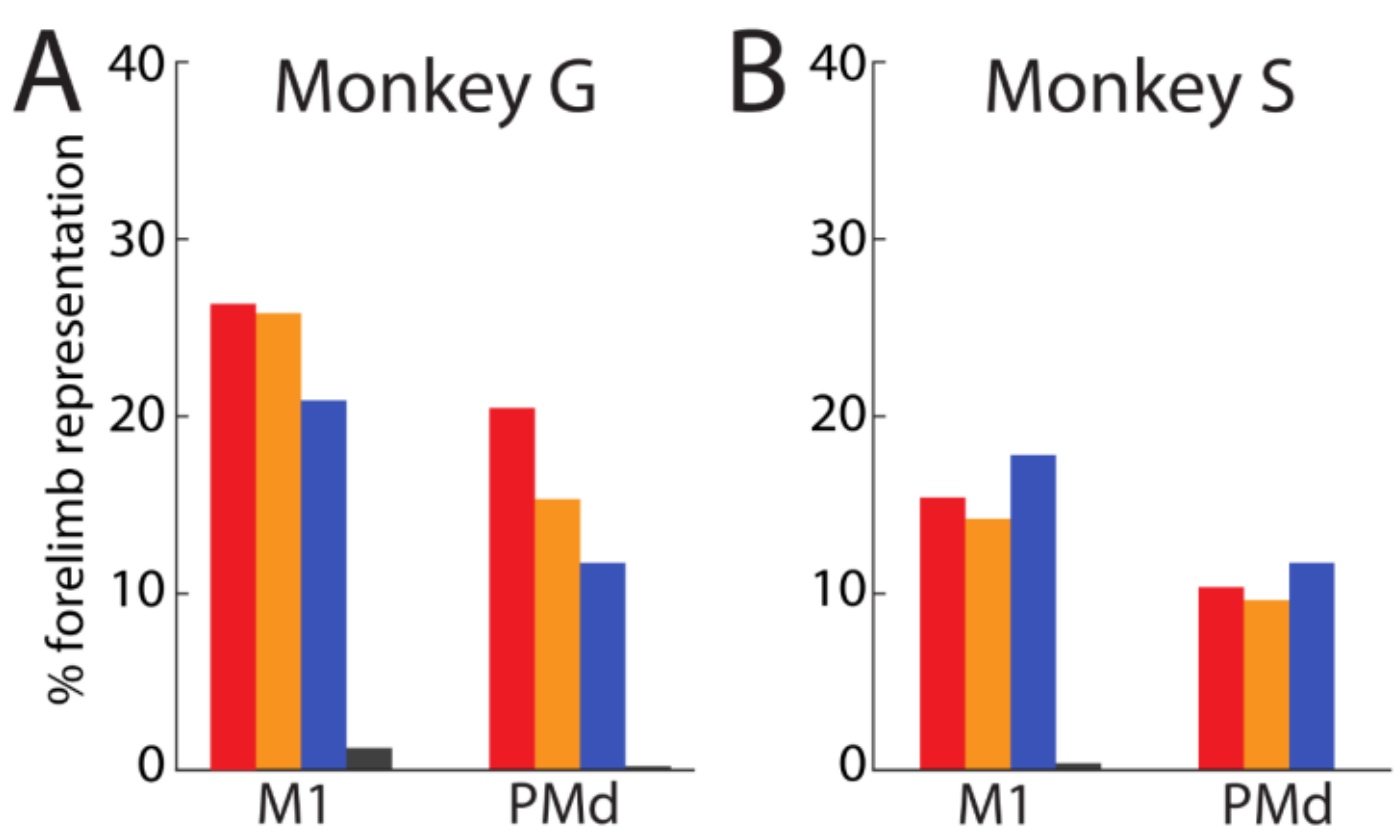

**Figure 7. Thresholded maps overlap small proportion of the forelimb representations (A-B) Same as Fig 5H & 5K but quantified here from a more conservative threshold (p<1e-7).**

For context on map sizes reported thus far, we redid the overlap measurements at every time point. Thus, we generated a thresholded map (t-test, p<0.0001) for every frame and measured its overlap with the M1 and PMd forelimb representations (Figure 6I & 6L). A consistent feature across movement conditions, cortical areas, and monkeys, was that the size of thresholded maps started to rapidly increase from ~3 s from Cue, which nearly coincided with the end of movement in the precision and power conditions. The size of the maps plateaued, or even decreased, by ~5s from Cue. At peak map expansion, overlap with the forelimb representations was 38% (median, IQR = 35-44%). Thresholded maps in Figures 6 and 8 were therefore only slightly smaller than peak maps sizes. Thus, the two strategies that we adopted for selecting movement frames (5-frame and 39-frame averages) did not underreport the size of the thresholded maps.

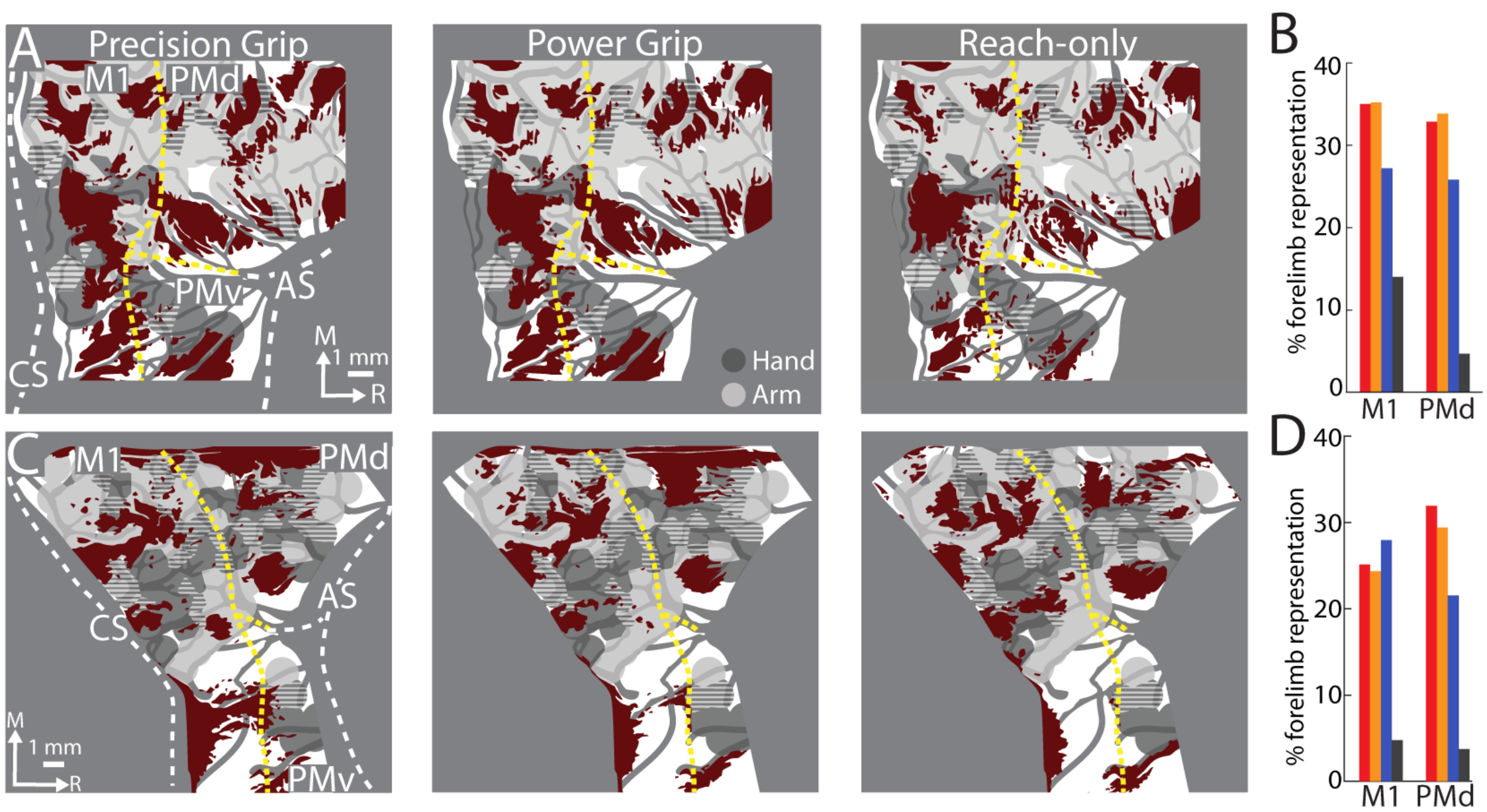

**Figure 8. Thresholded maps overlap small proportion of the forelimb representation Same as Fig 5G-H & 5J-K, but the movement frames here are generated from an average of the frames collected from +2.2 s to +6.0 s from Cue.**

### 2.3.6 Condition differences in cortical activity patterns

The spatial organization and size of thresholded maps suggested similar cortical activity across conditions. Nevertheless, time courses (Figure 6C-E) and non-binarized average maps (Figure 9A) showed greater magnitudes of activity in the reach-to-grasp conditions as compared to the reach-only condition. This motivated us to directly compare cortical activity between conditions, which we achieved in two ways.

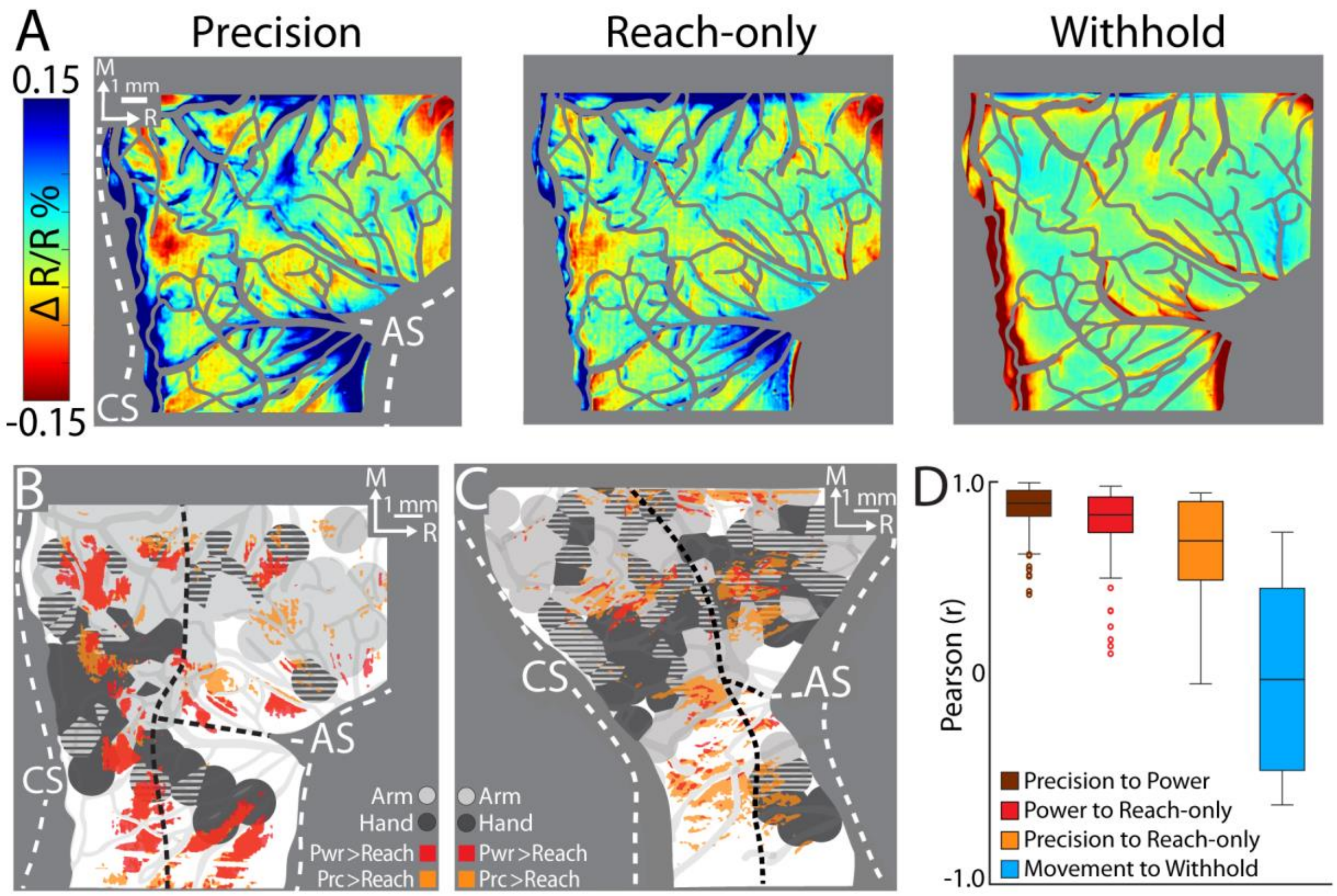

**Figure 9. Reflectance maps of movement conditions across and within sessions (A) Movement frames (+2.2 to +6.0 s from Cue) averaged across conditions (n=326-358 trials, monkey G). Reflectance change was clipped to the median pixel value $\pm$ 1.5 SD. (B) Thresholded maps from 2 paired-comparisons. Red and orange pixels flag locations that darkened significantly in the power (pwr) and precision (prc) conditions, respectively, over the reach-only (t-test, $p<0.001$). Dashed black lines mark cortical area borders. (C) Same as (B) for monkey S. (D) Cross-correlation comparisons of average time series from conditions pairs. A data point here is the coefficient from cross-correlating (zero lag) one frame from a condition time series with the time-matched frame from a different condition. Each box is therefore comprised of 78 points (39 frames x 2 monkeys). The top and bottom of each box are the first and third quartiles of the data; whiskers span 1.5x the interquartile**

**range. The blue box includes 234 points from comparing three movement conditions to the withhold condition (39 frames x 3 comparisons x 2 monkeys).**

First, we conducted t-test comparisons on movement frames (39-frame average) from condition pairs (Figure 9B-C). Our objective was to generate thresholded maps of pixels that darkened significantly in one movement condition relative to another movement condition. Maps thresholded at $p<0.0001$ did not reveal distinguishable patches. After relaxing the threshold to $p<0.001$, we found several patches in the M1 and PMd forelimb representations that darkened more in the reach-to-grasp conditions as compared to the reach-only condition (Figure 9B-C). In contrast, fewer and smaller patches reported differences between the precision and power conditions (not shown). Thus, the thresholded maps in Figure 9B-C flag cortical zones that were more active in the reach-to-grasp conditions as compared to the reach-only condition.

Second, we used cross-correlation at zero lag to compare average time series from pairs of conditions. We focused on the post-movement frames and correlated each frame from one condition with the time-matched frame in the paired condition. The analysis was limited to pixels within the forelimb representations. Correlation coefficients were pooled across monkeys such that each condition pair returned a distribution of 78 points (39 frames x 2 monkeys).  A one-way ANOVA showed an effect of condition pairs on the correlation coefficients ($F(2,231) = 12.98$, $p <0.01$; Figure 9D). Post hoc tests corrected for multiple comparison ($n=3$; $p<0.01$), showed that the correlation coefficients for precision X reach-only was lower than the other two condition pairs. The cross-correlations therefore show that the spatiotemporal patterns of cortical activity were most dissimilar between the precision and reach-only conditions. For context, cross-correlating movement conditions with the withhold condition returned a much lower distribution of coefficients (Figure 9D). Cross-correlation was therefore sensitive enough to detect condition differences in the time series. In sum, the two approaches that we used for direct condition comparisons indicate that condition differences were small but most pronounced between the precision and reach-only conditions.

### 2.3.7 Forelimb use differences across conditions

We examined whether forelimb use could shed light on the cortical activity patterns observed in the movement conditions. From the consistent sizes of thresholded maps (Figure 6H & 6K), we expected similar forelimb use across conditions. In contrast, from the negative peak magnitudes in Figure 6C-E and the condition differences in Figure 8B-D, we expected greater forelimb activity in the reach-to-grasp conditions as compared to the reach-only condition. To test these possibilities, we measured forelimb kinematics for 12 degrees of freedom and measured EMG activity from 7 forelimb muscles.

Figure 10A shows average time courses for 8 degrees of freedom in arm and hand joints. Shoulder and elbow time courses were similar in the precision and power conditions. In the precision condition, however, digit and wrist angles were sustained throughout the hold period. The reach-only condition differed from the other two conditions in having truncated time courses and nearly absent flexion/extension of the digits. To quantify condition differences, we measured the area under the curve (AUC) from Cue to Replace. AUC values were standardized for each joint to facilitate comparisons across conditions. The reach-only condition stood out for having the smallest AUC (Figure 10B) in all degrees of freedom. To simplify condition comparisons (Figure 10C), we first combined the AUC calculated for arm joints (shoulder and elbow) and separately combined the AUC for hand joints (wrist and digits). For statistical testing, we then combined the AUC for arm and hand into a single group. A one-way ANOVA on those AUC distributions showed a main effect of movement condition ($F(2,4332) = 6194$, $p < 0.0001$). Post hoc tests corrected for multiple comparison (n=3; p<0.01) showed that the reach-only condition had a smaller AUC than the other two conditions and that the power condition was smaller than the precision condition.

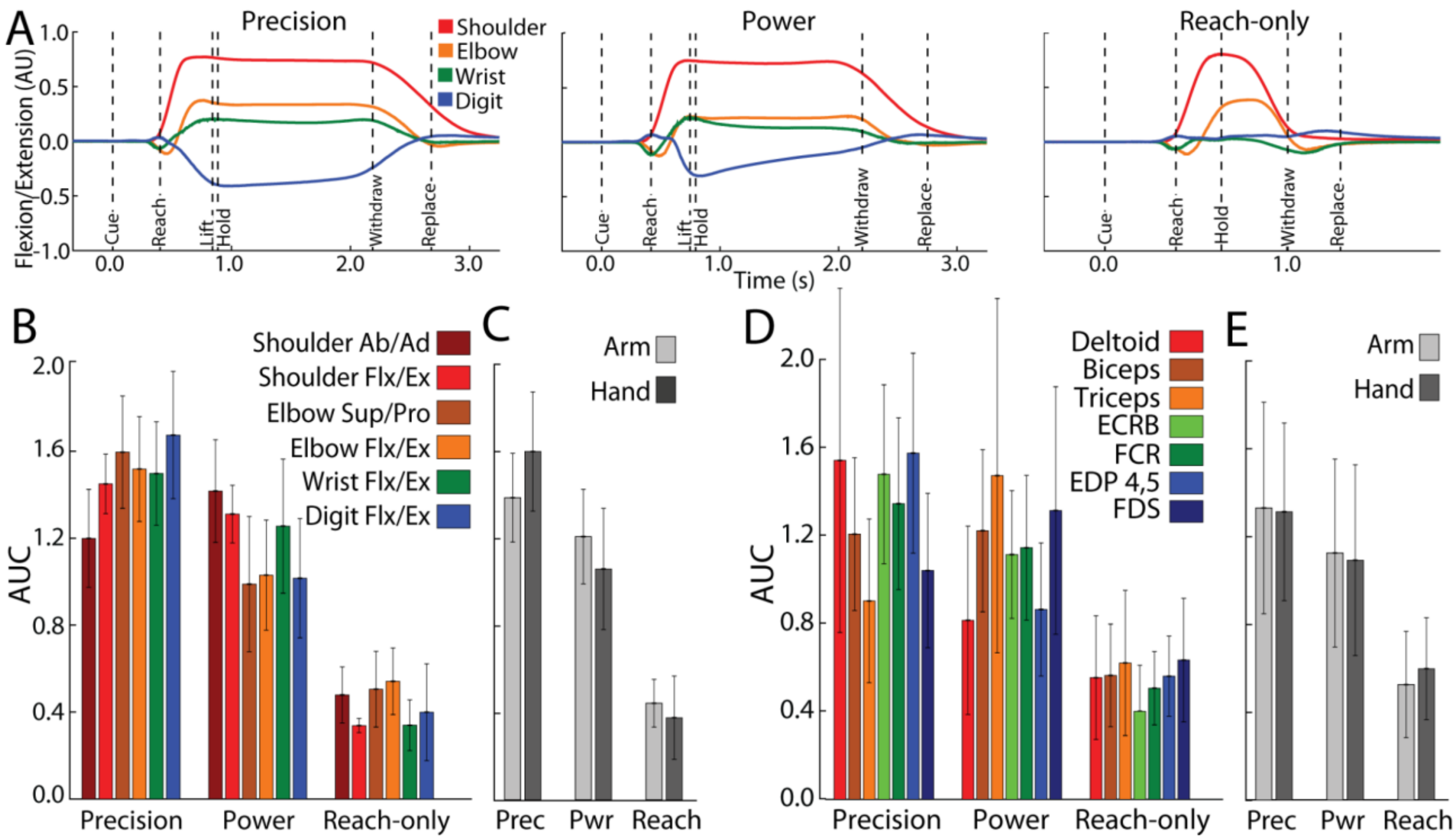

**Figure 10. Forelimb activity scales across conditions (A) Mean flexion/extension changes in 2 arm joints and 2 hand joints as a function of time in the 3 movement conditions (5–15 sessions, 143–405 trials, monkey G). Y-axis scale is arbitrary units (AU). (B) Area under curve (AUC, mean ± SD) for joint activity from trials in (A). (C) Pooled AUC (mean ± SD) for joint activity in the three conditions. Shoulder and elbow joints were classified as arm. Wrist and digit joints were classified as hand. (D-E) Same as (B-C), but for muscle activity (6-14 sessions, 308-595 trials, monkey G). ECRB: extensor carpi radialis brevis. EDP: extensor digitorum profondus. FCR: flexor carpi radialis. FDS: flexor digitorum superficialis.**

The same condition differences in joint activity were evident in the activity of individual arm and hand muscles (Figure 10B). We combined the muscle AUC measurements into groups as was had done for the joints. We found the same main effect and post hoc condition differences in the muscle AUC measurements [Figure 10D; ANOVA, $F(2,9794) = 3,633$, $p < 0.0001$]. Thus, joint angle kinematics and EMG indicate that arm and hand activity systematically differed across conditions: precision > power > reach-only. These forelimb use differences across conditions are consistent with the dissimilarities in cortical activity between the precision and reach-only conditions (Figure 6C-E and 8B-D).

## 2.4 Discussion

We investigated the relationship between movement-related cortical activity and motor maps in motor (M1) and dorsal premotor (PMd) cortex. Intrinsic signal optical imaging (ISOI) showed that cortical activity that supports reaching and grasping is concentrated in patches within the forelimb representations. The spatial configuration of the patches was consistent across task conditions despite variations in the reach-to-grasp movements. Nevertheless, the magnitude of activity within the patches scaled with forelimb use as measured from joint and muscle activity. The spatial organization of thresholded maps, and relative invariance across conditions, suggests that neural activity for arm and hand functions is not uniformly distributed throughout the forelimb representations. Instead, our observations are consistent with a functional organization wherein subzones of the forelimb representations are preferentially tuned for certain actions.

### 2.4.1 Measuring activity in M1 & PMd

Two provisions facilitated our findings. First, we used ISOI to measure cortical activity. ISOI is well-established for studying sensory cortex (Bonhoeffer & Grinvald, 1991; Chen et al., 2001; Malonek & Grinvald, 1996; Ts'o et al., 1990), but few studies have leveraged it for cortical control of movement (Friedman, Chehade, et al., 2020a; Siegel et al., 2007). Functional MRI has been used more extensively to investigate reaching and grasping in monkeys and humans (e.g., Cavina-Pratesi et al., 2010; Gallivan & Culham, 2015; Nelissen & Vanduffel, 2011b). Our objective, however, was better served with the higher contrast and spatial resolution afforded from ISOI. Similarly, calcium imaging has gained traction in cortical control of movement and brain-computer paradigms because it provides single cell resolution and better temporal fidelity to neural signals. But in monkeys, the field-of-view (FOV) in mini-scopes and 2-photon microscopes, covers only a small fraction of one forelimb representation (Bollimunta et al., 2021; Ebina et al., 2018b; Kondo et al., 2018b; Trautmann et al., 2021). In contrast, the present FOV was large enough for both M1 and PMd forelimb representations. Second, we obtained motor maps from the same chronic chambers and used them to quantify the spatial extent of the cortical activity recorded with ISOI. Without the present high density motor maps, it would not have been possible to distinguish between cortical activity in the forelimb representations and the counterpart in the surrounding trunk and face zones. Thus, conducting ISOI and motor mapping in the same hemispheres was central to our results and interpretations.

Nevertheless, our approach had limitations that must be considered for proper interpretation of the results. First, our monkeys performed a reaching task that explored only a narrow range of their repertoire of arm and hand actions. Diversifying the task conditions would have expanded the range of arm and hand movements, which could have provided additional insight into the organization of cortical activity that supports forelimb actions. Second, the motor maps were based on low dimensional classifications of the ICMS-evoked movements. Although our motor mapping approach was sufficient for distinguishing the forelimb representations from the surrounding zones, it may have oversimplified the spatial organization of the arm and hand zones relative to each other. For that reason, we measured the spatial extent of the cortical activity with respect to the forelimb representations as a whole and did not examine potential differences between arm and hand zones.

### 2.4.2 Delayed negative peaks are locked to movement

Pixel darkening peaked several seconds after movement onset. Our three control experiments and the withhold indicated that most of the present pixel darkening was predicated on movement execution. The reflectance change time courses were consistent with those reported in the few studies that used ISOI with forelimb tasks (Friedman, Chehade, et al., 2020a; Heider et al., 2010). Nevertheless, the lag between movement onset and peak pixel darkening was 2-3 times longer than expected from stimulus onset (peripheral or ICMS) and peak darkening in sensorimotor cortex (e.g., Card & Gharbawie, 2020; Chen et al., 2001; Friedman, Morone, et al., 2020). Slower intrinsic signals have been reported in other behavioral paradigms (Grinvald et al., 1991; Tanigawa et al., 2010), which raises the possibility of time course differences between awake and anesthetized ISOI. Fundamental differences could also exist between intrinsic signals evoked from movement (active) and those from stimulation (passive). In our task, the arm and hand were active for ~2 s per trial. In contrast, sensory stimuli are typically more confined in both space and time to optimize focal activation in cortex (e.g., S1 barrel, V1 orientation column).

The protracted peak darkening should be considered in relation to the triphasic time course of reflectance change in ISOI (Chen-Bee et al., 2007; Sirotin et al., 2009). Studies with stimulus evoked responses typically focus on the initial dip, which is the first dark peak and occurs 1-2 s after stimulus onset. The subsequent rebound (bright peak) and undershoot (second dark peak) unfold sequentially over several seconds. In our time courses, the timing of the darkest peak would have coincided with the rebound in stimulus-evoked responses. The timing of the dark peak reported here can potentially fit with the triphasic response in three ways. (1) Our dark peaks could have been initial dips locked to movement onset. In this interpretation our movement-evoked time courses would be slower iterations of stimulus-evoked responses. (2) Our dark peaks could have been initial dips locked to movement offset. The early negative peak of the reach-only condition does not provide clear insight into this point because this condition had an early offset time as well as a relatively small negative peak that could have peaked early. Thus, direct testing of the movement offset interpretation would require systematic manipulation of movement duration that we did not explore. (3) Our dark peaks could have been the undershoot locked to movement onset. In this case the initial dip may have been undetected because it was too small, obscured by motion artifact, or both. In this interpretation our movement-evoked time courses would be consistent with stimulus-evoked responses, albeit with a relatively brief rebound.

### 2.4.3 Movement activates subzones of the forelimb representations

Average thresholded maps reported cortical locations where task-related activity was present in hundreds of trials. Those maps overlapped <40% of the M1 and PMd forelimb representations. Expanding the number of post movement frames had no impact on the size of the thresholded maps and correcting the maps for multiple comparisons further diminished their size. Relatively small size was therefore a robust feature of the thresholded maps and we consider potential explanations for that result. (1) ISOI sensitivity. It is unlikely that ISOI lacked sensitivity for capturing the full extent of cortical activity and thereby underreported thresholded map sizes. If anything, ISOI sensitivity to subthreshold electrophysiological signals can lead to overestimation of the spatial extent of modulated neural activity (Frostig et al., 2017; Grinvald et al., 1994). The issue is minimized by focusing on early segments of the ISOI response, which reports cortical organization and connectivity at columnar resolution (Bonhoeffer & Grinvald, 1991; Card & Gharbawie, 2020, 2022; Friedman, Morone, et al., 2020; Lu & Roe, 2007; Vanzetta et al., 2004). (2) Task overtraining. We started ISOI after ~2 years of task training. Extensive training as such has been shown to refine network activity and reduce metabolic demand in M1 (Peters et al., 2014; Picard et al., 2013). Both factors could have shrunk the thresholded maps from larger sizes in earlier training stages to the sizes reported here. (3) Functional organization. The small size of thresholded maps may reflect the functional organization of M1 and PMd. Specifically, neural activity that supports reaching and grasping could be largely concentrated in subzones of the forelimb representations. In this organizational framework, cortex in between those subzones may be better tuned for forelimb actions other than reaching and grasping. Testing this possibility would require training monkeys to perform tasks that probe a wide range of arm and hand actions, which was not done here. Nevertheless, the notion that actions may be spatially organized in cortex has support in long-train ICMS (500 ms), which maps multi-joint actions (e.g., reach, manipulate, climb) to contiguous zones in M1 and PMd (Baldwin et al., 2017; Gharbawie et al., 2011b; M. S. A. Graziano et al., 2002b; Kaas et al., 2013).

### 2.4.4 Functional differences between somatotopic zones

Patches that overlapped arm zones could have been tuned to different task phases than patches that overlapped hand zones. This possibility is supported by distinct time course profiles from ROIs in arm and hand zones. Nevertheless, the temporal resolution of intrinsic signals is not sufficient for relating reflectance change to behavioral events that occur in close succession (e.g., reach and grasp). We are currently conducting high-density electrophysiological recordings to interrogate relationships between movement and spiking activity in constituents of the thresholded maps. Our report on a small sample of single units recorded in other hemispheres, suggested functional differentiation between patches (Friedman, Chehade, et al., 2020a). Specifically, we found that a medial M1 patch contained more neurons tuned for reaching than neurons tuned for grasping. We found an opposite neural tuning distribution in a lateral M1 patch. Other recordings, however, from caudal M1 (central sulcus) and from rostral M1 (precentral gyrus) have reported spatially co-extensive encoding of reaching and grasping (Rouse & Schieber, 2016a; Saleh et al., 2012a; Vargas-Irwin et al., 2010b). Thus, the temporal dimension provides mixed evidence about functional differences across the spatial dimension (i.e., planar axis) of the forelimb representation.

Condition differences provide insight into the functional organization of the recorded cortical activity. For example, Nelissen & Vanduffel, 2011b defined grasp zones as regions that were more active (BOLD fMRI) in reach-to-grasp conditions as compared to a reach-only condition. A similar analysis on our data revealed patches in M1 and PMd where activity in the reach-to-grasp conditions exceeded activity in the reach-only condition (Figure 8). Nevertheless, these condition differences were less apparent in the spatial size and organization of the maps, which is consistent with other studies that used univariate analyses on fMRI measurements (Fiave et al., 2018; Gallivan et al., 2011; Nelissen et al., 2018). In contrast, multi-variate analyses (e.g., multi-voxel pattern analysis) on the same data showed that condition identities were embedded in the spatial dimensions of activity maps.

### 2.4.5 Differentiating between M1 and PMd activity

Functional differences between M1 and PMd were most apparent in the observation experiment. Time courses from the observation condition showed that there was no activity in M1. In contrast, comparable activity was recorded in PMd for the performed and observed conditions, which is consistent with the higher concentrations of mirror neurons in premotor areas than in M1 (Grèzes et al., 2003; Papadourakis & Raos, 2019; Raos et al., 2007). These results suggest that PMd activity could have been related to motor cognitive process as well as movement, whereas M1 activity was more closely associated with movement only.

## 2.5 Conclusion

We showed that in M1 and PMd, activity related to reaching and grasping was concentrated in subzones of the forelimb representation. The results collectivity indicate that the spatial dimension of cortical motor areas is central to their functional organization. Taking this feature into consideration in electrophysiological studies can shed new light on the relationships between neural signals and movement control.

# 3.0 The Spatiotemporal Organization of Single Unit Activity in Motor Cortex

## 3.1 Introduction

Primary motor cortex (M1) is indispensable to achieve voluntary movements. Early studies of M1 used intracortical microstimulation (ICMS) to establish a somatotopic organization of motor output. Stimulation injected at medial M1 evoked movements of the foot, followed by the trunk, arm, hand, and face at lateral M1 (Penfield & Boldrey, 1937; Woolsey, 1958). Subsequent study of the M1 forelimb representation identified overlapping populations of cells with direct projections to muscles of the arm and hand using transneuronal tracers injected into forelimb muscles (Rathelot & Strick, 2009a). Stimulus-triggered averaging of forelimb muscle activity revealed the M1 hand representation clustered in a central core surrounded by a horseshoe-shaped representation of the arm (Park, Belhaj-Saïf, et al., 2001a). Additionally, the hand representation was separated from the arm representation by a region comprising both. Thus, although not spatially discrete, various methodologies have corroborated an anatomical organization of the arm and hand in M1 (Strick et al., 2021).

Spatial intermingling of the M1 arm and hand representation is speculated to facilitate neural activation of forelimb muscle synergies to achieve coordinated control of the arm and hand (McKiernan et al., 1998; Overduin et al., 2015). Recent investigations of the kinematics and muscle activity during coordinated forelimb movements revealed that both the arm and hand varied in response to changes of the reach trajectory and grip posture (Rouse & Schieber, 2015, 2016c). Furthermore, effects pertaining to reach and grasp modifications were realized in distinguishable phases; reach-related effects preceded grasp-related effects. In M1, unit activity recorded from the forelimb representation comodulates with various parameters related to reach and grasp, such as reach direction or hand posture (Georgopoulos et al., 1986; Goodman et al., 2019). Yet, few groups have investigated the spatial organization of functional M1 unit activity.

Of the few studies that address the functional organization of M1 activity, investigators note that units encoding reach were spatially intermingled with units encoding grasp throughout the forelimb representation (Rouse & Schieber, 2016a; Saleh et al., 2012a). However, unit activity was recorded from implanted multielectrode(s) that spanned a fraction of the M1 forelimb representation. Restricted sampling of unit activity potentially misrepresents the overall functional organization in the M1 forelimb representation. In fact, imaging M1 while monkeys completed a reach-to-grasp task indicated that distinct zones in cortex increased activity to achieve movements. Furthermore, these zones of activity comprised less than half of the total surface area in M1 that represented the forelimb (Chehade & Gharbawie, 2023). Recording unit activity in these zones revealed preferential encoding of grasp within the hand representation compared to a zone in the arm representation (Friedman, Chehade, et al., 2020a). These studies imply a spatiotemporal organization of M1 activity that may have been undetectable using implanted multielectrode arrays.

We aim to clarify how M1 activity is functionally organized by comprehensively sampling unit activity in the M1 forelimb representation, as defined by motor mapping with ICMS. Given that reach and grasp effects are temporally separable in the kinematic and muscle activity of the forelimb, we suspect that M1 units will selectively modulate activity during the reach or grasp phase (Figure 11I: red and blue traces). On the other hand, units may modulate activity evenly during the reach and grasp phase (Figure 11I: yellow trace). We defined units according to activity selectivity in the reach and grasp phase to quantify the proportion of phase-selective and non-selective units in the M1 forelimb representation. We then explored the spatial organization of phase selective units to understand how function relates to structure in M1. If structure and function are especially interrelated, we expect grasp encoding units clustered in the M1 hand representation and reach encoding units concentrated in the M1 arm representation. Alternatively, units encoding reach and grasp may be intermingled all throughout the M1 forelimb representation.

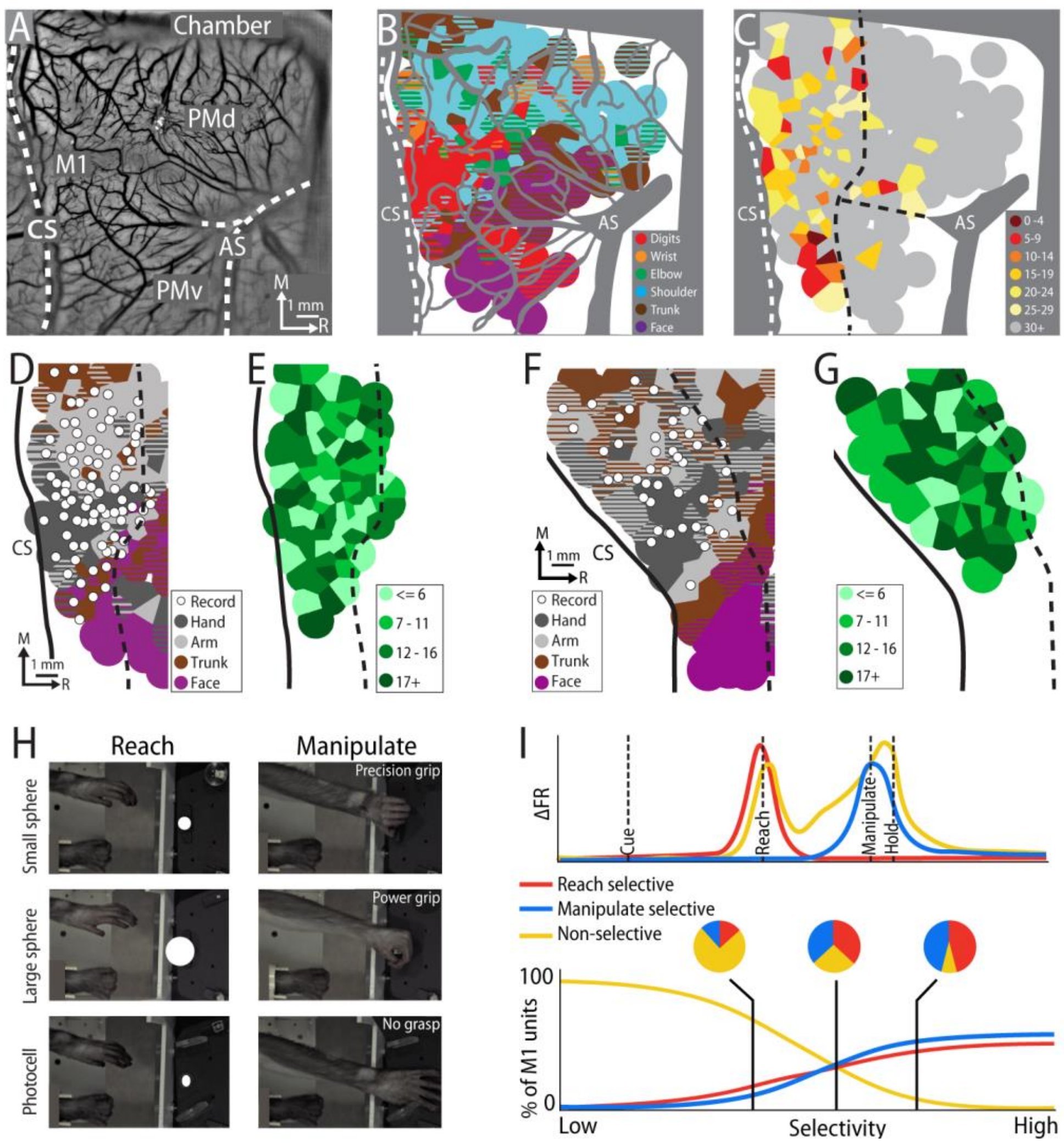

**Figure 11. Investigating the spatiotemporal dynamics of single unit (SU) activity recorded from a laminar multielectrode in M1 (A) Image of the cranial chamber from monkey G. Native dura was surgically replaced with transparent membrane. Major landmarks are depicted with dashed lines: central sulcus (CS), intraparietal sulcus (IPS), and arcuate sulcus (AS). (B) Intracortical microstimulation sites (n = 217) classified**

**according to ICMS-evoked movement. Major blood vessels and edge of the chamber are masked in gray. Voronoi tiles (0.75 mm radius) are color-coded according to movement. Striped tiles represent dual movements. (C) Same motor map in (B) color-coded according to the minimum current amplitude (μA) required to evoke movement. Estimated M1/premotor border (dashed line) drawn at the transition from low (< 30 μA) to high (>= 30 μA) current thresholds. (D) Same motor map as in (A), but wrist and digit sites are classified as hand (dark gray); shoulder and elbow sites are classified as arm (light gray). White dots depict microelectrode recording sites (n = 84). (E) Voronoi tiles (0.75 mm radius) of recording sites are color-coded according to the number of single units observed (SU = 800). (F)-(G) Same maps for monkey S (n = 53; SU = 795). (H) Still frames captured from video recordings taken during the reach (left column) and manipulate (right column) phase of movement for the small sphere (top row), large sphere (middle row), and photocell (bottom row). The manipulate posture was held for 1000 ms (small and large sphere) or 250 ms (photocell). The task is performed with the left forelimb and the right forelimb is restrained. (I) Top. The firing rate of three representative units with distinct temporal profiles in relation to Cue, reach and hold onset. The red and blue trace represent units with selective spiking to the reach and manipulate phase, respectively. The yellow trace exhibits non-specific modulation. Bottom. A continuum of hypothetical distributions of M1 units classified according to the temporal profiles established in the top panel. Three points along the continuum are marked with solid black lines; the pie graphs visualize the corresponding distribution of unit types.**

## 3.2 Methods

### 3.2.1 Animals

The right hemisphere was studied in two male macaque monkeys (Macaca mulatta). Monkeys were 9-10 years old during data collection and weighed 9-11 kg. All procedures were approved by the University of Pittsburgh Animal Care and Use Committees and followed the guidelines of the National Institutes of Health guide for the care and use of laboratory animals.

### 3.2.2 Head post and recording chamber

After an animal acclimated to the primate chair and training environment, a head-fixation device was secured to the occipital bone and caudal parts of the parietal bone. Task training with head-fixation started after ~1 month (monkey G) or ~9 months (monkey S) and lasted for ~22 months. At the end of this training period, the monkey was considered ready for neural recordings. A craniotomy was performed for implanting a chronic recording chamber (30.5 x 25.5 mm internal dimensions) over motor and somatosensory cortical areas. The chamber was secured to the skull with ceramic screws and dental cement. Within the recording chamber, native dura was resected and replaced with a transparent silicone membrane (500 µm thickness) that we fabricated from a mold. Protocols for the artificial dura have been previously described in detail (Arieli & Grinvald, 2002; Ruiz et al., 2013). The walls of the artificial dura lined the walls of the recording chamber. The floor of the artificial dura (i.e., the optical window) was flush with the surface of cortex and facilitated visualization of cortical blood vessels and landmarks (Figure 11A). The walls and floor of the artificial dura delayed regrowing tissues from encroaching underneath the optical window. Single electrodes and linear electrode arrays were readily driven through the optical window without permanent deformation.

### 3.2.3 Reach-to-grasp task

Monkeys performed a reach-to-grasp task while head fixed in a primate chair. The left forelimb was used, and the right forelimb was secured to the waist plate. The task apparatus was positioned in front of the animal. A stepper motor rotated the carousel in between trials to present a target ~200 mm from the start position of the left hand. The relative position of the target made it directly visible and was also positioned to encourage consistent reach trajectories across conditions and trials. Task instruction was provided with LEDs mounted above the target. Photocells were embedded in multiple locations within the apparatus to monitor hand location and target manipulation. An Arduino board (Arduino Mega 2560, www.arduino.cc) running a custom script (1 kHz) controlled task parameters, timing, and logged the monkey's performance on each trial. The task involved four conditions presented in an event-related design (1 successful trial/condition/block). Condition order was randomized across blocks.

#### 3.2.3.1 Two reach-to-grasp conditions.

In a successful trial, the animal had to reach, grasp, lift, and hold a sphere. The small sphere condition (12.7 mm diameter) and the large sphere condition (31.8 mm diameter) were used to motivate precision and power grips, respectively (Figure 11H, first two rows). Spheres were attached to rods that moved in a vertical axis only. Task rules were identical for both conditions and are therefore described once. To initiate a trial, an animal placed its left hand over a photocell embedded in the waist plate. Covering the photocell for 300 ms turned on an LED, which signaled the start of the trial. Holding this start position for 5000 ms triggered the Cue, which was a blinking LED. The animal had 2400 ms (monkey G) or 2550 ms (monkey S) to reach, grasp, and lift the sphere; time limits were also set for each phase. Lifting the sphere by 15 mm turned the blinking LED solid, which signaled the beginning of the hold phase. Maintaining the lifted position for 1000 ms turned off the LED, which instructed the animal to release the object and withdraw its hand back to the start position within 900 ms. Maintaining the start position for 5000 ms triggered a tone and LED blinking. After an additional 2000 ms in the start position the trial was considered successful; tone and LEDs turned off and water reward was delivered. The animal could not initiate a new trial for another 3000 ms. Failure to complete any step within the allotted time window resulted in an incorrect trial signaled by a 1500 ms tone and a 5000 ms timeout in which the apparatus was unresponsive to the monkey's actions. After the timeout, a new trial could be initiated with hand placement in the start position. Across both monkeys, the median failure rate per session was 15% (IQR = 6-32%) in the precision grip condition and 10% (IQR = 5-23%) in the power grip condition.

#### 3.2.3.2 Reach-only condition.

The target was a photocell embedded into the surface of the carousel (Figure 11H, last row). The photocell was visible to the monkey but was not graspable. The Go Cue was the same as the one used in the reach-to-grasp conditions, but here it prompted the monkey to reach and place its hand over the photocell. The hand had to cover the photocell for >220 ms (monkey G) or >320 ms (monkey S). All other task rules and steps were identical to the reach-to-grasp condition. Across both monkeys the median failure rate per session was 19% (IQR = 7-40%).

#### 3.2.3.3 Withhold condition.

In a successful trial, the monkey had to maintain its hand in the start position for ~10 s. Trial initiation was identical to the other conditions. Holding the start position for 5000 ms triggered the Withhold Cue, which was distinctly different from the Go Cue in the movement conditions. Maintaining the start position for another 2800 ms triggered a tone and LED blinking. After an additional 2000 ms in the start position the trial was considered successful and rewarded. Removing the hand from the start position at any time resulted in an incorrect trial and the same consequences described in a failed reach-to-grasp trial. Across both monkeys the median failure rate was 2% (IQR = 0-5%).

#### 3.2.3.4 Task phases

To relate the recorded neural activity to behavior, we focus on four key phases of the movement conditions. (1) *Cue*: 0 to 200 ms from cue onset. (2) *Reach*: -150 to 50 ms from reach onset. (3) *Manipulate*: -200 to 0 ms from target hold onset. (4) *Withdraw*: -150 to 50 ms from withdraw onset. We limited the phases to 200 ms windows to minimize overlap between the reach and manipulate phases, which occur in close succession.

### 3.2.4 Motor mapping

We used intracortical microstimulation (ICMS) to map the somatotopic organization of frontal motor areas. In monkey S, all sites (n=158) were investigated with a microelectrode in dedicated motor mapping sessions. We used the same approach in >50% of the sites (n=118) in monkey G. The remaining sites (n=99) were mapped with a linear electrode array at the end of recording sessions. In the dedicated motor mapping sessions, the monkey was head-fixed in the primate chair and sedated (ketamine, 2-3 mg/kg, IM, every 60-90 minutes). This mild sedation reduced voluntary movements but did not suppress reflexes or muscle tone.

A hydraulic microdrive (Narishige MO-10) connected to a customized 3-axis micromanipulator was attached to the recording chamber for positioning a tungsten microelectrode [250 μm shaft diameter, impedance = 850 ± 97 kΩ (mean + SD)] or a platinum/iridium microelectrode [250 μm shaft diameter, impedance = 660 ± 153 kΩ (mean + SD)]. A surgical microscope aided with microelectrode placement in relation to cortical microvessels. The microelectrode was in recording mode at the start of every penetration. Voltage differential was amplified (10,000×) and filtered (bandpass 300–5000 Hz) using an AC Amplifier (Model 2800, AM Systems, Sequim, WA). The signal was passed through a 50/60 Hz noise eliminator (HumBug, Quest Scientific Instruments Inc.) and monitored with an oscilloscope and a loudspeaker. As the electrode was lowered, the first evidence of neural activity was considered 500 μm below the pial surface.

The microelectrode was then switched to stimulation mode and the effects of ICMS were evaluated at >4 depths (500, 1000, 1500, 2000 μm). Microstimulation trains (18 monophasic, cathodal pulses, 0.2 ms pulse width, 300 Hz) were delivered from an 8-Channel Stimulator (model 3800, AM Systems). Current amplitude, controlled with a stimulus isolation unit (model BSI-2A, BAK Electronics), was increased until a movement was evoked (max 300 μA). The stimulation threshold for each depth was the current amplitude that evoked movement on 50% of stimulation trains.

One experimenter controlled the location and depth of the microelectrode. A second experimenter, blind to microelectrode location, controlled the microstimulation. Both experimenters inspected the evoked response and discussed their observations to agree on the active joints (i.e., digits, elbow, etc.) and movement type (flexion, extension, etc.). Movement classification was not cross-checked against EMG recording or motion tracking. The overall classification for a given penetration included all movements evoked within 30% of the lowest threshold across depths; striped tiles indicate multiple joints were evoked within that range. The location of each penetration (500-1000 μm apart) was recorded in relation to cortical microvessels. Color-coded maps were generated from this data using a voronoi diagram (MATLAB *voronoi* function) with a maximum tile radius of 750 μm (Figure 11B). Striped tiles are sites that evoked movements around multiple segments of the forelimb or multiple body parts. The rostral border of M1 was marked to separate sites with thresholds <30 μA from higher threshold sites (Figure 11C). Motor maps were further simplified by consolidating site classifications into broader categories (e.g., elbow and shoulder became arm) (Figure 11D & 8F).

The mapping protocol was similar for penetration sites stimulated with a linear electrode array (32 or 24 channels, 15 µm contact diameter, 100 µm inter-contact distance, 210-260 µm probe diameter; V-Probe, Plexon). Each penetration was mapped ~2.5 hours after the linear array was inserted into cortex, which was also the end of the neural recordings during task performance. Only 1 penetration was mapped per session. Microstimulation parameters were identical to the ones used with the microelectrode but were controlled here using a Ripple Neuro (Scout model, Salt Lake City, UT). Channels were stimulated one at a time and every other channel was used. One experimenter controlled the microstimulation and classified the evoked movements.

### 3.2.5 Neural recordings

We recorded neurophysiological signals from M1 zones that are on the dorsal surface of the precentral gyrus (i.e., old M1). In every recording session, a linear electrode array was acutely inserted orthogonally to the cortical surface. Two types of arrays were used. (1) V-probe: 32 or 24 channels, 15 µm contact diameter, 100 µm inter-contact distance, 210-260 µm shaft diameter (Plexon Inc). (2) LMA-v2: 32-channel, 12.5 – 15 µm contact diameter, 100 µm inter-contact distance, 300 µm shaft diameter (Microprobes). Array position was controlled with a custom 3-axis micromanipulator outfitted with a hydraulic microdrive (Narishige MO-10). We used a surgical microscope to guide each penetration in relation to cortical microvessels, which were landmarks for registering these penetrations to the motor map (Figure 11D & 11F). At the first sign of neural activity in the deepest recording channel, we relied exclusively on the hydraulic microdrive to advance the linear array. Once ~90% of channels were in cortex, we allowed ~45 minutes for the array to settle. The monkey sat quietly in the dark during that period and minor adjustments were made to the depth of the array to optimize single unit isolation across channels. Raw signals were filtered (bandpass 0.3 – 7,500 Hz) and recorded at 30 kHz with a Ripple Neuro system (Scout model, Salt Lake City, UT). In a separate data stream, raw signals were filtered (bandpass 250 – 7,500 Hz) for spike extraction. A threshold (typically 2-3 σ) was manually set for each channel, and distinguishable waveforms were flagged as units using time-amplitude window discriminators. Spike-waveforms, spike-times, and event-times were recorded at 30 kHz. Throughout each recording, micro adjustments were made with the hydraulic microdrive to limit waveform drift across recording channels.

### 3.2.6 Spike sorting

The spike waveforms recorded for each channel were sorted using Offline Sorter (Plexon Inc). Two criteria had to be met for us to classify an isolated single unit. (1) Peak-to-trough amplitude of the average waveform was >5x the mean voltage from the entire recording from the same channel. (2) Violations of the inter-spike interval (1.8 ms) were <0.5% of the waveforms assigned to a unit. Only single units (n=1,573) were considered for analyses. Those units were recorded from 136 penetrations (84 from monkey G; 52 from monkey S) that we placed in the arm, hand, and trunk zones of M1. For reference, the spatial organization of the units is shown side-by-side with the motor maps (Figure 11E & 11G).

### 3.2.7 Spike density function

Only successful trials were analyzed. For a given unit, the spike times in a trial were aligned to the onset of three behavioral events: reach, hold, and withdraw. Rasters were plotted for each unit to show trial spike trains in 10 ms bins (Figure 12A & 12C). The peristimulus time histogram (PSTH) of each trial was approximated from the spike train using a gaussian kernel density estimate with a bandwidth of 15 (i.e., 150 ms). PSTHs were averaged across trials to generate unit PSTHs for each condition (Figure 12B & 12D).

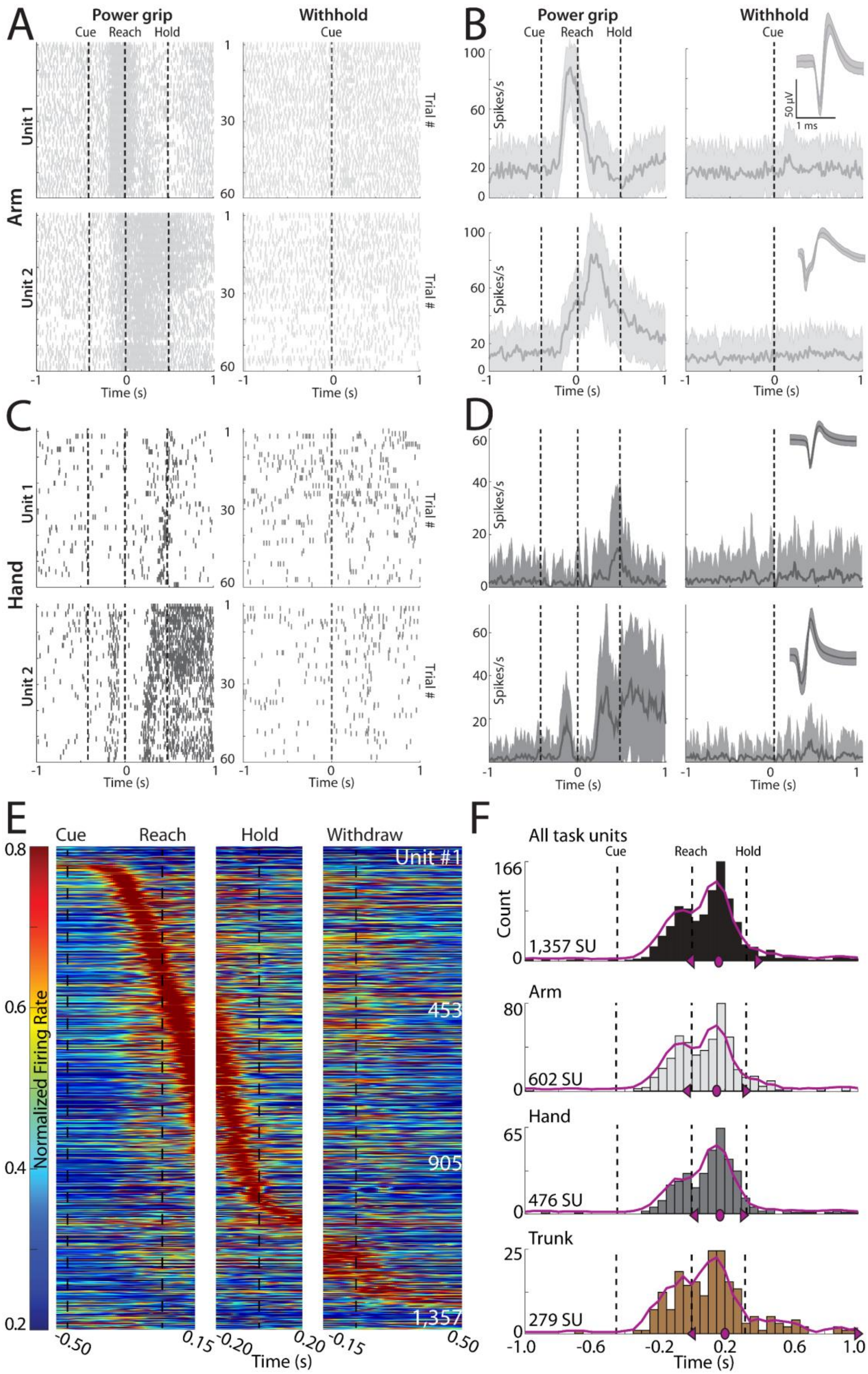

A
Power grip
Cue
Reach
Hold
Withhold
Unit 1
Unit 2
Arm
Trial #
Time (s)
B
Spikes/s
50 µV
1 ms
C
Hand
D
E
Reach
Withdraw
Unit #1
453
905
1,357
Normalized Firing Rate
F
All task units
Count
1,357 SU
602 SU
476 SU
279 SU
Trunk

**Figure 12. M1 units peak all throughout movement regardless of somatotopic representation Raster plots for 2 example units recorded from the M1 arm representation. Spike times of individual trials (rows) are accumulated into 10 ms bins from 1 s before to 1 s after reach onset. The Cue, reach and hold onset are marked as dashed black lines for the power grip condition (left). The Withhold Cue onset is marked in the withhold condition (right). (B) Peristimulus time histograms (PSTHs) of the 2 example units in (A). Firing rate was estimated in each trial using a gaussian kernel density estimate with a kernel bandwidth of 15 (150 ms). The mean trial PSTH is plotted as a solid grey line and bounded by one standard deviation (shaded area). Insets show the mean waveform (solid line) of the spikes from the recorded unit. Shaded area captures one standard deviation above and below the mean. (C)-(D) Same as (A)-(B) for units that were recorded in a separate session from the M1 hand representation. (E) Firing rates of 1,357 SUs for the power grip condition. Each row in the heatmap visualizes the average firing rate of an individual unit. The firing rate of each unit was normalized by the peak rate during movement. Hot colors indicate unit firing near its maximum rate. To avoid jitter, PSTHs were aligned and visualized at three time points: reach, hold, and withdraw onset (dashed black line, separated by white bars). Units were sorted from top to bottom according to the time that each unit achieved maximum firing. (F) Histograms tally the units that peaked in 50 ms bins from 1 s before to 1 s after reach onset. The top histogram was generated from all task units. Subsequent histograms separate the unit peak times by somatotopy. The probability density of each distribution was estimated using a 4-bin moving mean kernel (purple lines). The median peak and 25th and 75th percentile times of the distribution are marked in purple along the x-axis.**

### 3.2.8 Task modulated units

To determine if a unit was task modulated, we compared its firing rates during *movement* to those at *baseline*. Trial PSTHs were first normalized by the mean firing rate at *baseline*. In the *baseline phase* [-4000 to -1000 ms from Cue], the monkey's hand was in the start position as the monkey had initiated a trial and was waiting for the next Cue. The *movement phase* [+200 ms from Cue through to hold onset] encompasses reaching and manipulation of the target object. For each trial, we calculated the mean firing rate of the *movement phase*. We also calculated the mean firing rate during the *baseline phase*, but here we averaged 12 bootstrapped samples trial. Each baseline sample was matched in duration to the *movement phase* from the same trial. Next, we directly compared (paired t-test) firing rates in the movement and baseline phases. For each single unit, the statistical comparison was done independently for each of the three movement conditions. A unit was considered task modulated if any of the tests reported significant difference between *movement* and *baseline* ($p<0.05$). Subsequent analyses are limited to the task modulated units (n = 1,357; 93% of total single units).

### 3.2.9 Task unit selectivity classification

Every task modulated unit was further classified according to its selectivity, if any, for the reach or manipulated phases. Thus, for every trial, we calculated the mean firing rate of the reach phase and of the manipulate phase. We then directly compared the two distributions (paired t-test). Units with significantly ($p<0.05$) higher firing rates in the reach phase were classified as reach selective. Units with significantly higher firing in the manipulate phase were classified as manipulate selective. Units with no statistical difference were classified as non-selective. Task phase selectivity was assessed independently for each movement condition. We then generated average spike density functions for the reach selective units, manipulate selective units, and the non-selective units.

### 3.2.10 Task unit selectivity index

For every task-modulated unit, we calculated a selectivity index (SI). Our objective was to report on a ratio scale a unit's preference, if any, for the reach or manipulate phases:

$$SI = \frac{FR_R - FR_M}{FR_R + FR_M}$$ Equation 1. Selectivity Index

$FR_R$ is the mean firing rate in the reach phase. $FR_M$ is the mean firing rate in the manipulate phase. The range of SI values is +1 to -1, where +1 indicates that firing was exclusive to the reach phase and -1 indicates that firing was exclusive to the manipulate phase. A value of 0 indicates that firing was equal in the reach and manipulate phases.

The SI of each unit was calculated independently for each condition. Input values were calculated from the average spike density function. The start and end times of each phase were determined from averaging those times across the trials of a given unit.

### 3.2.11 Muscle activity

Electromyography (EMG) was conducted in 7 forelimb muscles. EMG was recorded concurrently with neural recordings or in separate sessions. After head fixation, the monkey was lightly sedated with a single dose of ketamine (2-3 mg/kg, IM). Sedation was confirmed from a reduction in voluntary movements with the working forelimb. At that point, pairs of stainless-steel wires (27 gauge, AM Systems) were inserted percutaneously into each muscle (~15 mm below skin). Three arm muscles were targeted (1) deltoideus, (2) triceps brachii, and (3) biceps brachii. Four extrinsic hand muscles were targeted (1) extensor carpi radialis brevis (ECRB), (2) flexor carpi radialis (FCR), (3) extensor digitorum 4-5 (EDC), and (4) flexor digitorum superficialis (FDS). The task started 45-60 min after sedation. The monkey was fully alert by that point and showed no lingering effects of sedation. The non-working forelimb was restrained and therefore could not tamper with the EMG wires.

EMG signals were filtered (bandpass 15-350 Hz) and recorded at 2 kHz with the same Ripple Neuro system that we used for neural signal acquisition. Recorded signals were segmented into trials and their power spectral density was estimated with a discrete Fourier transform (MATLAB *fft* function, Natick, MA). Trials with power >7 $\mu V^2$ in the 1-14 Hz range were presumed to have artifact and were excluded from further analysis. EMG signals were rectified and smoothed with a 100 ms sliding window (MATLAB *filtfilt* function). Trials were normalized to baseline activity, which was the mean voltage in the baseline phase. Then, each trial was aligned to the onset of three behavioral events: reach, hold, and withdraw. For each movement condition, arm muscle trials and hand muscle trials were independently averaged to obtain summary reports on arm and hand activity. The averages were weighted by the number of trials to ensure that individual muscles contribute equally. Thus, for the group of arm muscles, each muscle was weighted as the ratio of the number of trials from the individual muscle to the total number of trials with EMG arm activity (2,489-2,938 trials/muscle). The same weighting strategy was applied to the group of hand muscles (2,482 – 2,556 trials/muscle).

### 3.2.12 Decoding

We investigated the extent to which movement conditions could be decoded from the recorded unit activity. We were particularly interested in relating classification accuracy to the somatotopic locations of the units (e.g., arm zone vs hand zone) and to the phase selectivity of the units (e.g., reach selective vs manipulate selective). Unit activity in each task phase was modeled using a Poisson distribution with the $\lambda$ parameter calculated from the average spike count in the phase (Shadlen & Newsome, 1998). Thus, in each phase, the likelihood of a unit spiking $x$ times was calculated as:

$$P(x;\lambda) = \frac{e^{-\lambda}\lambda^{x}}{x!}$$ Equation 2. Poisson Distribution

We used Naïve Bayes classifiers to predict task conditions from the spike counts of individual units (Lange et al., 2023). The principle underlying Bayesian inference here is that each observation of condition classification is determined from the maximum a posteriori probability (MAP) among conditions. The posterior probability of each condition is the product of the prior probability and the marginal likelihood that the observation occurs for each condition. In the context of our model, we calculate the posterior probability (i.e., $P(c|x)$) of a given trial with $x$ observed spikes to a given condition $(c)$ as proportional to the product of the prior probability ($P(c)$), and the likelihood that $x$ spikes are observed for the condition $P(x|c)$ (as calculated from Equation 1). From the set of conditions $(C)$, the posterior probability of observing $(x)$ spikes in condition $(c)$ is:

$$P(C = c|x) \propto P(C = c) \times P(x|C = c)$$ Equation 3. Posterior Probability

Note that the prior probability in our model is uniform across conditions (i.e., equal number of trails across conditions). Thus, prior probability would not add meaning here and is therefore not included in the calculations. To be comprehensive, the prior probability was included in Equation 2 but will be dropped in subsequent equations. Once we compute the posterior probabilities for each condition, we assign the classification to the condition with the highest probability.

#### 3.2.12.1 Classifying conditions using 1 unit

To provide intuition for classifier training and testing, we first describe the process for a single unit. For each condition, we generated 10 sets of trials (i.e., 10-fold cross-validation). Each fold consisted of 20 trials/condition that were randomly selected from all available trials. The 20 trials were then randomly assigned to the training set (n=18) or the testing set (n=2). The training trials were averaged to obtain an expected spike count ($\lambda$) for modeling the Poisson distribution for each condition. The 10-set cross validation resolves variance resulting from (1) trial selection for a fold, (2) trial assignment to training and testing sets.

Subsequently, the estimated expected spike count distributions of each condition were used to calculate the posterior probability of the 6 test trials (3 conditions x 2 trials/condition). The condition that returns the maximum posterior probability is considered as the classification of the test trial. This procedure is repeated for the 6 test trials to determine the classification accuracy of a fold. The mean accuracy across the 10 folds is considered the classification accuracy for a single iteration.

Five hundred iterations are performed to obtain a bootstrapped distribution of classification accuracy. Our rationale for this bootstrapping approach is to control the variance from randomly sampling the task-modulated units as inputs to the classifier. Note, the average classification accuracy obtained from the 500 iterations does *not* capture the accuracy for any one specific unit, but instead reports the average accuracy of a unit sampled from task-modulated units.

#### 3.2.12.2 Classifying conditions using multiple units

We scaled up the classification procedure so that it considered inputs from varying numbers of units [1, 2, 5, 10, 15, 25, 35, 50, 75, 100]. Modeling unit responses as probability distributions from the exponential family (e.g., Poisson) allows us to assume conditional independence between units (Ma et al., 2006). Conditional independence is what defines the Bayesian classifier as Naïve; the input from any given unit to be independent from all other units. As such, each unit comprised its own set of posterior probabilities corresponding to each condition. Because the classifier assumes conditional independence between units, the posterior probabilities are simply the product of posterior probabilities across units. Thus, we simplify Equation 2 by removing the uniform prior (constant) and calculate the posterior probability for a given condition that considers *N* units as:

$$P(c|x_N) = \prod_{u=1}^{N} P(x_u|c)$$

Equation 4. Maximum Likelihood Estimation

Note that unit responses were pooled from different recording sessions and monkeys. For a given test trial, we compute independent posterior probabilities for each unit. The posterior probability for a condition is estimated from the product of each unit's ($N$) posterior probabilities for that condition. Thus, each unit 'votes' with the vote weighted according to the likelihood that the current observation corresponded to the respective condition. The predicted condition is the one that yielded maximum posterior probability. To evaluate classification accuracy for specific populations of single units, we preselected units based on three criteria (1) *somatotopic zone:* trunk, arm, hand, face; (2) *task phase selectivity:* reach-selective, manipulate-selective, non-selective; (3) *cortical depth:* superficial layers, and deep layers).

### 3.2.13 Statistical Analyses

We used paired t-tests (MATLAB, *ttest*) for within unit analyses. Specifically, to determine if unit activity was task modulated and to determine if unit activity was selective for task phase. In those tests, trials supplied the distributions of values analyzed. We used repeated measures ANOVA (SPSS) to determine whether unit activity differed across somatotopic zones, task phases and conditions. Somatotopic zones were treated as the between subject variable, and task phases and conditions were treated as within subject variables. In this analysis, the average spike density functions of the individual units supplied the distributions of values analyzed. EMG activity was treated comparably to determine whether muscle activity differed between the proximal and distal forelimb, task phases and task conditions. Forelimb segments were treated as the between subject variable, and task phases and conditions were treated as within subject variables. Post-hoc tests for both models were corrected for multiple comparisons (Bonferroni). We used several one-way ANOVAs to assess differences between unit types in the classification accuracy of task conditions. For all analyses, post hoc tests were corrected for multiple comparisons (Bonferroni correction).

## 3.3 Results

Two monkeys performed an instructed forelimb movement task that consisted of four conditions: (1) reach-to-grasp with precision grip, (2) reach-to-grasp with power grip, (3) reach-only, and (4) withhold. Both monkeys completed a minimum of 20 successful trials/condition/session (median [IQR] = 60 [59-60] trials for monkey G; 31 [30-33] trials for monkey S). In both monkeys we obtained detailed motor maps to guide the neurophysiological recordings. The overall organization of the motor maps was consistent across monkey. The forelimb representation occupied most of the mapped territory and was flanked medially by trunk zones and laterally by trunk and face zones (Figure 11D & 11F). Within the M1 forelimb representation, the main hand zone (i.e., digits and wrist) was surrounded by an arm zone, or an arm and trunk zone. This nested organization is consistent with previous maps from macaque monkeys (Kwan, Mackay, et al., 1978; Sessle & Wiesendanger, 1982), squirrel monkeys (Card & Gharbawie, 2000), and even maps obtained with stimulus triggered averaging of EMG activity (Park, Belhaj-Saïf, et al., 2001a).

We leveraged the microvessel patterns visible through the artificial dura to plan our recording sites in relation to the motor map. In most recording sessions, we purposely targeted arm and hand zones in parts of M1 rostral to the central sulcus (Figurer 11D & 11F). Neurophysiological signals were recorded with an acute linear electrode array (1 penetration/ session; 32-channels in most penetrations) from 136 penetrations (monkey G= 84, monkey S=52). Our analyses of the neurophysiological recordings focused on single units, which we sorted offline. We identified 1,573 units (monkey G= 895, monkey S= 678) that met our criteria for single units (see Methods). Waveforms from four representative single units are shown in Figure 12B & 12D. Of the 1,573 units, 1507 (monkey G= 854, monkey S= 653 units) were modulated in at least one of the three movement conditions and were therefore considered task modulated. Figs 11E & 11G show the non-uniform spatial organization of the task-modulated units.

### 3.3.1 M1 single units encode all task phases

After inspecting the rasters and PSTHs of all task-modulated units, we made some general observations that we introduce with the activity profiles of representative task-modulated units. Firing rates were modulated from baseline in movement conditions, but not in the withhold condition (Figure 12A-D). Firing rates differed considerably across units but were relatively consistent and maintained the same temporal profile across trials of individual units. In some units, firing rate modulation was limited to a specific task phase. For the representative unit in Figure 12A-B (unit 1), activity increased sharply ~200 ms after the Go Cue, peaked during with reach onset, and returned to baseline before the object was grasped. This unit was therefore likely tuned to movements that occurred during reaching. We observed similarly narrow tuning in Figure 12C-D (unit 1), but here firing rate peaked on hold onset, which is when the hand stabilizes the grasped object after it has been lifted. Modulation was multi-phasic in the two other representative units. For unit 2 in Figure 12A-B, the firing rate began to increase before reach onset and peaked during object grasp (i.e., between the markers for reach and hold). For unit 2 in Figure 12C-D, firing rate modulation was more isolated for the reach and object grasp and was then sustained during the hold phase. We note that the different temporal profiles in Figure 12A-B were recorded from a single penetration in the M1 arm zone. Same applies to the differing profiles in Figure 12C-D, but the penetration here was in the M1 arm zone.

This heterogeneity in temporal profiles prompted us to investigate if firing rate peaks were organized across the population of M1 units. Thus, for every task modulated unit, we normalized the units average PSTH by the peak firing rate during movement. We normalized every unit's average PSTH by the peak firing rate during movement. We sorted the normalized PSTHs by latency of peak firing and plotted them into a heatmap (Figure 12E is for the power grip condition). The drifting red band indicates that across the population of M1 units (i.e., rows), firing rates peaks covered all movement phases in the task. Nevertheless, most of the peaks occurred between reach-onset and hold-onset, and therefore coincided with grasping of the target object.

A formal count of the number units as a function of peak firing time (50 ms bins) showed that the highest count was near the midpoint of reach and hold, coinciding with target grasp (Figure 12F, top row). We note the presence of an earlier, smaller, peak in unit count at ~200 ms before reach-onset and is therefore shaped by the number of units tuned for reaching. From the unit counts, we estimated the probability density function (PDF, 200 ms bins) of the time of peak firing rate (Figure 12F, purple). The estimated PDF more clearly captured the bimodal distribution of unit counts with a smaller peak preceding reach-onset and a larger peak coinciding with grasp. We then separated the unit counts by their spatial location within the motor map (Figure 12F) with an expectation of temporal shifts in peak counts across somatotopic zones. But we were surprised to find that the shape of the estimated PDF was largely conserved across arm, hand, and trunk. We noted only two subtle differences. First, the reach onset peak was slightly smaller in the hand zone than in the other two zones. Second, estimated PDF of the trunk zone had a rightward skew, which would have been shaped by the number of units that peaked during the hold phase (i.e., object grasped with the arm stretched and raised just below shoulder level).

### 3.3.2 Muscle activity is not a simple readout of M1 neural activity

To further understand temporal dynamics of M1 unit activity, we consider the average activity of the arm and hand representation. To generate the PSTH of a given unit, spike times from the beginning to the end of each trial were binned (10 ms) and smoothed using a gaussian kernel density estimate with a bandwidth of 15 (150 ms). Activity was baseline normalized and flagged as task-modulated if activity significantly deviated from baseline during the movement period. In each trial, signal was aligned to three behavioral time points: reach onset, object hold and withdraw onset. The trial PSTHs were then averaged together to produce the unit PSTH. All unit PSTHs were averaged together corresponding to the somatotopic representation of each unit.

The average PSTHs of units from the arm representation had similar temporal dynamics across movement condition (Figure 13A). Namely, activity increased ~250 ms before reach onset and peaked ~100 ms to reach onset. After a 100 ms plateau during the reach, activity increased again to an overall maximum peak ~100 ms before object hold, which coincided with the object grasping (power and precision conditions) or hand placement on the target (reach-only condition). During the hold phase, activity declined to a plateau that was sustained through the withdrawal phase, which returned the hand to the start position. Principally, activity modulates in two phases: (1) activity increases and peaks prior to reach and (2) increases and peaks post reach and pre object contact, respectively. These results corroborate the bimodal distribution of unit peak times discussed in Figure 12F. Thus, considering the equivalency between peak time distributions of the arm and hand representation, the temporal dynamics in the hand representation are similar. Indeed, the average PSTHs of units from the hand representation exhibit two phases of activity (Figure 13B). Following reach onset, movement conditions are distinguishable in the PSTHs of units from the hand representation compared to those from the arm representation. To further synthesize the temporal dynamics with respect to somatotopy, we define four time intervals to summarize activity.

Using the average PSTH of each unit, we averaged the activity in four 200 ms windows beginning at: Cue (C), 150 ms preceding reach onset (R), 200 ms preceding object hold (M), and 150 ms preceding withdraw onset (W). The span of each window is visualized in relation to the PSTHs in Figure 13A. For each condition, we visualized the distribution of unit activity from the arm and hand representation in each of the four phases (Figure 13C) and used a mixed model (see Methods) to test for effects. Note that the Cue phase was excluded from the model. The mixed model reported a main effect of phase ($F(2, 2{,}106) = 52.89$, $p < 0.001$) and condition (F(2, 2,106) = 8.69 p < 0.001) and an interaction effect of phase by condition (F(4, 4,212) = 6.54, p<0.001). Surprisingly, there was no main effect of somatotopy found for units from the arm and hand representation (F(1,1053) = 0.02, p = 0.90). However, there were interaction effects of phase by somatotopy (F(2, 2,106) = 6.19, p = 0.002), condition by somatotopy (F(2, 2,106) = 9.31, p<0.001), and phase by condition by somatotopy (F(2, 2,106) = 5.06, p = 0.006). As there was no effect of somatotopy independently, we sought to further define behavior on a temporal scale equivalent to the activity, so that we could be confident the results were in fact a fair depiction of the resulting behavior.

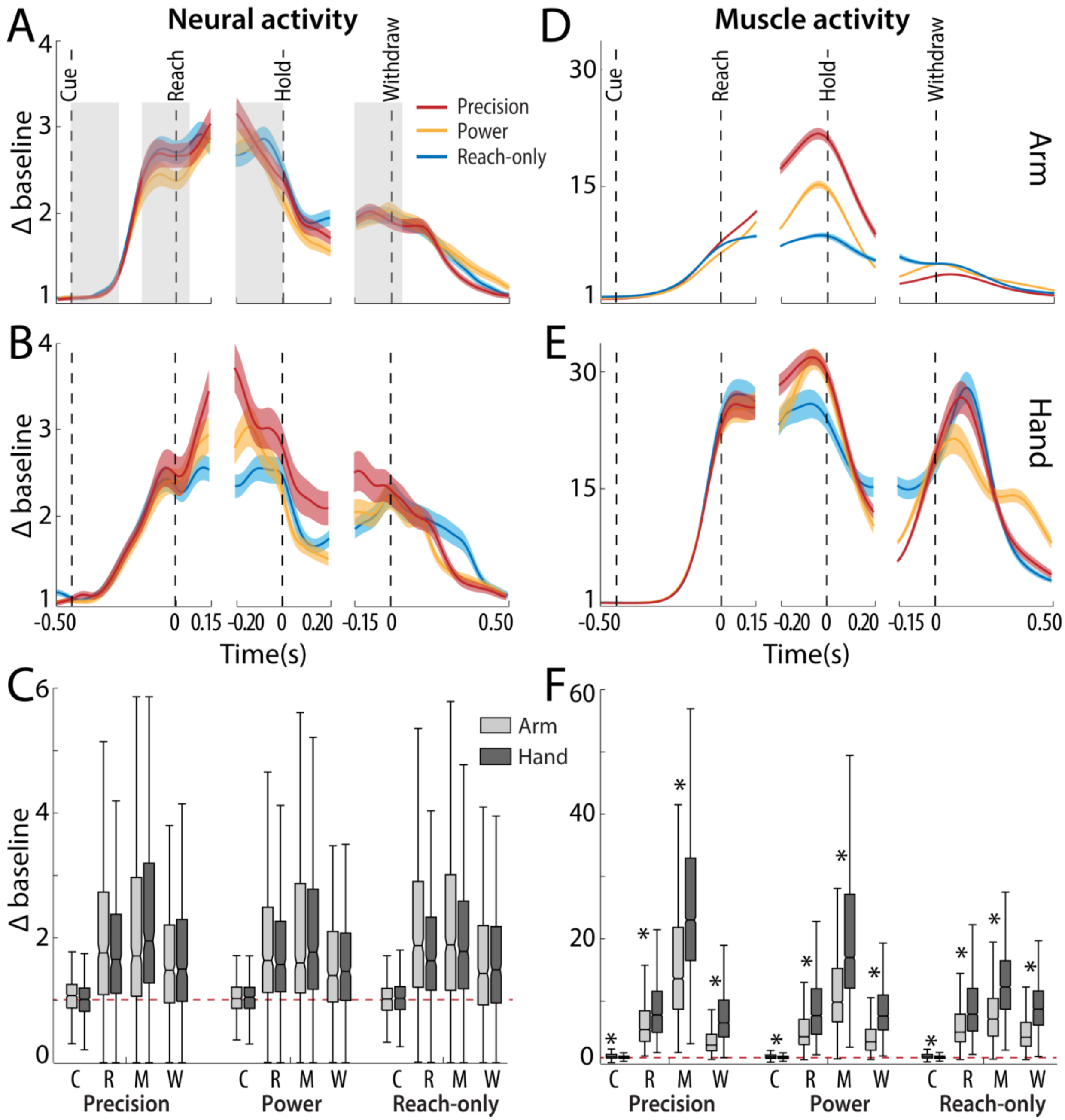

**Figure 13. Activity differences between arm and hand muscles are not reproduced in the neural activity of M1 arm and hand units (A) Baseline normalized average PSTHs of units recorded from the M1 arm representation. Shaded areas for the precision, power, and reach-only condition (red, orange, and blue, respectively) indicate the weighted standard deviation from each arm muscle. The time course of activity is visualized in three separate time windows (separated by breaks in x-axis), which correspond to three voltage trace alignments: the reach, hold, and withdraw onset. Grey shaded boxes depict 200 ms windows that define**

**phase boundaries used to calculate the mean activity across time of the cue, reach, manipulate, and withdraw phase in (C) and (F). (B) Baseline normalized average PSTHs of units recorded from the M1 hand representation (C) Box plots of the averaged neural activity from units in the M1 arm and hand representation for each movement condition in the four phases visualized in (A): cue, reach, manipulate, and withdraw (C, R, M, W). The solid black line in each box marks the median value of the distribution; box edges represent the 25th and 75th percentiles of the distribution. Whiskers span from the box edges to 1.5 times the interquartile range. Red dashed line indicates no change in activity from baseline (∆baseline = 1). (D) The baseline normalized average activity of arm muscles (deltoid, biceps, and triceps) calculated as the mean across trials weighted by the ratio of the total trials per muscle. respectively. (E) The normalized average time course of recorded hand muscles (extensor carpi radialis brevis, flexor carpi radialis, extensor digitorum 4-5, and flexor digitorum superficialis) calculated as in (D). (F) Box plots of the averaged arm and hand (light and dark grey) muscle activity for each movement condition in the four phases. Asterisks indicate significant differences (Bonforonni corrected, $p < 0.001$) between arm and hand activity.**

To model behavior with a spatial resolution comparable to the noted unit activity, we recorded EMG activity from three arm muscles (deltoideus, triceps brachii, biceps brachii) while monkeys performed the precision, power, and reach-only condition. The EMG signal for any individual muscle was segmented into individual trials, rectified, smoothed (100 ms moving mean), and normalized to baseline activity. For each condition, the muscle activity in the arm was summarized as the trial weighted mean from the deltoideus, triceps brachii, and biceps brachii (Figure 13D). The summarized arm muscle activity demonstrates an approximate version of the noted bi-phasic modulation of unit activity. Furthermore, the initial peak in activity of arm muscles is not as pronounced as the activity of units from the arm representation. In fact, arm muscle activity is distinguishable between conditions and scales according to the dexterity. Upon observing the disparity of arm muscle activity between conditions, we considered the activity of hand muscles.

We recorded and processed the EMG signal of four hand muscles (extensor carpi radialis brevis, flexor carpi radialis, extensor digitorum 4-5, and flexor digitorum superficialis) as specified for the arm muscle activity (Figure 13E). Again, we observe a bi-phasic response in the hand muscle activity for each condition. While not as discernable as the arm muscle activity between conditions, the second phase of hand muscle activity demonstrates separability of the grasp conditions from the reach-only condition. Consistent with statistical analysis of unit activity, there was a main effect of phase ($F(2, 10{,}932) = 1{,}041.79$ , $p<0.001$) and condition ($F(2, 10{,}932) = 6.79$, $p<0.001$) and an interaction effect of phase by condition ($F(4, 21{,}864) = 67.93$, $p<0.001$). However, in contrast to the undifferentiable unit activity between the arm and hand representation, there was a main effect of forelimb segment (e.g., activity from the arm or hand) on EMG activity (Figure 13F; $F(1,5466) = 428.17$, $p < 0.001$). In fact, post hoc tests revealed greater muscle activity in the hand than the arm in the reach and manipulate phase of every condition. The distinction between arm and hand muscle activity is incongruent with the equivalency between unit activity in the arm and hand representation. Therefore, we consider further classifying the organization of unit activity according to laminar depth.

It is possible that the discrepancy between the activity of forelimb muscles and units with respect to forelimb somatotopy derives from obfuscating unit activity. Considering units were recorded from various layers of cortex, unit activity could be confounded when averaging if temporal dynamics differ with respect to lamina. Indeed, organizing neurons and/or neural activity by way of cortical depth is a tenet in cortex (Hawken et al., 1988; Paulignan, MacKenzie, et al., 1991; Weiler et al., 2008). Therefore, we examined the average PSTHs of M1 arm and hand units separated by cortical depth (Figure 14). We classified units recorded from the first 16 channels of the electrode as units from superficial cortical layers and the last 16 channels as units from deeper cortical layers. In comparison to the previously discussed PSTHs, the activity between conditions is slightly more separable in both arm (Figure 14A & B) and hand (Figure 14C & D) units when further classifying units according to the cortical depth. Additionally, activity recorded deeper in cortex (Figure 14B & D) conveys greater modulation overall—albeit this distinction is more pronounced in the activity of units from the hand representation. The general resemblance among PSTHs generated according to both somatotopy and/or laminar depth led us to consider alternative organizations that could exist in M1.

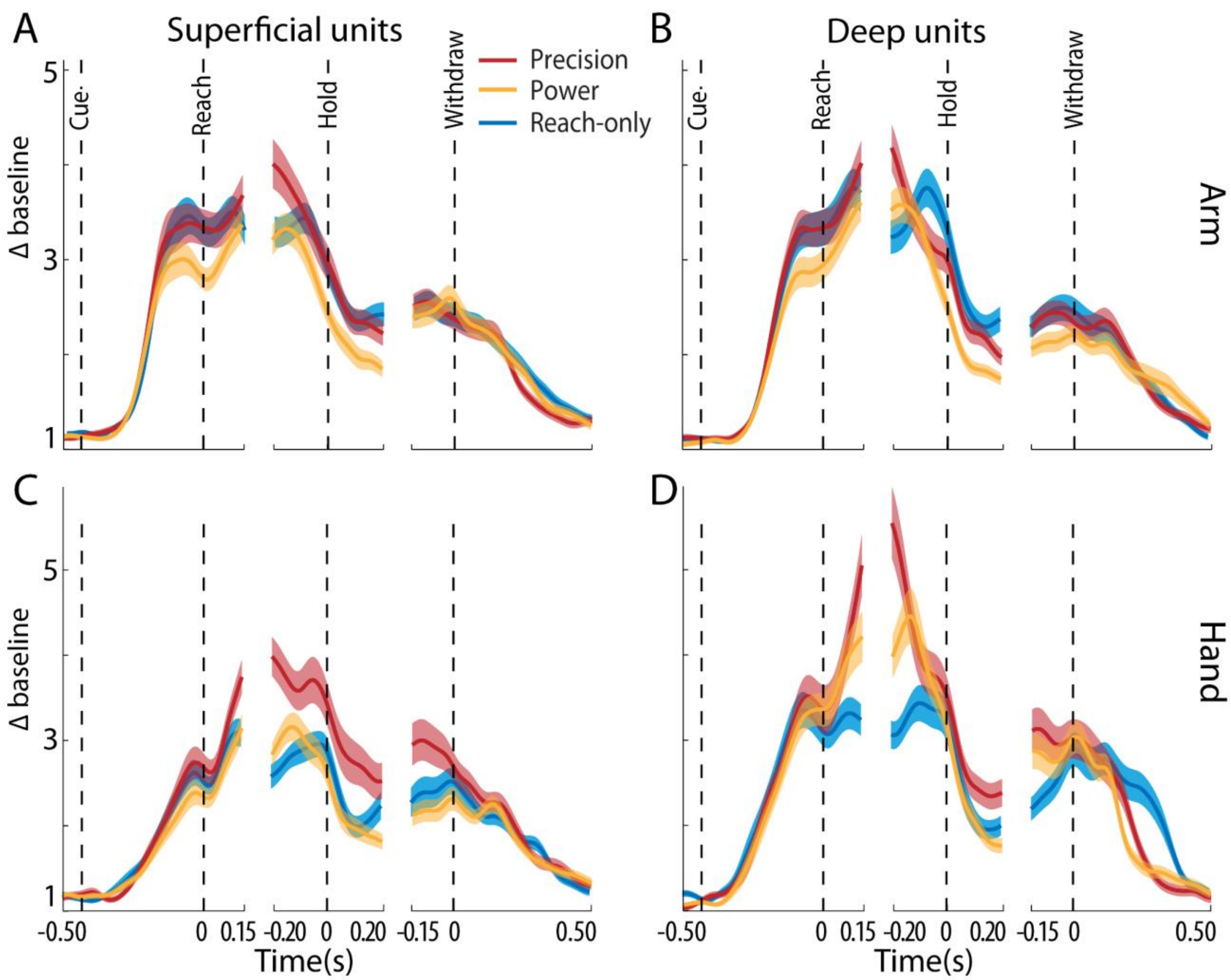

**Figure 14. Activity differences between conditions are expressed distinctively according to somatotopy and laminar depth (A)-(B) Baseline normalized average PSTHs of units recorded from superficial (1-16, A) or deep channels (17-32, B) of the linear multielectrode in the arm representation. Shaded areas for the**

**precision, power, and reach-only condition (red, orange, and blue, respectively) indicate one standard deviation from the mean unit activity. The time course is visualized in three separate time windows (separated by breaks in x-axis), which correspond to the three different spike-alignment points used to generate the PSTHs (reach, hold and withdraw alignment) (C)-(D) The same PSTHs of units recorded from the hand representation**

### 3.3.3 Using the M1 spatial organization to understand spatiotemporal dynamics of M1 activity

In the reach and manipulate phase of each movement condition, there is no difference in the modulation of units from the M1 arm representation and hand representation, despite marked differences in the activity of arm and hand muscles. Fortunately, our methodology enables us to investigate unit activity dynamics at a spatial resolution that exceeds the somatotopy and/or laminar depth. To that end, we visualized the activity of each recording site in the M1 forelimb representation in the different behavioral phases. Each site was summarized as the mean unit modulation from baseline of all units recorded at that site. The modulation in the M1 recording sites of monkey G is visualized in the cue, reach, manipulate and withdraw phase for the precision grip condition in Figure 15A. The dotted black line indicates the M1/pre-motor area border, and the solid black trace overlaid on the recording site tiles indicates the perimeter of the M1 hand representation. Contrary to the equivalency ascertained among the distributions of unit peak times grouped according to somatotopy, the M1 forelimb representation modulates heterogeneously in space and time.

In the previously discussed results when activity was summarized according to somatotopy, there were no differences detected in the modulation of M1 arm and hand units in any phase for the precision grip condition (Figure 13F). However, these maps clearly demonstrate that modulation is not distributed evenly throughout the M1 forelimb representation (Figure 15A). Furthermore, certain sites in both the arm and hand representation demonstrate temporal dynamics that contradict the previously noted pattern (i.e., greater modulation in the manipulate than the reach phase). Notably, there is a strong response found in a cluster of sites immediately rostral to the central sulcus within the hand representation during the manipulate phase. Interestingly, this localized response is diminished in the manipulate phase of the reach-only condition (Figure 15B). Thus, despite the homogeneity of M1 activity found when summarized according to somatotopy, the site maps capture unique dynamics in activity both spatially and temporally.

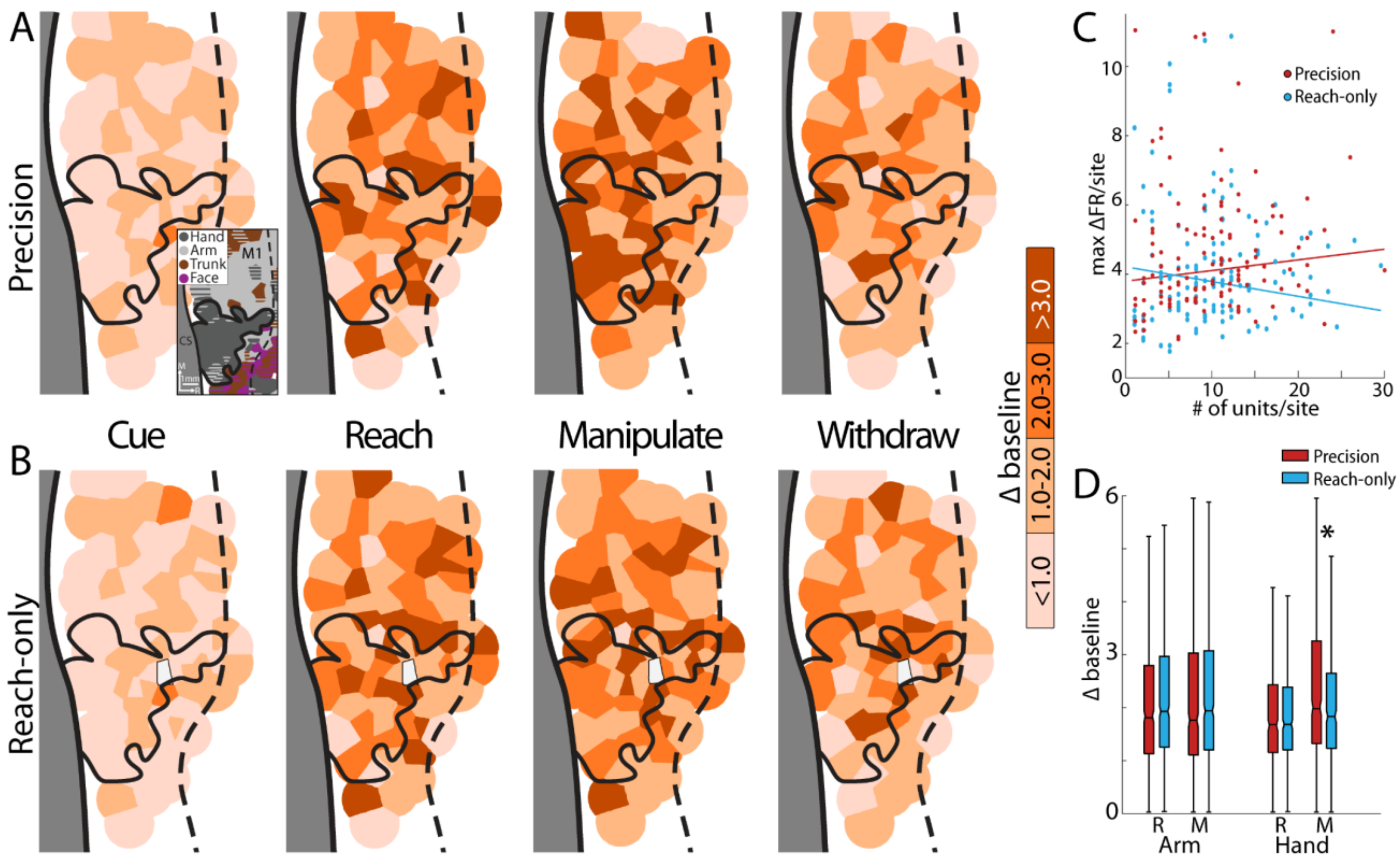

**Figure 15. Average modulation of units per recording site respond heterogeneously throughout M1 revealing condition differences related to somatotopy (A) The average firing rate change of recording sites (Voronoi tiles with 0.75 mm radius) for the precision condition from monkey G. Tiles are color coded according to the average firing rate of all SUs per site in the cue, reach, grasp and withdraw phase (columns). Darker colors**

**indicate high firing rates. Inset shows the simplified motor map for monkey G. The central sulcus is marked in black along the grey cortical mask. The dashed black line indicates the border between M1 and frontal premotor areas. The black trace over recording sites outlines the M1 hand representation. (B) The average firing rate change of recording sites visualized with Voronoi tiles for the reach-only condition from monkey G. The light grey tile indicates the site had no task-modulated units in the reach-only condition. (C) The maximum modulation (throughout movement) plotted as a function of the number of units per site (both monkeys) for the precision (red) and reach-only condition (blue). Linear regression analyses of each set of points generated non-significant models, with adjusted R2= 0.002 and R2= -0.005 for the precision and reach-only condition, respectively. (D) Box plots of the average firing rates of units (both monkeys) in the arm and hand representation in the reach and manipulate phase. The solid black line in each box marks the median value and the box edges represent the 25th and 75th percentiles of the distribution. Whiskers span from the box edges to 1.5 times the interquartile range. The asterisk marks a significant difference (Bonforonni corrected post-hoc analysis) in the activity modulation between the precision and reach-only condition.**

To ensure that the heterogeneity observed among the activity of M1 recording sites is, in fact, indicative of varied spatiotemporal dynamics, we plotted the modulation strength of a site in relation to the number of recorded units per site. In Figure 15C, the maximal modulation achieved during the movement period is plotted against the number of units recorded per site. The linear regression is modeled for both the precision grip and reach-only condition. In either case, the adjusted $R^2$= 0.0023 (F(1,133)= 1.3, p=0.25) and $R^2$= -0.0048 (F(1,122)= .41, p=0.52) values indicate no significant association between the number of units per site and the activity modulation. After establishing the unlikelihood of spatiotemporal heterogeneity among M1 sites resulting due to irregular unit sampling, we sought to establish quantifiable differences in the spatiotemporal dynamics of M1 units.

As noted previously, a cluster of sites in the M1 hand representation had a strong response in the manipulate phase of the precision condition but was absent in the reach-only condition (Figure 15A & B). To quantify this visual distinction, we aggregated the unit responses from both monkeys and compared the distributions of activity modulation for the precision grip and reach-only condition with respect to somatotopy (Figure 15D). Note that these distributions and the statistical model are identical to those in Figure 13, excluding the power grip condition and the withdraw phase. Post-hoc Bonferroni corrected t-tests indicated that units from the arm representation had no effect of condition in neither the reach nor manipulate phase. However, units from the hand representation increased activity more in the manipulate phase of the precision grip condition (1.89±4.44) than the reach-only (1.78±2.61). Furthermore, units from the hand representation had no difference in activity between the precision grip and reach-only condition. These results exemplify unique spatiotemporal signatures of activity that differentiate between movements and, thus, imply a potential organization of movement in the M1 forelimb representation.

To further explore potential distinctions in the modulation of M1 units from the forelimb representation, we consider the temporal specificity of units to the reach or manipulate phase. The selectivity strength of a unit to the reach or manipulate phase is defined as the modulation in the respective phase normalized by the summed modulation of both phases. The average magnitude of selectivity to the reach phase for the precision grip condition is visualized for M1 sites from monkey G (Figure 16A). As previously noted, M1 sites convey heterogeneity in the selectivity to the reach phase. Sites with greater selectivity to the reach phase appear outside of the hand representation. Indeed, the cluster of M1 hand sites that had a heightened response in the manipulate phase of the precision condition visually coincide with the cluster of sites that convey strong selectivity to the manipulate phase (Figure 16B). Hence, for the precision condition, a cluster of sites in the M1 hand representation fires substantially and selectivity in the manipulate phase. In contrast, the selectivity of M1 sites for the reach-only condition is spatially more homogenous in both the reach (Figure 16C) and manipulate phase (Figure 16D). Furthermore, M1 sites with strong selectivity to either phase are sparse in the reach-only condition. These maps further support a functional organization of movement found in both the M1 arm and hand representation.

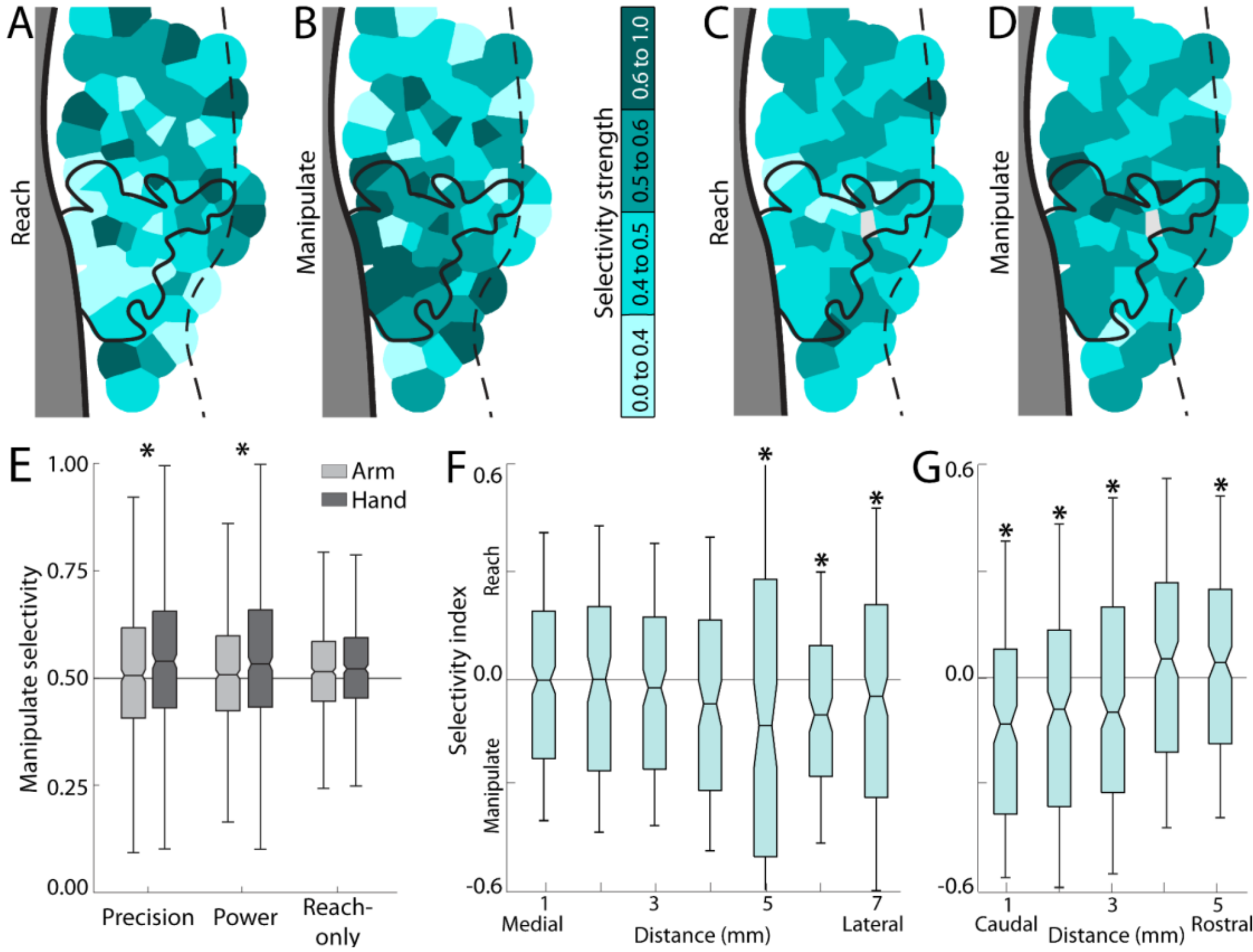

**Figure 16. Preferential modulation to the reach and manipulate phase is discernable between arm and hand units for conditions involving object grip (A) The average reach selectivity of units per recording site (Voronoi tiles with 0.75 mm radius) of the precision condition from monkey G. The reach selectivity of a unit is the activity in the reach phase normalized by the sum of the activity in the reach and manipulate phase. The central sulcus is marked in black along the grey cortical mask. The dashed black line indicates the border between M1 and frontal premotor areas. The black contour over recording sites outlines the M1 hand representation. (B) The average manipulate selectivity of units per recording site of the precision condition from monkey G. The manipulate selectivity of a unit is the activity in the manipulate phase normalized by the sum of activity in the reach and manipulate phase. (C)-(D) The average reach (C) and manipulate (D) selectivity of units per recording site of the reach-only condition from monkey G. The light grey tile indicates**

**the site had no task-modulated units in the reach-only condition. (E) Box plots of the manipulate selectivity from units in the arm and hand representation (both monkeys) for each condition. Note that, for any given unit, the reach selectivity is equal to the difference between 1 and the manipulate selectivity. The solid black line in each box marks the median value of the distribution and the box edges represent the 25th and 75th percentiles of the distribution. Whiskers span from the box edges to 1.5 times the interquartile range. Asterisks mark greater modulated in hand units than arm units according to Bonferroni corrected post hoc analyses. (F) Box plots of the selectivity index from all task-modulated units (both monkeys) for the precision condition. The location of the most medial recording site from each monkey and 1 mm along the lateral direction from that site define the bounds of the first group. Subsequent groups are bounded incrementally by 1 mm steps. The solid black line in each box marks the median value of the distribution and the box edges represent the 25th and 75th percentiles of the distribution. Whiskers span from the box edges to the 5th and 95th percentile of the distribution. Asterisks indicate that the mean selectivity of units in the respective distribution significantly modulates more in the respective phase (one-sample t-test, $p<0.05$). (G) The same conventions in (F) for the selectivity index of all task-modulated units from the most caudal recording site to 5 mm along the rostral direction.**

After visually identifying clustering in the temporal selectivity of M1 sites, we sought to establish quantifiable differences in the selectivity of M1 units from the arm and hand representation. We plotted the selectivity of units to the manipulate phase grouped according to somatotopy aggregated from both monkeys (Figure 16E). Note that we do not show the selectivity to the reach phase because, for any given unit, the selectivity to the reach phase is simply the difference between 1 and the selectivity to the manipulate phase. A 2-way analysis of variance indicated an main effect of somatotopy ($F(2, 5{,}466) = 44.67$, $p<0.001$) and an interaction effect between somatotopy and condition ($F(4, 5{,}466) = 10.04$, $p<0.001$). Consistent with the selectivity maps, post hoc Bonferroni corrected t-tests reported significantly greater selectivity to the manipulate phase in units from the hand representation than units from the arm representation for both grasp conditions. However, in the reach-only condition, there was no difference in selectivity between units from the arm and hand representation. Again, the results further support a functional organization of movement that is represented in the M1 forelimb representation.

The clustered response of M1 hand units identified for both the modulation and selectivity strength in monkey G suggests that activity may be organized at a level beyond the somatotopy. To simultaneously probe unit selectivity to both the reach and manipulate phase, we defined the selectivity index as the difference between the activity in reach and manipulate phase normalized by the sum of both phases of activity. Thus, positive values indicate selectivity to the reach phase and negative values indicate selectivity to the manipulate phase; a value of 0 indicates equal activity in the reach and manipulate phase. Then, to investigate the spatial organization of unit selectivity agnostic to somatotopy, we binned units in 1 mm intervals along the caudal/rostral and medial/lateral axes. The most medial site (i.e., comprises the M1 arm/trunk border) and the most caudal site (bounded by the central sulcus) in the M1 forelimb representation establishes the zero point in each axis for both monkeys. Distributions marked with an asterisk indicate that the units which comprise the 1 mm slice of cortex comprise a selectivity that significantly responds to either the reach or manipulate phase (student t-test, $p<0.05$). Considering that somatotopic differences were found for the grasp conditions but not the reach-only condition, we present the results from the precision grip condition, for simplicity (Figure 16F).

The selectivity of units is comparable between the reach and grasp phase for the 3 mm most medial in the M1 forelimb representation ($t(294) = -0.20$, $p=0.842$; $t(130) = -0.32$, $p = 0.753$; $t(176) = -0.883$, $p = 0.378$). Selectivity to the manipulate phase gradually increased ($t(137) = -1.31$, $p=0.194$) and subsequently peaked approximately 5 mm from the medial edge of the M1 forelimb representation ($t(107) = -2.40$, $p = 0.018$). Subsequently, units are less selective but still significantly tuned to the manipulate phase ($t(146) = -3.35$, $p=0.001$; $t(231) = -2.33$, $p=0.021$). None of the distributions created along the medial/lateral axis indicated significant selectivity to the reach phase. We repeated this analysis along the caudal/rostral axis. Interestingly, units recorded from the caudal edge of the M1 forelimb representation to approximately 3 mm rostral of the central sulcus demonstrate significant selectivity to the manipulate phase ($t(193) = -4.84$, $p<0.001$; $t(249) = -3.22$, $p<0.001$; $t(319) = -2.78$, $p=0.006$). The subsequent 2 mm of the M1 forelimb representation comprise units with no selectivity and selectivity to the reach phase, respectively ($t(232) = 0.45$, $p = .652$; $t(230) = 1.98$, $p=0.049$). These results establish spatial trends in the temporal selectivity of M1 unit activity, which recapitulates features of the somatotopy (e.g., increased selectivity to the manipulate phase in the lateral direction) and additionally establishes characteristics not necessarily inherent in the somatotopy (e.g., graded decrease in selectivity to the manipulate phase in the rostral direction).

### 3.3.4 Using M1 temporal dynamics to understand spatiotemporal dynamics of M1 activity

Now that we have established spatiotemporal features of M1 activity with respect to the somatotopic organization, we aimed to investigate spatiotemporal features of M1 activity with respect to temporal dynamics. We previously illustrated that units from the M1 forelimb representation peak all throughout the reach and manipulate phase (Figure 12E). Thus, we classified units according to their temporal profiles to further examine the dynamics of M1 spatiotemporal activity. Units from the M1 forelimb representation of both monkeys were grouped into three categories that demonstrated distinct temporal profiles: reach, manipulate and non- selective units. The phase selective units (reach or manipulate) modulated significantly more in the phase corresponding to the label name (paired t-test, $p<0.05$). Units that demonstrated non-significant differences in the modulation of the reach and manipulate phase were classified as non-selective units. After establishing temporal definitions of unit activity, we explored the PSTHs of the phase selective and non-selective units in both the M1 arm and hand representation.

Classifying units according to temporal dynamics demonstrate features that were not found in the previously discussed PSTHs. Reach selective units in the M1 arm and hand representation (Figure 17A & D) demonstrate condition separability in the activity magnitude. Furthermore, the magnitude of the activity scales in accordance with dexterity (e.g., greatest activity for precision condition). Manipulate selective units in both the M1 arm and hand representation (Figure 17B & E) peak prior to object contact with peaks varying according to condition. Explicitly, the activity of manipulate selective units peak earliest in the precision condition, followed by the power condition and finally by the reach-only condition. It is worth noting that the manner of condition classification is different for the reach and manipulate selective units; reach selective units demonstrate magnitude variation between conditions whereas manipulate selective units convey temporal variation between conditions. Furthermore, even PSTHs generated from non-selective units appear to distinguish between grasp conditions and the reach-only condition (Figure 17C & F). Following this classification scheme, we pursued investigating the spatial organization of units with respect to selectivity classification.

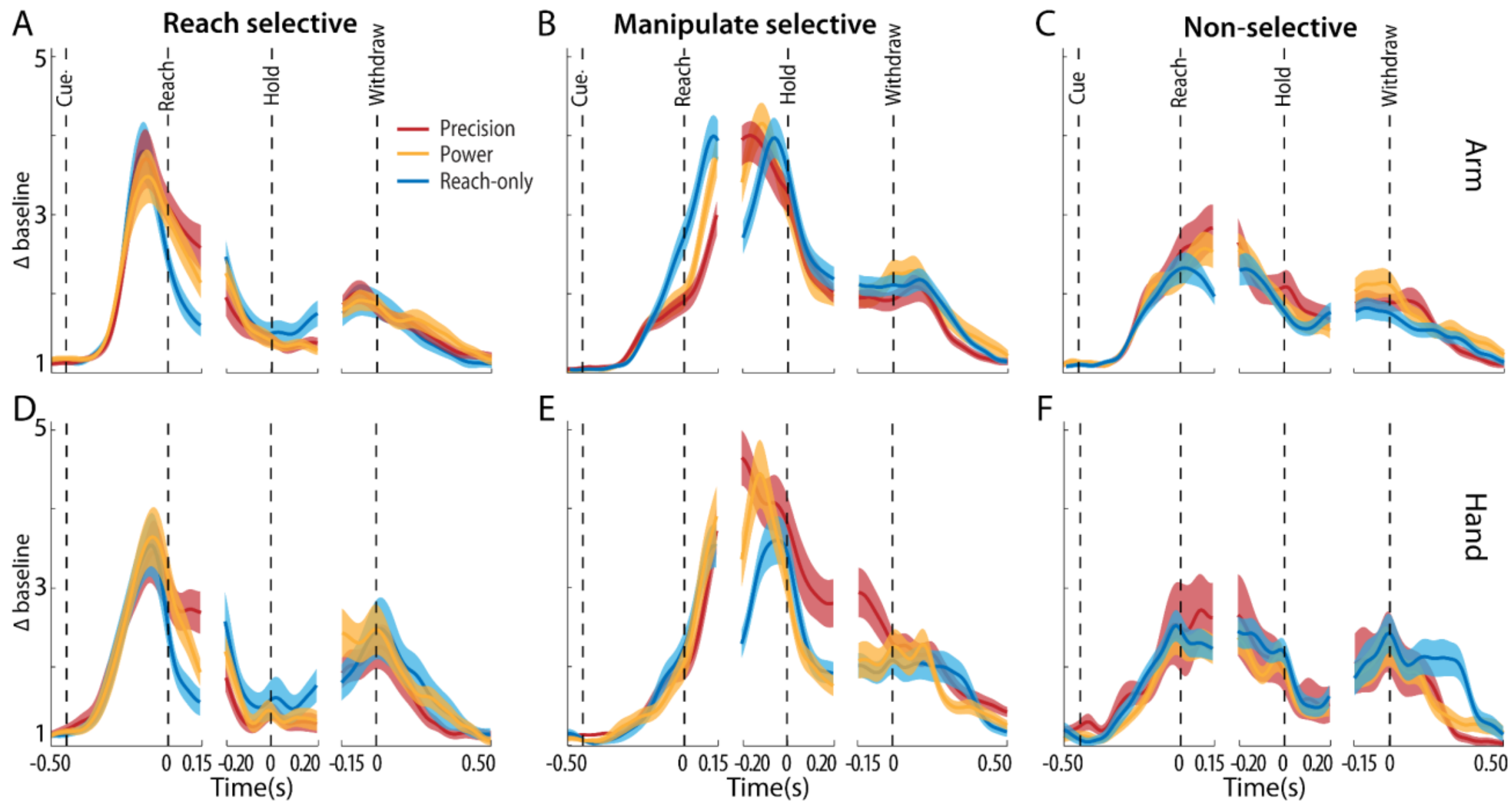

**Figure 17. Additional differences revealed in the activity between conditions when grouped according to somatotopy and phase selectivity (A)-(C) Baseline normalized average PSTHs of units recorded from the arm representation that statistically modulated more in the reach (A) or manipulate (B) phase. Units that modulate comparably in the reach and manipulate phase are plotted in (C). Shaded areas for the precision, power, and reach-only condition (red, orange, and blue, respectively) indicate the standard deviation from the average activity across units. The time course is visualized in three separate time windows (separated by breaks in x-axis), which correspond to the three different alignment points used to generate the PSTHs (reach, hold and withdraw point alignment) (D)-(F) The same conventions in (A)-(C) for the PSTHs of units from the hand representation.**

Next, we sought to understand the spatial organization of the newly classified units with respect to somatotopy. Initially, we calculated the ratio of unit selectivity types in each M1 somatotopic zone for the precision grip condition (Figure 18A, left). For reference, the assumed null distribution is visualized in the central bar. To determine whether the ratio of unit phase selectivity types was distributed equally in each somatotopic representation, we used a chi-square goodness of fit test. For the precision grip condition, we found that unit selectivity types were not equally distributed in the arm nor hand representation of M1 ($X^2(2)= 15.82$, $p<0.001$;

$X^2(2) = 21.32$, $p<0.001$). In both the arm and hand representation, the ratio of manipulate selective units is larger than expected from the null distribution. Furthermore, the ratio of manipulate selective units is greater in the hand compared to the arm representation, whereas the ratio of reach selective units is greater in the arm compared to the hand representation. Consistent with distribution of unit selectivity types for the precision grip condition, both the arm and hand representation have a larger ratio of manipulate selective units compared to the null distribution for the reach-only condition (Figure 18A, right). Interestingly, despite the reach-only condition free of any evoked grip posture, the manipulate selective units outnumber the reach selective units in both the arm and hand representation. To comprehensively understand the organization of unit selectivity types, we subsequently calculated the ratio of somatotopic representations for each unit selectivity type.

We examined the ratio of units according to somatotopic representation for each unit selectivity type (defined from the precision grip condition; Figure 18, left). The null distribution is visualized in the central bar as the sampled ratio of units for each somatotopic zone. In relation to the sampled distribution of the unit somatotopic representations, none of the unit selectivity types were distributed in such a way that deviated significantly from the sampled distribution ($p>0.05$). This result is replicated in the equivalent distributions for the reach-only condition (Figure 18B, right). Overall, the ratio of somatotopic representations per unit selectivity types conforms more to the null distribution as opposed to the noted variations in the ratio of unit selectivity types per somatotopy. Therefore, we subsequently summarized the spatial organization of the unit selectivity types agnostic to somatotopy.

To understand the spatial organization of the selectivity unit types, we constructed the empirical cumulative distribution function (eCDF) for each type along the medial/lateral axis in the precision and reach-only condition (Figure 18C & D). The point along the medial/lateral axis at which half of the unit population is represented is marked along the x-axis with the corresponding color. A two-sample Kolmogrov-Smirnov indicated that manipulate selective units are spatially distributed distinctly from the reach (D(822)= 0.10, p =0.045) and non-selective units (D(859)= 0.12, p = 0.009) along the medial/lateral axis for the precision grip, but not the reach-only condition. No other pairs of selectivity types demonstrated spatially distinct distributions along the medial/lateral axis in either condition. The lateral shift in the manipulate selective units for the precision grip condition is potentially a depiction of the somatotopic organization (i.e., hand representation is lateral to the arm representation). Thus, we next investigated whether there were any distinctions of unit selectivity types in the spatial distribution along the rostral/caudal axis.

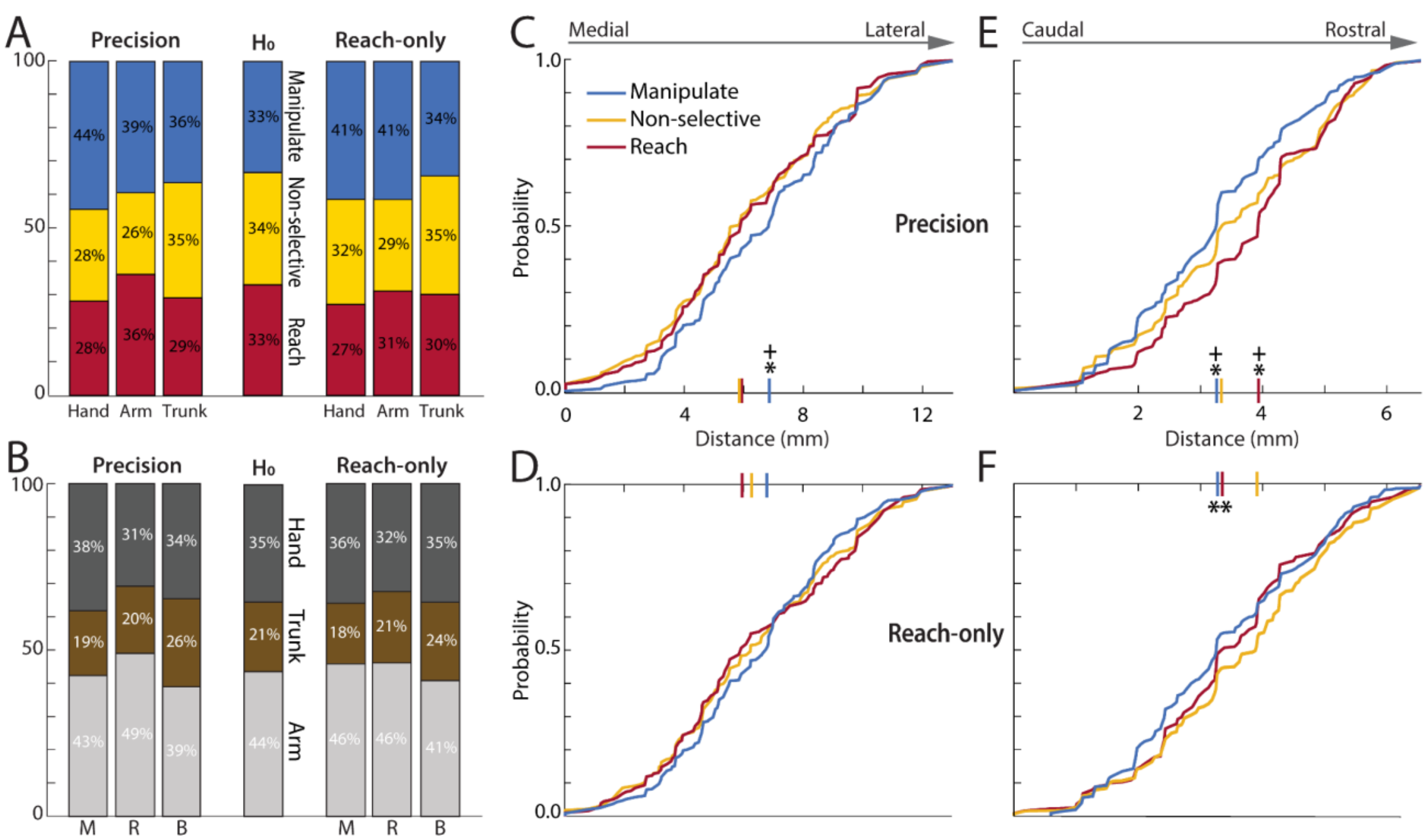

**Figure 18. Phase selective units are distributed unevenly through the M1 arm and hand representation (A) Percentage of task-modulated units per somatotopic representation that were classified as manipulate selective (blue), reach selective (red), or non-selective (yellow) for the precision (left) and reach-only (right)**

**conditions. The bar in the middle visualizes the null hypothesis, namely, the selectivity types are equally distributed within each somatotopic zone. (B) Percentage of task-modulated units per selectivity type that were recorded from the hand, arm, or trunk representation for the precision (left) and reach-only (right) conditions. The bar in the middle visualizes the null hypothesis, namely, the somatotopic representations are proportionally represented for each selectivity type. (C) The empirical cumulative distribution function (eCDF) of each selectivity type along the medial/lateral axis for the precision. The median medial/lateral location of each selectivity type is marked on the x-axis. Asterisks indicate that a two-sample Kolmogorov-Smirnov test ($p<0.05$) found the respective distribution differs from the distribution of the non-selective unit types. Crosses indicate that a two-sample Kolmogorov-Smirnov ($p<0.05$) test found the respective phase selective distribution differs from the distribution of the contrary phase selective type. (D) The same eCDFs of each selectivity type for the reach-only condition along the medial/lateral axis. (E)-(F) The same ecDFs of each selectivity type for the precision (E) and reach-only (F) condition along the rostral/caudal axis.**

Along the rostral/caudal axis, the precision grip and reach-only condition both convey unique characteristics in the spatial distribution of unit selectivity types (Figure 18E & F). Primarily, the manipulate units were more caudal than the non-selective units in both the precision grip ($D(775)= 0.12$ , $p =0.010$) and reach-only ($D(962)= 0.12$, $p=0.002$) conditions. Furthermore, reach selective units were distributed rostral to the non-selective units in the precision condition ($D(859)= 0.12$, $p=0.005$) and caudal to the non-selective units in the reach-only condition ($D(820)= 0.12$ , $p=0.007$). Finally, the reach and manipulate selective units are spatially distributed distinctly between each other in the precision condition ($D(822)= 0.21$, $p <0.001$). These results elucidate spatial organizations of the temporal dynamics from M1 unit activity and, in tandem with somatotopy, help characterize functionally distinct dynamics in M1.

### 3.3.5 Using spatiotemporal characteristics of M1 activity to decode movement conditions

By classifying units according to anatomical spatial definitions and to temporal dynamics, we found unique spatiotemporal dynamics in M1 unit activity to implement different reach-to-grasp movements. We evaluated the effectiveness of these dynamics in classifying the different conditions to understand the extent that movement (function) is encoded among the various somatotopic representations and phase selectivity types. To that end, we implemented a Naïve Bayes classifier using the activity of task-modulated units from the Cue, reach and manipulate time ranges and evaluated classification accuracy grouping upon the various spatiotemporal characteristics outlined throughout the results (see Methods for further details). To begin, we evaluated condition classification using unit activity from the Cue, reach and manipulate window as a function of the number of units. Note that for each evaluated number of units, units were randomly sampled and evaluated 500 times.

As expected, classifiers that were trained and tested with activity from the same phase asymptotically improved decoding as units were added (Figure 19A, solid lines). Classifiers evaluated with activity from untrained phases achieved chance level decoding (Figure 19A, dashed and dotted lines). The sole exception was from classifying manipulate phase activity using activity from the reach phase. Moreover, decoding manipulate phase activity from reach phase activity was consistently more accurate than classifiers that trained and tested with activity from the Cue window. Additionally, the classifier that trained and tested with activity from the manipulate phase outperformed the other classifiers with matched training and testing phases. Only ~15 units were necessary as input for the classifier to achieve 90% decoding accuracy. Over 100 units were required to achieve equivalent performance from the reach window and the activity in the Cue window lacks enough information to achieve 90% accuracy. Next, we separated units according to somatotopy and evaluated the performance in the Cue, reach and manipulate phase to evaluate the usefulness of somatotopy in distinguishing movement conditions (Figure 19B).

51 units from the face representation (lateral border of the forelimb representation) were included to serve as a reference for accuracy. Indeed, among all somatotopic zones, the classifier does not surpass 60% accuracy with the maximal number of evaluated units (n = 100). Using activity from the reach phase, all somatotopic zones, save the face representation, demonstrate comparable decoding with respect to the number of units. Finally, in the manipulate phase, all somatotopic zones, including the face representation, achieve >90% accuracy with ~20 units. Because the activity from the reach phase demonstrates appropriate discrimination among the performance of somatotopic zones as it relates to reach-to-grasp movements (i.e., activity related to movements of the face are not suitable for discerning forelimb movements), we primarily focus on the results that use the reach phase for decoding movement conditions. Furthermore, to avoid overfitting the classifier and obtain distributions with meaningful variation, we assessed subsequent classifiers using 35 units—approximately halfway chance level and 90% decoding for the appropriate trend lines in the reach window.

We first considered the performance of the classifier by categorizing unit subpopulations using spatial/anatomical categorizations. Evaluating classifiers using the activity from the manipulate phase demonstrates an issue in using this phase of activity to evaluate performance; units from all somatotopic zones achieve more than 90% classification accuracy and demonstrate no distinctions between them. These results corroborate elements discussed in the PSTHs separated according to unit depth and somatotopy and support the number of units used to evaluate the classifiers. We finally consider the classification differences of the phase selectivity units.

To be comprehensive, we show the accuracy distribution of classifiers that trained and tested from activity in the Cue, reach and manipulate phase according to somatotopy and unit selectivity type (Figure 19C). The classifiers that used activity from the reach phase recapitulate the separation in accuracy between the units from the face representation and units from the other somatotopic representations. The following results were determined from an analysis of variance on the accuracy according to unit selectivity type. ANOVAs were independently calculated for each group of unit selectivity types per phase and somatotopic representation and reported as significant using Bonferroni t-test corrections ($p<0.05$). Each somatotopic representation demonstrates greater accuracy in at least one of the phase selective unit types. Furthermore, unique to the performance of units from the arm and hand representation, both phase selective unit types outperform the non-selective unit types. Interestingly, the reach selective units from the hand representation additionally outperforms the manipulate selective units. This result suggests that defining M1 units in relation to their temporal dynamics demonstrates an additional tier of organization in M1 activity. Indeed, classifying conditions using activity from the manipulate phase showed that the reach-selective and manipulate-selective units consistently outperformed the non-selective units. We posit that by examining spatiotemporal dynamics of M1 activity, we have established evidence of a functional organization further embedded in the somatotopic representation of M1.

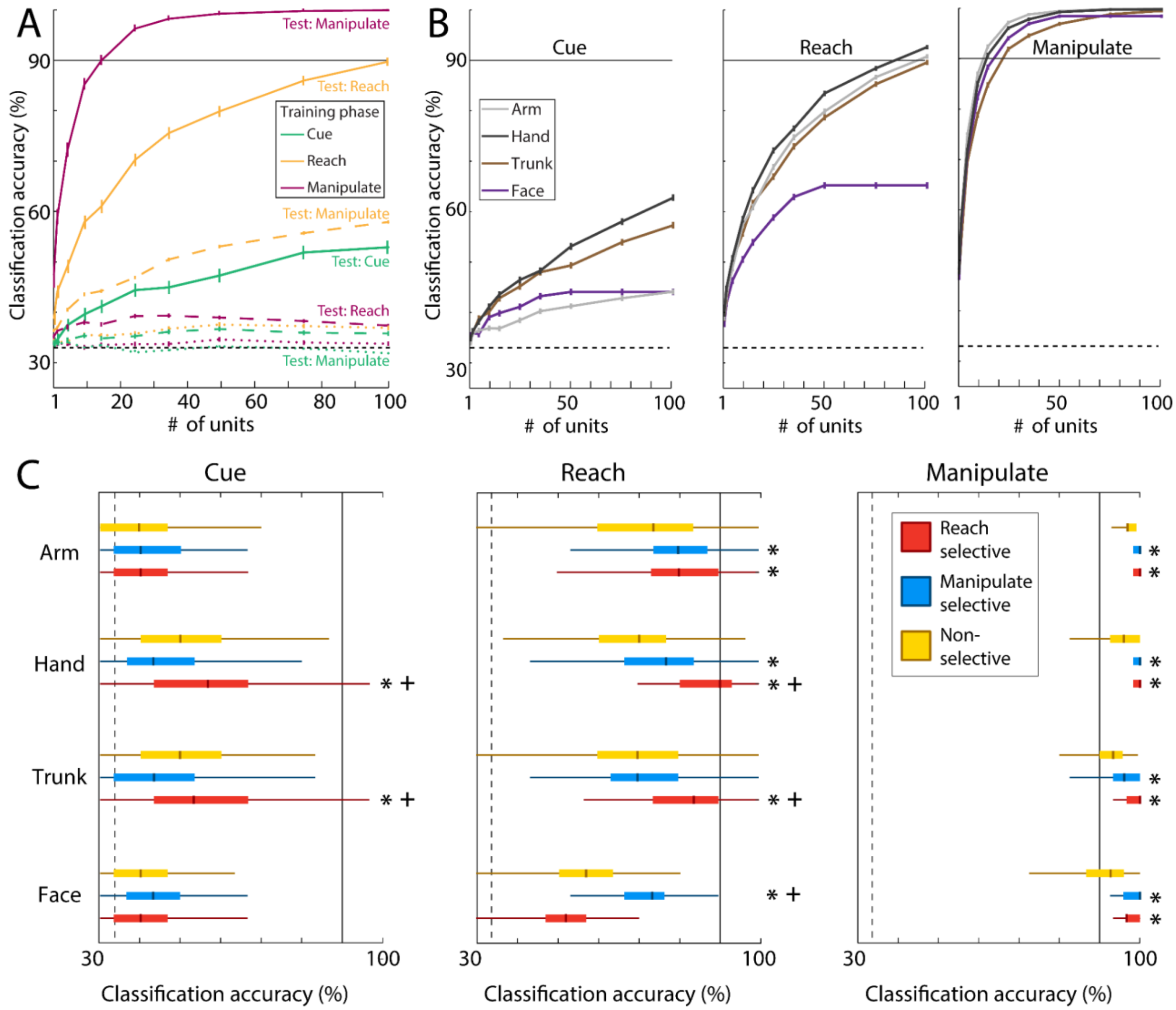

**Figure 19. Naïve Bayes classifiers trained with activity from the reach phase decode movement conditions better from units that are phase selective (A) Accuracy of naïve bayes classifier plotted against the number of units used in the classifier. Classifiers that were trained and tested using activity in either the Cue, reach, or manipulate phase are plotted as sold lines. Dashed and dotted lines plot decoding accuracy from classifiers trained and tested using activity in different phases. For each number of units (N) considered, 500 samples of N units were randomly chosen. All task-modulated units were considered for testing. Error bars at each test point mark the standard error from the mean over the 500 samples. Dashed line signifies chance decoding level (33%); solid line marks 90% decoding accuracy. (B) Same results from (A) separated according to somatotopy using activity in the Cue (left), reach (middle), and manipulate (right) phase. (C) Box plots of classification accuracy for activity in the Cue (left), reach (middle), and manipulate (right) phase. Classifiers**

**were assessed using 35 units from each somatotopic representation. 500 samples were tested using the reach, manipulate, and broad tuned units. Bonforonni corrected t-tests with significant differences in accuracy between the non-selective and reach or manipulate selective units are marked with a single asterisk; significant differences in accuracy between reach and manipulate selective units are marked with a cross. The median value of the distribution is marked as a dark line in the box, whiskers indicate the standard error of the mean.**

## 3.4 Discussion

We investigated the relationship between M1 structure and function while monkeys performed a reach-to-grasp task. We defined M1 structure according to somatotopic representation, which was evaluated by motor mapping with ICMS. We characterized M1 function from unit activity in two time periods—one preceding movement onset and the other preceding object contact—to summarize reach and grasp, respectively. We hypothesized that functional activity would intuitively map with somatotopy; namely, units from the M1 arm representation fundamentally modulate activity in the reach phase and units from the M1 hand representation principally modulate activity in the manipulate phase. We found that the average unit activity from the M1 arm and hand representation increased in both the reach and manipulate phase. Yet, when we categorized units according to modulation selectivity, we found that over half of the units from the arm and hand representation selectively modulated activity to either the reach or manipulate phase. Furthermore, phase selective units were spatially distributed in discrete territories of the M1 forelimb representation. The spatiotemporal activity of units in M1 revealed functional organizing that is complementary, although distinct, from the somatotopy.

### 3.4.1 Arm and hand muscle activity modulated in the reach and manipulate phase.

We summarized arm and hand muscle activity as the average EMG activity of 3 proximal forelimb muscles (deltoid, biceps, triceps) and 4 distal forelimb muscles (ECRB, FCR, EDC, FDS), respectively. We found that arm and hand muscles were both involved with reaching and grasping. These results were surprising, as we expected the activity of arm muscles to increase around reach onset and the activity of hand muscles to increase around object contact. However, as noted in previous investigations, the average arm and hand muscle activity increased prior to reach onset and peaked close to object contact (Jeannerod, 1986; Rouse & Schieber, 2016c).

We found that both arm and hand muscle activity of the grasp conditions were separable from the reach-only condition in the manipulate phase. Though expected for the muscle activity of the hand, we were surprised that arm activity was larger in the manipulate phase than the reach phase. Considering that the task was designed to maintain object location and vary object size, we expected minor variations in arm activity between conditions (Paulignan, Jeannerod, et al., 1991). Rather, we expected that any condition-dependent differences in arm muscle activity would appear around movement onset (Martelloni et al., 2009). We suspect that arm activity was larger in the manipulate compared to the reach phase, particularly for the grasp conditions, as a result of lifting, not gripping, the objects.

Although arm and hand muscle activity were unexpectedly similar throughout reach-to-grasp movement, we found a consistent discrepancy between hand and arm muscle activity. Activity was reliably larger from the hand than the arm in each movement phase for all conditions. This potentially results from stronger and/or more terminals in motoneuron pools of distal forelimb muscles than those of proximal forelimb muscles (Jankowska et al., 1975; McKiernan et al., 1998), resulting in larger activity modulation from baseline.The prominent distinction between muscles of the arm and hand provided a feature to investigate correspondence in the M1 neural activity from the arm and hand representation.

### 3.4.2 Average unit activity is indistinguishable between the M1 arm and hand representation.

Unit activity within M1 covaries with various kinetic and kinematic properties (Kakei et al., 1999; Suminski et al., 2015). Therefore, we anticipated that unit activity from the M1 arm and hand representation would mirror temporal activity profiles observed from the muscle activity of the arm and hand. Indeed, we found unit activity from the arm and hand representation increased prior to reach onset and subsequently peaked prior to object hold. Unit activity between the arm and hand representation was even more alike than the muscle activity between the arm and hand. In fact, the greater activity of muscles in the hand compared to the arm was not recapitulated in the unit activity from the respective cortical representations. The coherence in unit activity from the arm and hand representation corroborates previous work that found reach and grasp encoding units all throughout the M1 forelimb representation (Rouse & Schieber, 2016a; Saleh et al., 2012a; Stark et al., 2007). Indeed, had we not analyzed unit activity beyond the average of each somatotopic representation, our results would also corroborate that reach and grasp encoding is distributed throughout the M1 forelimb representation.

We examined whether averaging activity from hundreds of individual units masked potential activity differences between somatotopic representations. Averaging the activity of individual units within a somatotopic representation presents challenges that are immaterial when averaging muscle activity. EMG activity recorded from session-to-session targeted the same muscles and, thus, maintained consistency in the activity profiles of each muscle. However, unit activity is averaged from hundreds of single units. Furthermore, recordings were intentionally dispersed throughout the M1 forelimb representation to comprehensively sample the diverse activity profiles of M1 units. Thus, the sample of units in each somatotopic representation likely comprised unique activity profiles and obscured neural activity when averaged according to somatotopic representations.

Consider the possibility that units in the M1 forelimb representation exclusively modulate activity to either the reach or manipulate phase (Figure 11E: red & blue units). The average activity of the M1 forelimb representation, consisting of these temporally discrete units, would suggest equivalent activity in the reach and manipulate phase (Figure 11E: yellow unit). We suspected that differences in the neural activity between the arm and hand representation were concealed by averaging units together. To explore whether analogous activity from the M1 arm and hand representation was a result of indiscriminate unit encoding of reach and grasp or confounding activity ensuing from averaging among phase-selective units, we considered categorizing units to additional characteristics.

We considered categorizing units according to cortical recording depth to understand if activity differences were discernable using an additional anatomical classification. A recent investigation reported movement-specific responses were differentially expressed in units from superficial layers in motor cortex (Currie et al., 2022). However, the average activity of units from superficial or deep cortex representation revealed minor variation from the average activity previously observed from the arm and hand representation. We did note that, especially for hand units, condition-specific activity was more apparent after further separating units by cortical depth. Therefore, we contemplated that averaging activity from hundreds of units masked distinctions in the activity between somatotopic representations. We reasoned that by reducing the number of units grouped and averaged together, we could achieve a more representative summary of unit activity. Though defining units based on cortical recording depth did not notably distinguish between activity from the arm and hand representation, we explore the spatiotemporal activity in M1 according to the unit's location within the forelimb representation.

### 3.4.3 Movement-specific activity was localized in distinct regions within the M1 forelimb representation.

Laminar multielectrode recordings enabled us to average unit activity within a given recording site (tens of units) rather than within a somatotopic representation (hundreds of units). Previous work has reported strong correspondence in the encoding of units from the same electrode (Ben-Shaul et al., 2003; Hatsopoulos, 2010). Thus, by reducing the number of units averaged together and by averaging among units with more congruous activity profiles, we can better characterize unit activity according to recording locations. Immediately, it was apparent that activity is spatially irregular throughout M1—even within the same somatotopic representation. Within the M1 hand representation of monkey G, we observed neighboring recording sites in the caudal territory with pronounced activity modulation and phase selectivity for the precision and power grip conditions (Figure 15A & 16B). The same response was spatially absent for the reach-only condition (Figure 15B & 16D). This distinction of activity between conditions and somatotopic representations was statistically validated by averaging units according to somatotopy. We clarify that, although we could distinguish unit activity between somatotopic representations, the recording site maps indicated that the spatiotemporal activity in M1 did not strictly adhere to the somatotopic organization.

The greater activity and selectivity of the manipulate phase from units in the hand representation was not expressed broadly throughout the hand representation. Only a few adjacent recording sites in the hand representation produced the difference in the activity between the hand and arm representation. We proceeded to characterize (without confines of somatotopic boundaries) spatial territories within the M1 forelimb representation that selectively modulated activity to the reach or grasp. Naturally, we found greater selectivity to the grasp in units from the most lateral 3 mm of the M1 forelimb representation. The lateral concentration of manipulate selective units corresponds with the somatotopic organization; the hand representation is approximated lateral to the arm representation (Penfield & Boldrey, 1937; Woolsey, 1958).

From caudal to rostral in the M1 forelimb representation there was a gradation of unit selectivity for the manipulate to the reach phase. Reach selectivity in units most rostral in the M1 forelimb representation may also correspond with the somatotopic organization. Namely, the M1 arm representation encloses the M1 hand representation in a concentric arrangement and comprises the predominant representation in the rostral M1 forelimb representation (He et al., 1995; Park, Belhaj-Saïf, et al., 2001a). The units from the M1 arm and hand representation are equivalently distributed in the most caudal portions of the M1 forelimb representation. Yet, the 3 mm most caudal in the M1 forelimb representation consisted of units with greater activity modulation to the manipulate phase. We wanted to further leverage the temporal dynamics of unit activity to investigate functional organization within the M1 forelimb representation.

### 3.4.4 Grouping units by phase selectivity revealed functional organization distinct from somatotopic organization.

Prior to this point, we examined temporal differences in the activity of units throughout the M1 forelimb representation. Now, we consider spatial differences in the distribution of units defined according to phase-selectivity (e.g., what is the ratio of units in the M1 forelimb representation that are reach phase selective compared to manipulate phase selective?). To understand possible organizing principles in M1 activity corresponding to temporal dynamics, we assessed the hypothesis that reach and manipulate units would map to the arm and hand representation, respectively.

In accordance with previous work, we found considerable heterogeneity in the temporal activity dynamics of M1 units during reach-to-grasp movements; neurons modulated activity in the reach and/or grasp with temporally narrow and/or broad responses (Reimer & Hatsopoulos, 2010; Yallapu et al., 2012). Studies have found that units recorded from cortical motor areas in the arm representation modulated activity during reach and units from the hand representation modulated activity to grip type (Friedman, Chehade, et al., 2020a; Gentilucci et al., 1991). Other investigations reported mixing of reach and grasp units throughout the arm and hand representation (Saleh et al., 2012a; Stark et al., 2007). We sought to resolve these competing outcomes by investigating the spatial organization of units with selective modulation of activity to the reach and/or grasp.

Grouping units according to phase selectivity produced average activity profiles that were more separable between conditions than when grouped according to cortical depth. Moreover, condition-specific activity was more apparent from phase selective units rather than non-selective units. Indeed, Naïve Bayes classifiers trained from the activity of phase selective units consistently decoded between movement conditions better than classifiers trained from non-selective units. Interestingly, the activity modulation profiles of non-selective units most resembled the average unit activity from the arm and hand representation. Yet, the arm and hand representation predominately comprised units that were selective to either the reach or manipulate phase. In fact, over two-thirds of units from the M1 arm and hand representation selectively modulated activity to the reach or manipulate phase (Figure 18). This supports our speculation that averaging units according to somatotopic representation produced activity profiles that were non-representative of typical unit responses.

Units from the arm and hand representation were more likely to modulate activity selectively to the reach or manipulate phase, whereas units from the trunk representation were equally likely to modulate activity selectively and non-selectively (Figure 19A). Whereas phase selective units were disproportionately expressed in the M1 forelimb representation, phase selective and non-selective units were equally distributed within the arm, hand, and trunk representation. Categorizing units according to somatotopy repeatedly muddied functional activity differences that were better distinguishable when categorized by phase selectivity types.

Phase selective units were localized in discrete territories within the M1 forelimb representation. Consistent with spatial gradients found when analyzing the selectivity index along the rostral/caudal and medial/lateral axes, manipulate selective units were localized more laterally and caudally in the M1 forelimb representation than the reach selective and non-selective units. Additionally, reach selective units were localized more rostral in the M1 forelimb representation than manipulate selective and non-selective units. In contrast, the reach and manipulate selective units were not spatially separable for the reach-only condition. Again, the spatial distributions of phase selective units corresponded to the somatotopic organization (e.g., medial to lateral gradient in arm/hand and reach/manipulate selectivity) but revealed nuances that would not be appreciable at the level of somatotopic organization. In general, we consistently found spatial organizations of function that were further refined upon defining units according to their functional, rather than somatotopic, definition.

### 3.5 Conclusion

In conclusion, we observed functional differences in the unit activity from the M1 arm and hand representation for reach-to-grasp movements. We determined that categorizing units according to somatotopic representation produced average profiles with limited differences in activity between somatotopic representations, the reach and manipulate phases, and movement conditions. However, averaging units per recording site illustrated substantial heterogeneity in the spatiotemporal activity within any given somatotopic representation. We then evaluated unit activity differences grouped according to phase selectivity. Phase-selective units comprised most units in the M1 arm and hand representation. Furthermore, reach and grasp selective units were spatially distributed in distinct territories along the rostral/caudal and medial/lateral axes. We conclude that, though reach and grasp encoding is distributed throughout the M1 arm and hand representation, it does not imply non-specific activity modulation throughout the M1 forelimb representation. We emphasize that intermingling of reach and grasp encoding units throughout the M1 forelimb representation results from functional organization that is spatially unique, though related to the somatotopic organization.

# 4.0 Summary and Conclusions

## 4.1 Overall findings in relation to existing work

This dissertation was motivated by the aim of understanding the spatiotemporal organization of M1 activity that supports dexterous movements. In my initial experiments, the use of ISOI enabled continuous spatial evaluation of neural activity in the M1 forelimb representation while NHPs completed reach-to-grasp movements. Results indicated that patches of activity comprised less than half of the total M1 forelimb representation territory. Correspondingly, units recorded from sites throughout the M1 forelimb representation modulated activity differentially in both space and time. Both imaging and neurophysiology recordings suggested that dexterous movement is coded throughout M1 in discrete spatial clusters. This finding is corroborated by other observations in M1 that observe patchy movement encoding in M1. For example, Card & Gharbawie found that the intrinsic connections in the M1 forelimb representation establish networks of connecting patches (0.25 – 1.00 mm radius) that are preferentially matched according to somatotopic representation. Additionally, these results align with those of Rouse & Schieber, 2016a and Saleh et al., 2012a as both investigations noted intermingling of reach and grasp function in the M1 forelimb representation. Though ISOI (Figure 6B & C) and neurophysiology results (Figure 18C & E) produced a spatial separation between reach and grasp neural activity, the subtlety of these results would certainly be undetected without a spatial resolution exceeding what was used in the discussed results.

In fact, the idea that results potentially vary according to the spatial resolution of the investigative tool is illustrated directly in Figure 12F & 17A. When the peak times of units were grouped according to somatotopy, there was no apparent distinction among somatotopic representations. Without further inspection, this could lead to the incorrect conclusion that M1 unit encoding is similar irrespective of the spatial location. However, this is refuted according to the FR maps that convey diverse activity modulation throughout the spatial extent of the M1 forelimb representation. Additionally, the bearing of M1 location on the neural response is demonstrated by the spatial consistency of neural activity between movement conditions in the imaging results. The similarity supports a recent hypothesis of M1 organization that defines ethologically relevant movements mapped within M1 as a mosaic of masses (Aflalo & Graziano, 2007) Furthermore, less than half of the entire forelimb representation is engaged to implement reach-to-grasp movements, which further enforces the idea that neural activity is not homogenously distributed throughout M1.

Throughout the results from both sets of experiments, the spatially non-homogenous response of neural activity in M1 is consistent. It is most readily visualized in Figure 12F where the firing rate of M1 units are summarized in the heatmap. But it is also apparent in more unassuming outcomes. For example, the intrinsic signal time course differs between the M1 hand representation and M1 arm representation. Additionally, there is a stronger activity selectivity to the grasp than the reach in caudal M1 which is further augmented by the concentration of units that selectivity fire to the grasp. This caudal to rostral gradient of dexterity aligns with other investigations that delineate between caudal and rostral M1 as new and old M1, respectively, marked by the presence and absence of CM cells, respectively (Rathelot & Strick, 2009a). Though most of new M1 is buried in the sulcus and the analyzed units were sampled from the cortical surface, the gradient along the surface may represent the transition from old M1 to new M1 approaching the central sulcus. I reinforce this M1 spatial distinction of dexterity to highlight how easily a result as seemingly solid and previously validated can be masked contingent on analytical approaches.

One of the more unexpected findings from this work was the equivalent neural responses between somatotopic representations. It is tempting to take this conclusion at face value and discount somatotopy as a contending factor in understanding the neural basis of movement. However, as discussed in the Introduction, neuroscientists have noted the utility of examining structure/function relationships to aid in understanding the processes underlying our behaviors. And, indeed, when M1 units are analyzed according to site location, as opposed to somatotopy, it is immediately clear that the spatiotemporal activity in M1 is, in fact, largely diverse. I raise this point not to declare a proper perspective for analyzing neural activity, but rather, to consider the implications of ignoring certain features in the analysis of neural activity.

The analyses conducted in this dissertation have demonstrated that the functional coding of dexterous movements is indeed distributed throughout the M1 somatotopic representation of the arm and the hand. Additionally, subregions within the M1 forelimb representation appear to code for central reach-to-grasp movements. Future experiments that assess the spatiotemporal organization of M1 activity supporting a diverse set of movements (e.g., hand-to-mouth movement) would assist in corroborating this speculation. Nevertheless, the outcome of these experiments ascertains the importance of considering the location(s) in M1 which are sampled when analyzing the neural activity underlying movements. As the functional code underlying dexterous movements is spatially distributed in an irregular manner throughout M1, I believe that the spatial locations of the recorded units should be integrated into neural data analyses.

For example, a recent study in human M1 (Kunigk et al., 2024) found that multielectrode arrays recording activity from the medial M1 forelimb representation had greater activity and better decoding of object translation when subjects attempted/imagined proximal forelimb movements than did lateral arrays. Conversely, lateral arrays had greater activity and better decoding of object click/release when subjects attempted/imagined distal forelimb movements than did medial arrays. This work highlights the importance of the somatotopy for the encoding of coordinated forelimb movements. Considering that the discussed outcomes in this dissertation note discrete zones *within* M1 arm and hand representation that functionally code reach-to-grasp movements, it is important in future works to employ methods that achieve a spatial resolution high enough to delineate within somatotopic representations. Ultimately, maintaining a perspective that relates structure and function is crucial for investigating the neural basis of voluntary movements.

## 4.2 Limitations and potential future directions

Both studies directly related results to the motor map, which was established by defining evoked movements via ICMS. Motor maps were obtained using ICMS at various sites throughout M1. The experimenter(s) classified each site according to the observed evoked movement(s) in response to ICMS. There are more accurate methods to reveal the motor map, such as with stimulus-triggered averaging (Park, Belhaj-Saïf, et al., 2001a). Additionally, ICMS can indirectly stimulate other projections via synaptic afferents and ultimately obscure the evoked movement at a given site in M1 (Hussin et al., 2015). The motor maps I obtained are consistent with those of recent primate studies (Hudson et al., 2017; Park, Belhaj-Saïf, et al., 2001a; Schieber, 2001), though, it remains possible that more precise mapping could yield additional interactions to functional activity that were not obvious from the discussed results.

The temporal activity of M1 was captured from laminar multielectrode recordings that were acquired from ~150 separate sessions. Units were recorded exclusively from the dorsal surface of M1, which is referred to as old M1 (Rathelot & Strick, 2009b). Considering that CM cells are principally located in the anterior bank of the central sulcus (new M1), the analyzed unit activity from this dissertation unlikely includes any CM cells. However, despite a lack of CM cells, electrical stimulation in old M1 still evokes movement, including individual finger movements. Therefore, the analyzed unit activity in this dissertation still comprises units with a direct role in controlling movements. Additionally, the noted organizing principles are theoretically applicable to other species of mammals and not limited to primates because old M1 is evolutionarily persevered, whereas new M1 is limited to primates. If activity was recorded from units in the anterior bank of the central sulcus, it may have revealed organizations conditioned on different grips (e.g., precision vs power grip). Furthermore, units were not obtained simultaneously, which somewhat limits analyzing the neural activity as a population response.

The presented analyses do not explicitly implement any population analyses. Unit activity was averaged together according to different variables (e.g., somatotopy, location), but was never utilized to model movements from all unit responses. Though unit activity was aligned to behavioral events and the behavior itself was stereotyped, the total unit response would not be appropriately modeled as a population response to movements as all interactions between units from different sessions would be uninterpretable. However, units were ultimately summarized by averaging activity, which maintains other consequences. The implication of averaging was demonstrated from the similarity of unit responses when grouped according to somatotopy in contrast to the spatial distinction of unit responses when grouped according to recording location within M1. Even still, activity had to be averaged among units recorded within the same site. Laminar differences were briefly explored by grouping units based on recording channel (e.g., first and last half of channels). Possible analyses in the future could expand on laminar differences by increasing the discretization across the recording electrode.

The spatiotemporal dynamics of M1 activity was considered by investigating the intrinsic signal and unit responses in M1. However, the relationship between the results obtained from imaging and electrophysiology were not explicitly related to each other. Analyses exploring the activity of units in relation to the zones that were marked as active from imaging experiments would aid in understanding possible functional organizing principles in M1. For example, if ethologically relevant movements are represented in an organized manner within M1 (M. Graziano, 2006; M. S. A. Graziano, 2016), the activity of units in active zones indicated from imaging may modulate for reach-to-grasp in central space, but not to hand-to-mouth movements. The behavioral repertoire utilized in this dissertation was limited and may not have induced enough behavioral variation to assess different M1 spatiotemporal dynamics for different movements. Future investigations could also consider spatiotemporal dynamics of reach and grasp independently.

The goal in assessing spatiotemporal dynamics of M1 activity in relation to reach-to-grasp movements was to study neural activity that drove naturalistic movements. Reaching and grasping are often studied independently due to the complexity inherent in coordinated arm and hand movements to successfully execute reach-to-grasp movement. In the last set of analyses in this dissertation, units were defined as reach or grasp modulated. Although the work presented in this dissertation and others investigating forelimb kinematic/kinetic dynamics (Rouse & Schieber, 2015, 2016c) indicate that reach and grasp are predominately separated in time, it remains complicated to distinguish between reaching and grasp. Future experiments could isolate reaching and grasping between imaging sessions and could deduce if previously identified active zones in M1 are driven by reaching or grasping. Alternatively, potentially there is no spatial congruency between active zones of naturalistic reach-to-grasp and isolated reach/grasp, which would further support the idea that ethologically relevant behaviors are represented in an organized manner in M1. Subsequently, recording unit activity during isolated reaching and grasping would enable investigating different organizations of function (e.g., reaching and grasping) within M1. The presented data in this dissertation is vast and affords a multitude of possibilities for investigating functional organizing principles in M1.

## Appendix A Imaging Sessions Data

| Session # | Monkey | Correct trials | Total trials |
|---|---|---|---|
| 1 | Gilligan | 245 | 304 |
| 2 | Gilligan | 216 | 241 |
| 3 | Gilligan | 140 | 290 |
| 4 | Gilligan | 86 | 116 |
| 5 | Gilligan | 204 | 215 |
| 6 | Gilligan | 202 | 220 |
| 7 | Gilligan | 197 | 242 |
| 8 | Gilligan | 217 | 246 |
| 9 | Skipper | 91 | 117 |
| 10 | Skipper | 131 | 288 |
| 11 | Skipper | 196 | 267 |
| 12 | Skipper | 202 | 242 |
| 13 | Skipper | 200 | 230 |
| 14 | Skipper | 224 | 252 |
| 15 | Skipper | 178 | 327 |
| 16 | Skipper | 160 | 335 |
| 17 | Skipper | 202 | 273 |

# Appendix B Physiology Sessions Data

| UnitID | SiteID | Monkey | Somatotopy | X | Y | Channel |
|---|---|---|---|---|---|---|
| 1 | 1 | Gilligan | Hand | 168 | 332 | 15 |
| 2 | 1 | Gilligan | Hand | 168 | 332 | 17 |
| 3 | 1 | Gilligan | Hand | 168 | 332 | 19 |
| 4 | 1 | Gilligan | Hand | 168 | 332 | 23 |
| 5 | 1 | Gilligan | Hand | 168 | 332 | 25 |
| 6 | 1 | Gilligan | Hand | 168 | 332 | 4 |
| 7 | 1 | Gilligan | Hand | 168 | 332 | 4 |
| 8 | 1 | Gilligan | Hand | 168 | 332 | 8 |
| 9 | 1 | Gilligan | Hand | 168 | 332 | 14 |
| 10 | 1 | Gilligan | Hand | 168 | 332 | 16 |
| 11 | 1 | Gilligan | Hand | 168 | 332 | 18 |
| 12 | 1 | Gilligan | Hand | 168 | 332 | 18 |
| 13 | 1 | Gilligan | Hand | 168 | 332 | 22 |
| 14 | 2 | Gilligan | Hand | 185 | 282 | 28 |
| 15 | 3 | Gilligan | Arm | 237 | 291 | 7 |
| 16 | 3 | Gilligan | Arm | 237 | 291 | 13 |
| 17 | 3 | Gilligan | Arm | 237 | 291 | 15 |
| 18 | 3 | Gilligan | Arm | 237 | 291 | 19 |
| 19 | 3 | Gilligan | Arm | 237 | 291 | 27 |
| 20 | 3 | Gilligan | Arm | 237 | 291 | 6 |
| 21 | 3 | Gilligan | Arm | 237 | 291 | 10 |
| 22 | 3 | Gilligan | Arm | 237 | 291 | 14 |
| 23 | 3 | Gilligan | Arm | 237 | 291 | 18 |
| 24 | 3 | Gilligan | Arm | 237 | 291 | 22 |
| 25 | 3 | Gilligan | Arm | 237 | 291 | 24 |
| 26 | 3 | Gilligan | Arm | 237 | 291 | 28 |
| 27 | 4 | Gilligan | Arm | 321 | 299 | 1 |
| 28 | 4 | Gilligan | Arm | 321 | 299 | 15 |

| UnitID | SiteID | Monkey | Somatotopy | X | Y | Channel |
|---|---|---|---|---|---|---|
| 29 | 4 | Gilligan | Arm | 321 | 299 | 17 |
| 30 | 4 | Gilligan | Arm | 321 | 299 | 23 |
| 31 | 4 | Gilligan | Arm | 321 | 299 | 27 |
| 32 | 4 | Gilligan | Arm | 321 | 299 | 31 |
| 33 | 4 | Gilligan | Arm | 321 | 299 | 4 |
| 34 | 4 | Gilligan | Arm | 321 | 299 | 8 |
| 35 | 4 | Gilligan | Arm | 321 | 299 | 8 |
| 36 | 4 | Gilligan | Arm | 321 | 299 | 20 |
| 37 | 4 | Gilligan | Arm | 321 | 299 | 26 |
| 38 | 4 | Gilligan | Arm | 321 | 299 | 32 |
| 39 | 5 | Gilligan | Hand | 254 | 342 | 23 |
| 40 | 5 | Gilligan | Hand | 254 | 342 | 18 |
| 41 | 5 | Gilligan | Hand | 254 | 342 | 28 |
| 42 | 6 | Gilligan | Hand | 253 | 316 | 3 |
| 43 | 6 | Gilligan | Hand | 253 | 316 | 7 |
| 44 | 6 | Gilligan | Hand | 253 | 316 | 11 |
| 45 | 6 | Gilligan | Hand | 253 | 316 | 19 |
| 46 | 6 | Gilligan | Hand | 253 | 316 | 19 |
| 47 | 6 | Gilligan | Hand | 253 | 316 | 4 |
| 48 | 6 | Gilligan | Hand | 253 | 316 | 6 |
| 49 | 6 | Gilligan | Hand | 253 | 316 | 14 |
| 50 | 6 | Gilligan | Hand | 253 | 316 | 18 |
| 51 | 6 | Gilligan | Hand | 253 | 316 | 20 |
| 52 | 7 | Gilligan | Arm | 296 | 367 | 3 |
| 53 | 7 | Gilligan | Arm | 296 | 367 | 5 |
| 54 | 7 | Gilligan | Arm | 296 | 367 | 7 |
| 55 | 7 | Gilligan | Arm | 296 | 367 | 9 |
| 56 | 7 | Gilligan | Arm | 296 | 367 | 15 |
| 57 | 7 | Gilligan | Arm | 296 | 367 | 21 |
| 58 | 7 | Gilligan | Arm | 296 | 367 | 23 |
| 59 | 7 | Gilligan | Arm | 296 | 367 | 31 |
| 60 | 7 | Gilligan | Arm | 296 | 367 | 12 |

| UnitID | SiteID | Monkey | Somatotopy | X | Y | Channel |
|---|---|---|---|---|---|---|
| 61 | 7 | Gilligan | Arm | 296 | 367 | 14 |
| 62 | 7 | Gilligan | Arm | 296 | 367 | 32 |
| 63 | 8 | Gilligan | Arm | 139 | 226 | 3 |
| 64 | 8 | Gilligan | Arm | 139 | 226 | 5 |
| 65 | 8 | Gilligan | Arm | 139 | 226 | 15 |
| 66 | 8 | Gilligan | Arm | 139 | 226 | 17 |
| 68 | 8 | Gilligan | Arm | 139 | 226 | 29 |
| 69 | 8 | Gilligan | Arm | 139 | 226 | 14 |
| 70 | 8 | Gilligan | Arm | 139 | 226 | 22 |
| 71 | 8 | Gilligan | Arm | 139 | 226 | 28 |
| 72 | 8 | Gilligan | Arm | 139 | 226 | 28 |
| 73 | 9 | Gilligan | Arm | 119 | 136 | 1 |
| 74 | 9 | Gilligan | Arm | 119 | 136 | 1 |
| 75 | 9 | Gilligan | Arm | 119 | 136 | 3 |
| 76 | 9 | Gilligan | Arm | 119 | 136 | 3 |
| 77 | 9 | Gilligan | Arm | 119 | 136 | 17 |
| 78 | 9 | Gilligan | Arm | 119 | 136 | 19 |
| 79 | 9 | Gilligan | Arm | 119 | 136 | 21 |
| 80 | 9 | Gilligan | Arm | 119 | 136 | 21 |
| 81 | 9 | Gilligan | Arm | 119 | 136 | 23 |
| 83 | 9 | Gilligan | Arm | 119 | 136 | 25 |
| 84 | 9 | Gilligan | Arm | 119 | 136 | 4 |
| 85 | 9 | Gilligan | Arm | 119 | 136 | 4 |
| 86 | 9 | Gilligan | Arm | 119 | 136 | 4 |
| 87 | 9 | Gilligan | Arm | 119 | 136 | 6 |
| 88 | 9 | Gilligan | Arm | 119 | 136 | 8 |
| 89 | 9 | Gilligan | Arm | 119 | 136 | 10 |
| 90 | 9 | Gilligan | Arm | 119 | 136 | 12 |
| 91 | 9 | Gilligan | Arm | 119 | 136 | 14 |
| 93 | 9 | Gilligan | Arm | 119 | 136 | 22 |
| 94 | 9 | Gilligan | Arm | 119 | 136 | 24 |
| 95 | 9 | Gilligan | Arm | 119 | 136 | 24 |

| UnitID | SiteID | Monkey | Somatotopy | X | Y | Channel |
|---|---|---|---|---|---|---|
| 96 | 9 | Gilligan | Arm | 119 | 136 | 26 |
| 97 | 9 | Gilligan | Arm | 119 | 136 | 26 |
| 98 | 9 | Gilligan | Arm | 119 | 136 | 28 |
| 99 | 9 | Gilligan | Arm | 119 | 136 | 28 |
| 100 | 9 | Gilligan | Arm | 119 | 136 | 30 |
| 101 | 10 | Gilligan | Arm | 181 | 176 | 1 |
| 102 | 10 | Gilligan | Arm | 181 | 176 | 15 |
| 103 | 10 | Gilligan | Arm | 181 | 176 | 17 |
| 104 | 10 | Gilligan | Arm | 181 | 176 | 19 |
| 105 | 10 | Gilligan | Arm | 181 | 176 | 21 |
| 107 | 10 | Gilligan | Arm | 181 | 176 | 23 |
| 108 | 10 | Gilligan | Arm | 181 | 176 | 27 |
| 109 | 10 | Gilligan | Arm | 181 | 176 | 27 |
| 110 | 10 | Gilligan | Arm | 181 | 176 | 29 |
| 111 | 10 | Gilligan | Arm | 181 | 176 | 18 |
| 112 | 10 | Gilligan | Arm | 181 | 176 | 22 |
| 113 | 10 | Gilligan | Arm | 181 | 176 | 22 |
| 114 | 10 | Gilligan | Arm | 181 | 176 | 26 |
| 115 | 11 | Gilligan | Arm | 174 | 224 | 3 |
| 117 | 11 | Gilligan | Arm | 174 | 224 | 11 |
| 118 | 11 | Gilligan | Arm | 174 | 224 | 11 |
| 119 | 11 | Gilligan | Arm | 174 | 224 | 17 |
| 120 | 11 | Gilligan | Arm | 174 | 224 | 19 |
| 121 | 11 | Gilligan | Arm | 174 | 224 | 23 |
| 122 | 11 | Gilligan | Arm | 174 | 224 | 25 |
| 123 | 11 | Gilligan | Arm | 174 | 224 | 27 |
| 124 | 11 | Gilligan | Arm | 174 | 224 | 27 |
| 125 | 11 | Gilligan | Arm | 174 | 224 | 29 |
| 126 | 11 | Gilligan | Arm | 174 | 224 | 31 |
| 127 | 11 | Gilligan | Arm | 174 | 224 | 6 |
| 128 | 11 | Gilligan | Arm | 174 | 224 | 8 |
| 129 | 11 | Gilligan | Arm | 174 | 224 | 12 |

| UnitID | SiteID | Monkey | Somatotopy | X | Y | Channel |
|---|---|---|---|---|---|---|
| 131 | 11 | Gilligan | Arm | 174 | 224 | 24 |
| 132 | 11 | Gilligan | Arm | 174 | 224 | 26 |
| 133 | 12 | Gilligan | Trunk | 276 | 224 | 21 |
| 134 | 12 | Gilligan | Trunk | 276 | 224 | 21 |
| 135 | 12 | Gilligan | Trunk | 276 | 224 | 25 |
| 136 | 12 | Gilligan | Trunk | 276 | 224 | 25 |
| 137 | 12 | Gilligan | Trunk | 276 | 224 | 4 |
| 138 | 12 | Gilligan | Trunk | 276 | 224 | 6 |
| 140 | 12 | Gilligan | Trunk | 276 | 224 | 8 |
| 141 | 12 | Gilligan | Trunk | 276 | 224 | 8 |
| 142 | 12 | Gilligan | Trunk | 276 | 224 | 14 |
| 144 | 12 | Gilligan | Trunk | 276 | 224 | 18 |
| 145 | 12 | Gilligan | Trunk | 276 | 224 | 20 |
| 146 | 12 | Gilligan | Trunk | 276 | 224 | 22 |
| 148 | 12 | Gilligan | Trunk | 276 | 224 | 24 |
| 150 | 12 | Gilligan | Trunk | 276 | 224 | 30 |
| 151 | 13 | Gilligan | Hand | 236 | 404 | 3 |
| 152 | 13 | Gilligan | Hand | 236 | 404 | 7 |
| 153 | 13 | Gilligan | Hand | 236 | 404 | 7 |
| 154 | 13 | Gilligan | Hand | 236 | 404 | 11 |
| 155 | 13 | Gilligan | Hand | 236 | 404 | 31 |
| 156 | 13 | Gilligan | Hand | 236 | 404 | 4 |
| 157 | 13 | Gilligan | Hand | 236 | 404 | 8 |
| 158 | 13 | Gilligan | Hand | 236 | 404 | 14 |
| 159 | 13 | Gilligan | Hand | 236 | 404 | 18 |
| 160 | 13 | Gilligan | Hand | 236 | 404 | 18 |
| 161 | 13 | Gilligan | Hand | 236 | 404 | 22 |
| 162 | 13 | Gilligan | Hand | 236 | 404 | 26 |
| 163 | 13 | Gilligan | Hand | 236 | 404 | 30 |
| 164 | 14 | Gilligan | Hand | 303 | 323 | 19 |
| 165 | 14 | Gilligan | Hand | 303 | 323 | 23 |
| 166 | 14 | Gilligan | Hand | 303 | 323 | 23 |

| UnitID | SiteID | Monkey | Somatotopy | X | Y | Channel |
|---|---|---|---|---|---|---|
| 167 | 14 | Gilligan | Hand | 303 | 323 | 25 |
| 168 | 14 | Gilligan | Hand | 303 | 323 | 27 |
| 169 | 14 | Gilligan | Hand | 303 | 323 | 27 |
| 170 | 14 | Gilligan | Hand | 303 | 323 | 29 |
| 171 | 14 | Gilligan | Hand | 303 | 323 | 6 |
| 172 | 14 | Gilligan | Hand | 303 | 323 | 6 |
| 173 | 14 | Gilligan | Hand | 303 | 323 | 8 |
| 174 | 14 | Gilligan | Hand | 303 | 323 | 8 |
| 175 | 14 | Gilligan | Hand | 303 | 323 | 8 |
| 176 | 14 | Gilligan | Hand | 303 | 323 | 10 |
| 177 | 14 | Gilligan | Hand | 303 | 323 | 10 |
| 178 | 14 | Gilligan | Hand | 303 | 323 | 18 |
| 179 | 14 | Gilligan | Hand | 303 | 323 | 20 |
| 180 | 14 | Gilligan | Hand | 303 | 323 | 20 |
| 181 | 15 | Gilligan | Hand | 235 | 345 | 1 |
| 183 | 15 | Gilligan | Hand | 235 | 345 | 9 |
| 184 | 15 | Gilligan | Hand | 235 | 345 | 11 |
| 185 | 15 | Gilligan | Hand | 235 | 345 | 17 |
| 186 | 15 | Gilligan | Hand | 235 | 345 | 23 |
| 187 | 15 | Gilligan | Hand | 235 | 345 | 25 |
| 188 | 15 | Gilligan | Hand | 235 | 345 | 27 |
| 189 | 15 | Gilligan | Hand | 235 | 345 | 31 |
| 190 | 15 | Gilligan | Hand | 235 | 345 | 4 |
| 191 | 15 | Gilligan | Hand | 235 | 345 | 12 |
| 192 | 15 | Gilligan | Hand | 235 | 345 | 16 |
| 193 | 15 | Gilligan | Hand | 235 | 345 | 26 |
| 194 | 15 | Gilligan | Hand | 235 | 345 | 30 |
| 195 | 15 | Gilligan | Hand | 235 | 345 | 32 |
| 196 | 16 | Gilligan | Hand | 120 | 297 | 1 |
| 197 | 16 | Gilligan | Hand | 120 | 297 | 7 |
| 198 | 16 | Gilligan | Hand | 120 | 297 | 11 |
| 199 | 16 | Gilligan | Hand | 120 | 297 | 13 |

| UnitID | SiteID | Monkey | Somatotopy | X | Y | Channel |
|---|---|---|---|---|---|---|
| 200 | 16 | Gilligan | Hand | 120 | 297 | 15 |
| 201 | 16 | Gilligan | Hand | 120 | 297 | 17 |
| 202 | 16 | Gilligan | Hand | 120 | 297 | 19 |
| 203 | 16 | Gilligan | Hand | 120 | 297 | 23 |
| 204 | 16 | Gilligan | Hand | 120 | 297 | 25 |
| 205 | 16 | Gilligan | Hand | 120 | 297 | 29 |
| 206 | 16 | Gilligan | Hand | 120 | 297 | 31 |
| 207 | 16 | Gilligan | Hand | 120 | 297 | 4 |
| 208 | 16 | Gilligan | Hand | 120 | 297 | 6 |
| 209 | 16 | Gilligan | Hand | 120 | 297 | 8 |
| 210 | 16 | Gilligan | Hand | 120 | 297 | 14 |
| 211 | 16 | Gilligan | Hand | 120 | 297 | 18 |
| 212 | 16 | Gilligan | Hand | 120 | 297 | 24 |
| 213 | 16 | Gilligan | Hand | 120 | 297 | 26 |
| 214 | 16 | Gilligan | Hand | 120 | 297 | 28 |
| 215 | 16 | Gilligan | Hand | 120 | 297 | 30 |
| 216 | 17 | Gilligan | Arm | 230 | 435 | 7 |
| 217 | 17 | Gilligan | Arm | 230 | 435 | 13 |
| 218 | 17 | Gilligan | Arm | 230 | 435 | 17 |
| 219 | 17 | Gilligan | Arm | 230 | 435 | 27 |
| 220 | 17 | Gilligan | Arm | 230 | 435 | 29 |
| 221 | 17 | Gilligan | Arm | 230 | 435 | 4 |
| 222 | 17 | Gilligan | Arm | 230 | 435 | 6 |
| 223 | 17 | Gilligan | Arm | 230 | 435 | 8 |
| 224 | 17 | Gilligan | Arm | 230 | 435 | 26 |
| 225 | 17 | Gilligan | Arm | 230 | 435 | 28 |
| 226 | 17 | Gilligan | Arm | 230 | 435 | 30 |
| 228 | 18 | Gilligan | Hand | 164 | 370 | 1 |
| 229 | 18 | Gilligan | Hand | 164 | 370 | 3 |
| 230 | 18 | Gilligan | Hand | 164 | 370 | 9 |
| 231 | 18 | Gilligan | Hand | 164 | 370 | 11 |
| 232 | 18 | Gilligan | Hand | 164 | 370 | 13 |

| UnitID | SiteID | Monkey | Somatotopy | X | Y | Channel |
|---|---|---|---|---|---|---|
| 233 | 18 | Gilligan | Hand | 164 | 370 | 25 |
| 234 | 18 | Gilligan | Hand | 164 | 370 | 27 |
| 235 | 18 | Gilligan | Hand | 164 | 370 | 31 |
| 236 | 18 | Gilligan | Hand | 164 | 370 | 4 |
| 237 | 18 | Gilligan | Hand | 164 | 370 | 6 |
| 238 | 18 | Gilligan | Hand | 164 | 370 | 8 |
| 239 | 18 | Gilligan | Hand | 164 | 370 | 10 |
| 240 | 18 | Gilligan | Hand | 164 | 370 | 12 |
| 241 | 18 | Gilligan | Hand | 164 | 370 | 14 |
| 242 | 18 | Gilligan | Hand | 164 | 370 | 22 |
| 243 | 18 | Gilligan | Hand | 164 | 370 | 24 |
| 244 | 18 | Gilligan | Hand | 164 | 370 | 32 |
| 245 | 19 | Gilligan | Hand | 188 | 349 | 1 |
| 246 | 19 | Gilligan | Hand | 188 | 349 | 3 |
| 247 | 19 | Gilligan | Hand | 188 | 349 | 5 |
| 248 | 19 | Gilligan | Hand | 188 | 349 | 9 |
| 249 | 19 | Gilligan | Hand | 188 | 349 | 11 |
| 250 | 19 | Gilligan | Hand | 188 | 349 | 13 |
| 251 | 19 | Gilligan | Hand | 188 | 349 | 15 |
| 252 | 19 | Gilligan | Hand | 188 | 349 | 17 |
| 253 | 19 | Gilligan | Hand | 188 | 349 | 21 |
| 254 | 19 | Gilligan | Hand | 188 | 349 | 14 |
| 255 | 19 | Gilligan | Hand | 188 | 349 | 24 |
| 256 | 19 | Gilligan | Hand | 188 | 349 | 26 |
| 257 | 20 | Gilligan | Arm | 218 | 439 | 1 |
| 258 | 20 | Gilligan | Arm | 218 | 439 | 5 |
| 259 | 20 | Gilligan | Arm | 218 | 439 | 9 |
| 260 | 20 | Gilligan | Arm | 218 | 439 | 11 |
| 261 | 20 | Gilligan | Arm | 218 | 439 | 17 |
| 262 | 20 | Gilligan | Arm | 218 | 439 | 21 |
| 263 | 20 | Gilligan | Arm | 218 | 439 | 25 |
| 264 | 20 | Gilligan | Arm | 218 | 439 | 27 |

| UnitID | SiteID | Monkey | Somatotopy | X | Y | Channel |
|---|---|---|---|---|---|---|
| 265 | 20 | Gilligan | Arm | 218 | 439 | 29 |
| 266 | 20 | Gilligan | Arm | 218 | 439 | 31 |
| 267 | 20 | Gilligan | Arm | 218 | 439 | 2 |
| 268 | 20 | Gilligan | Arm | 218 | 439 | 2 |
| 269 | 20 | Gilligan | Arm | 218 | 439 | 4 |
| 270 | 20 | Gilligan | Arm | 218 | 439 | 8 |
| 271 | 20 | Gilligan | Arm | 218 | 439 | 10 |
| 272 | 20 | Gilligan | Arm | 218 | 439 | 12 |
| 273 | 20 | Gilligan | Arm | 218 | 439 | 14 |
| 274 | 20 | Gilligan | Arm | 218 | 439 | 20 |
| 275 | 20 | Gilligan | Arm | 218 | 439 | 20 |
| 276 | 20 | Gilligan | Arm | 218 | 439 | 22 |
| 277 | 20 | Gilligan | Arm | 218 | 439 | 24 |
| 278 | 20 | Gilligan | Arm | 218 | 439 | 26 |
| 279 | 20 | Gilligan | Arm | 218 | 439 | 28 |
| 280 | 20 | Gilligan | Arm | 218 | 439 | 30 |
| 281 | 21 | Gilligan | Arm | 218 | 208 | 1 |
| 282 | 21 | Gilligan | Arm | 218 | 208 | 3 |
| 283 | 21 | Gilligan | Arm | 218 | 208 | 5 |
| 284 | 21 | Gilligan | Arm | 218 | 208 | 7 |
| 285 | 21 | Gilligan | Arm | 218 | 208 | 9 |
| 286 | 21 | Gilligan | Arm | 218 | 208 | 13 |
| 287 | 21 | Gilligan | Arm | 218 | 208 | 15 |
| 288 | 21 | Gilligan | Arm | 218 | 208 | 17 |
| 289 | 21 | Gilligan | Arm | 218 | 208 | 19 |
| 290 | 21 | Gilligan | Arm | 218 | 208 | 21 |
| 291 | 21 | Gilligan | Arm | 218 | 208 | 23 |
| 292 | 21 | Gilligan | Arm | 218 | 208 | 27 |
| 293 | 21 | Gilligan | Arm | 218 | 208 | 29 |
| 294 | 21 | Gilligan | Arm | 218 | 208 | 4 |
| 295 | 21 | Gilligan | Arm | 218 | 208 | 6 |
| 296 | 21 | Gilligan | Arm | 218 | 208 | 8 |

| UnitID | SiteID | Monkey | Somatotopy | X | Y | Channel |
|---|---|---|---|---|---|---|
| 297 | 21 | Gilligan | Arm | 218 | 208 | 10 |
| 298 | 21 | Gilligan | Arm | 218 | 208 | 12 |
| 299 | 21 | Gilligan | Arm | 218 | 208 | 14 |
| 300 | 21 | Gilligan | Arm | 218 | 208 | 18 |
| 301 | 21 | Gilligan | Arm | 218 | 208 | 20 |
| 302 | 21 | Gilligan | Arm | 218 | 208 | 22 |
| 303 | 21 | Gilligan | Arm | 218 | 208 | 24 |
| 304 | 21 | Gilligan | Arm | 218 | 208 | 26 |
| 305 | 21 | Gilligan | Arm | 218 | 208 | 28 |
| 306 | 22 | Gilligan | Trunk | 184 | 260 | 3 |
| 307 | 22 | Gilligan | Trunk | 184 | 260 | 5 |
| 308 | 22 | Gilligan | Trunk | 184 | 260 | 9 |
| 309 | 22 | Gilligan | Trunk | 184 | 260 | 11 |
| 310 | 22 | Gilligan | Trunk | 184 | 260 | 13 |
| 311 | 22 | Gilligan | Trunk | 184 | 260 | 23 |
| 312 | 22 | Gilligan | Trunk | 184 | 260 | 25 |
| 313 | 22 | Gilligan | Trunk | 184 | 260 | 29 |
| 314 | 22 | Gilligan | Trunk | 184 | 260 | 31 |
| 315 | 22 | Gilligan | Trunk | 184 | 260 | 12 |
| 316 | 22 | Gilligan | Trunk | 184 | 260 | 14 |
| 317 | 22 | Gilligan | Trunk | 184 | 260 | 18 |
| 318 | 22 | Gilligan | Trunk | 184 | 260 | 24 |
| 319 | 22 | Gilligan | Trunk | 184 | 260 | 28 |
| 320 | 23 | Gilligan | Hand | 185 | 483 | 1 |
| 321 | 23 | Gilligan | Hand | 185 | 483 | 3 |
| 322 | 23 | Gilligan | Hand | 185 | 483 | 9 |
| 323 | 23 | Gilligan | Hand | 185 | 483 | 11 |
| 324 | 23 | Gilligan | Hand | 185 | 483 | 13 |
| 325 | 23 | Gilligan | Hand | 185 | 483 | 15 |
| 326 | 23 | Gilligan | Hand | 185 | 483 | 19 |
| 327 | 23 | Gilligan | Hand | 185 | 483 | 23 |
| 328 | 23 | Gilligan | Hand | 185 | 483 | 25 |

| UnitID | SiteID | Monkey | Somatotopy | X | Y | Channel |
|---|---|---|---|---|---|---|
| 329 | 23 | Gilligan | Hand | 185 | 483 | 27 |
| 330 | 23 | Gilligan | Hand | 185 | 483 | 29 |
| 331 | 23 | Gilligan | Hand | 185 | 483 | 31 |
| 332 | 23 | Gilligan | Hand | 185 | 483 | 4 |
| 333 | 23 | Gilligan | Hand | 185 | 483 | 6 |
| 334 | 23 | Gilligan | Hand | 185 | 483 | 8 |
| 335 | 23 | Gilligan | Hand | 185 | 483 | 12 |
| 336 | 23 | Gilligan | Hand | 185 | 483 | 14 |
| 337 | 23 | Gilligan | Hand | 185 | 483 | 16 |
| 338 | 23 | Gilligan | Hand | 185 | 483 | 18 |
| 339 | 23 | Gilligan | Hand | 185 | 483 | 20 |
| 340 | 23 | Gilligan | Hand | 185 | 483 | 22 |
| 341 | 23 | Gilligan | Hand | 185 | 483 | 28 |
| 342 | 23 | Gilligan | Hand | 185 | 483 | 30 |
| 343 | 23 | Gilligan | Hand | 185 | 483 | 32 |
| 344 | 24 | Gilligan | Trunk | 217 | 474 | 3 |
| 345 | 24 | Gilligan | Trunk | 217 | 474 | 5 |
| 347 | 24 | Gilligan | Trunk | 217 | 474 | 9 |
| 348 | 24 | Gilligan | Trunk | 217 | 474 | 11 |
| 349 | 24 | Gilligan | Trunk | 217 | 474 | 13 |
| 350 | 24 | Gilligan | Trunk | 217 | 474 | 15 |
| 351 | 24 | Gilligan | Trunk | 217 | 474 | 21 |
| 354 | 24 | Gilligan | Trunk | 217 | 474 | 31 |
| 355 | 24 | Gilligan | Trunk | 217 | 474 | 4 |
| 356 | 24 | Gilligan | Trunk | 217 | 474 | 10 |
| 357 | 24 | Gilligan | Trunk | 217 | 474 | 14 |
| 359 | 24 | Gilligan | Trunk | 217 | 474 | 20 |
| 360 | 24 | Gilligan | Trunk | 217 | 474 | 22 |
| 361 | 25 | Gilligan | Hand | 140 | 425 | 9 |
| 362 | 25 | Gilligan | Hand | 140 | 425 | 17 |
| 363 | 25 | Gilligan | Hand | 140 | 425 | 2 |
| 364 | 25 | Gilligan | Hand | 140 | 425 | 6 |

| UnitID | SiteID | Monkey | Somatotopy | X | Y | Channel |
|---|---|---|---|---|---|---|
| 365 | 25 | Gilligan | Hand | 140 | 425 | 20 |
| 366 | 25 | Gilligan | Hand | 140 | 425 | 26 |
| 367 | 26 | Gilligan | Hand | 126 | 491 | 7 |
| 368 | 26 | Gilligan | Hand | 126 | 491 | 9 |
| 369 | 26 | Gilligan | Hand | 126 | 491 | 25 |
| 370 | 26 | Gilligan | Hand | 126 | 491 | 8 |
| 371 | 26 | Gilligan | Hand | 126 | 491 | 16 |
| 372 | 26 | Gilligan | Hand | 126 | 491 | 20 |
| 373 | 26 | Gilligan | Hand | 126 | 491 | 20 |
| 374 | 26 | Gilligan | Hand | 126 | 491 | 22 |
| 375 | 26 | Gilligan | Hand | 126 | 491 | 24 |
| 376 | 26 | Gilligan | Hand | 126 | 491 | 30 |
| 377 | 26 | Gilligan | Hand | 126 | 491 | 32 |
| 378 | 27 | Gilligan | Arm | 241 | 245 | 3 |
| 379 | 27 | Gilligan | Arm | 241 | 245 | 7 |
| 380 | 27 | Gilligan | Arm | 241 | 245 | 11 |
| 381 | 27 | Gilligan | Arm | 241 | 245 | 15 |
| 382 | 27 | Gilligan | Arm | 241 | 245 | 23 |
| 383 | 27 | Gilligan | Arm | 241 | 245 | 25 |
| 384 | 27 | Gilligan | Arm | 241 | 245 | 29 |
| 385 | 27 | Gilligan | Arm | 241 | 245 | 31 |
| 386 | 27 | Gilligan | Arm | 241 | 245 | 6 |
| 387 | 27 | Gilligan | Arm | 241 | 245 | 20 |
| 388 | 27 | Gilligan | Arm | 241 | 245 | 26 |
| 389 | 27 | Gilligan | Arm | 241 | 245 | 30 |
| 390 | 27 | Gilligan | Arm | 241 | 245 | 32 |
| 391 | 28 | Gilligan | Hand | 200 | 299 | 3 |
| 392 | 28 | Gilligan | Hand | 200 | 299 | 21 |
| 393 | 28 | Gilligan | Hand | 200 | 299 | 4 |
| 394 | 28 | Gilligan | Hand | 200 | 299 | 10 |
| 395 | 28 | Gilligan | Hand | 200 | 299 | 18 |
| 396 | 28 | Gilligan | Hand | 200 | 299 | 24 |

| UnitID | SiteID | Monkey | Somatotopy | X | Y | Channel |
|---|---|---|---|---|---|---|
| 397 | 28 | Gilligan | Hand | 200 | 299 | 28 |
| 398 | 28 | Gilligan | Hand | 200 | 299 | 30 |
| 399 | 28 | Gilligan | Hand | 200 | 299 | 32 |
| 400 | 29 | Gilligan | Hand | 282 | 289 | 3 |
| 401 | 29 | Gilligan | Hand | 282 | 289 | 15 |
| 402 | 29 | Gilligan | Hand | 282 | 289 | 21 |
| 403 | 29 | Gilligan | Hand | 282 | 289 | 25 |
| 405 | 29 | Gilligan | Hand | 282 | 289 | 29 |
| 406 | 29 | Gilligan | Hand | 282 | 289 | 31 |
| 407 | 29 | Gilligan | Hand | 282 | 289 | 2 |
| 408 | 29 | Gilligan | Hand | 282 | 289 | 6 |
| 409 | 29 | Gilligan | Hand | 282 | 289 | 10 |
| 410 | 29 | Gilligan | Hand | 282 | 289 | 14 |
| 411 | 29 | Gilligan | Hand | 282 | 289 | 30 |
| 412 | 30 | Gilligan | Arm | 143 | 253 | 3 |
| 413 | 30 | Gilligan | Arm | 143 | 253 | 3 |
| 414 | 30 | Gilligan | Arm | 143 | 253 | 11 |
| 415 | 30 | Gilligan | Arm | 143 | 253 | 19 |
| 416 | 30 | Gilligan | Arm | 143 | 253 | 27 |
| 417 | 30 | Gilligan | Arm | 143 | 253 | 27 |
| 418 | 30 | Gilligan | Arm | 143 | 253 | 31 |
| 419 | 30 | Gilligan | Arm | 143 | 253 | 4 |
| 420 | 30 | Gilligan | Arm | 143 | 253 | 4 |
| 421 | 30 | Gilligan | Arm | 143 | 253 | 8 |
| 422 | 30 | Gilligan | Arm | 143 | 253 | 8 |
| 423 | 30 | Gilligan | Arm | 143 | 253 | 12 |
| 424 | 30 | Gilligan | Arm | 143 | 253 | 16 |
| 425 | 30 | Gilligan | Arm | 143 | 253 | 26 |
| 426 | 30 | Gilligan | Arm | 143 | 253 | 28 |
| 427 | 31 | Gilligan | Arm | 181 | 286 | 5 |
| 428 | 31 | Gilligan | Arm | 181 | 286 | 21 |
| 429 | 31 | Gilligan | Arm | 181 | 286 | 29 |

| UnitID | SiteID | Monkey | Somatotopy | X | Y | Channel |
|---|---|---|---|---|---|---|
| 430 | 31 | Gilligan | Arm | 181 | 286 | 2 |
| 431 | 31 | Gilligan | Arm | 181 | 286 | 14 |
| 432 | 32 | Gilligan | Hand | 164 | 399 | 1 |
| 433 | 32 | Gilligan | Hand | 164 | 399 | 9 |
| 434 | 32 | Gilligan | Hand | 164 | 399 | 13 |
| 435 | 32 | Gilligan | Hand | 164 | 399 | 29 |
| 436 | 32 | Gilligan | Hand | 164 | 399 | 4 |
| 437 | 32 | Gilligan | Hand | 164 | 399 | 8 |
| 438 | 32 | Gilligan | Hand | 164 | 399 | 16 |
| 439 | 32 | Gilligan | Hand | 164 | 399 | 18 |
| 440 | 33 | Gilligan | Arm | 309 | 289 | 19 |
| 441 | 33 | Gilligan | Arm | 309 | 289 | 4 |
| 442 | 33 | Gilligan | Arm | 309 | 289 | 6 |
| 443 | 33 | Gilligan | Arm | 309 | 289 | 18 |
| 444 | 34 | Gilligan | Arm | 199 | 198 | 3 |
| 445 | 34 | Gilligan | Arm | 199 | 198 | 31 |
| 446 | 34 | Gilligan | Arm | 199 | 198 | 20 |
| 447 | 34 | Gilligan | Arm | 199 | 198 | 26 |
| 448 | 35 | Gilligan | Hand | 174 | 316 | 21 |
| 449 | 36 | Gilligan | Arm | 156 | 191 | 13 |
| 450 | 36 | Gilligan | Arm | 156 | 191 | 27 |
| 451 | 36 | Gilligan | Arm | 156 | 191 | 29 |
| 452 | 36 | Gilligan | Arm | 156 | 191 | 12 |
| 453 | 36 | Gilligan | Arm | 156 | 191 | 14 |
| 454 | 36 | Gilligan | Arm | 156 | 191 | 24 |
| 455 | 37 | Gilligan | Hand | 135 | 317 | 9 |
| 456 | 37 | Gilligan | Hand | 135 | 317 | 17 |
| 457 | 37 | Gilligan | Hand | 135 | 317 | 19 |
| 458 | 37 | Gilligan | Hand | 135 | 317 | 21 |
| 459 | 37 | Gilligan | Hand | 135 | 317 | 25 |
| 460 | 37 | Gilligan | Hand | 135 | 317 | 26 |
| 461 | 37 | Gilligan | Hand | 135 | 317 | 32 |

| UnitID | SiteID | Monkey | Somatotopy | X | Y | Channel |
|---|---|---|---|---|---|---|
| 462 | 38 | Gilligan | Arm | 84 | 174 | 3 |
| 463 | 38 | Gilligan | Arm | 84 | 174 | 11 |
| 464 | 38 | Gilligan | Arm | 84 | 174 | 27 |
| 465 | 38 | Gilligan | Arm | 84 | 174 | 8 |
| 466 | 38 | Gilligan | Arm | 84 | 174 | 12 |
| 467 | 38 | Gilligan | Arm | 84 | 174 | 16 |
| 468 | 38 | Gilligan | Arm | 84 | 174 | 16 |
| 469 | 38 | Gilligan | Arm | 84 | 174 | 20 |
| 470 | 38 | Gilligan | Arm | 84 | 174 | 28 |
| 471 | 38 | Gilligan | Arm | 84 | 174 | 28 |
| 472 | 39 | Gilligan | Trunk | 189 | 61 | 11 |
| 473 | 39 | Gilligan | Trunk | 189 | 61 | 26 |
| 474 | 39 | Gilligan | Trunk | 189 | 61 | 28 |
| 475 | 40 | Gilligan | Trunk | 140 | 557 | 31 |
| 476 | 40 | Gilligan | Trunk | 140 | 557 | 6 |
| 477 | 40 | Gilligan | Trunk | 140 | 557 | 10 |
| 478 | 40 | Gilligan | Trunk | 140 | 557 | 22 |
| 479 | 40 | Gilligan | Trunk | 140 | 557 | 32 |
| 480 | 41 | Gilligan | Arm | 205 | 146 | 3 |
| 481 | 41 | Gilligan | Arm | 205 | 146 | 5 |
| 482 | 41 | Gilligan | Arm | 205 | 146 | 11 |
| 483 | 41 | Gilligan | Arm | 205 | 146 | 19 |
| 484 | 41 | Gilligan | Arm | 205 | 146 | 21 |
| 485 | 41 | Gilligan | Arm | 205 | 146 | 27 |
| 486 | 41 | Gilligan | Arm | 205 | 146 | 2 |
| 487 | 41 | Gilligan | Arm | 205 | 146 | 4 |
| 488 | 41 | Gilligan | Arm | 205 | 146 | 6 |
| 489 | 41 | Gilligan | Arm | 205 | 146 | 14 |
| 490 | 41 | Gilligan | Arm | 205 | 146 | 18 |
| 491 | 41 | Gilligan | Arm | 205 | 146 | 20 |
| 492 | 42 | Gilligan | Trunk | 95 | 16 | 7 |
| 493 | 42 | Gilligan | Trunk | 95 | 16 | 25 |

| UnitID | SiteID | Monkey | Somatotopy | X | Y | Channel |
|---|---|---|---|---|---|---|
| 494 | 42 | Gilligan | Trunk | 95 | 16 | 8 |
| 495 | 42 | Gilligan | Trunk | 95 | 16 | 12 |
| 496 | 42 | Gilligan | Trunk | 95 | 16 | 22 |
| 497 | 42 | Gilligan | Trunk | 95 | 16 | 30 |
| 498 | 42 | Gilligan | Trunk | 95 | 16 | 32 |
| 499 | 43 | Gilligan | Arm | 86 | 77 | 2 |
| 500 | 43 | Gilligan | Arm | 86 | 77 | 4 |
| 501 | 43 | Gilligan | Arm | 86 | 77 | 6 |
| 502 | 44 | Gilligan | Hand | 148 | 365 | 26 |
| 503 | 45 | Gilligan | Arm | 207 | 424 | 5 |
| 504 | 45 | Gilligan | Arm | 207 | 424 | 15 |
| 505 | 45 | Gilligan | Arm | 207 | 424 | 19 |
| 506 | 45 | Gilligan | Arm | 207 | 424 | 21 |
| 507 | 45 | Gilligan | Arm | 207 | 424 | 18 |
| 508 | 46 | Gilligan | Arm | 216 | 379 | 10 |
| 509 | 47 | Gilligan | Arm | 247 | 154 | 15 |
| 510 | 47 | Gilligan | Arm | 247 | 154 | 19 |
| 511 | 47 | Gilligan | Arm | 247 | 154 | 23 |
| 512 | 47 | Gilligan | Arm | 247 | 154 | 4 |
| 514 | 48 | Gilligan | Arm | 185 | 112 | 5 |
| 515 | 48 | Gilligan | Arm | 185 | 112 | 9 |
| 517 | 48 | Gilligan | Arm | 185 | 112 | 19 |
| 518 | 48 | Gilligan | Arm | 185 | 112 | 27 |
| 519 | 48 | Gilligan | Arm | 185 | 112 | 27 |
| 520 | 48 | Gilligan | Arm | 185 | 112 | 29 |
| 521 | 48 | Gilligan | Arm | 185 | 112 | 31 |
| 522 | 48 | Gilligan | Arm | 185 | 112 | 31 |
| 523 | 48 | Gilligan | Arm | 185 | 112 | 10 |
| 524 | 48 | Gilligan | Arm | 185 | 112 | 12 |
| 525 | 48 | Gilligan | Arm | 185 | 112 | 14 |
| 526 | 48 | Gilligan | Arm | 185 | 112 | 20 |
| 527 | 48 | Gilligan | Arm | 185 | 112 | 24 |

| UnitID | SiteID | Monkey | Somatotopy | X | Y | Channel |
|---|---|---|---|---|---|---|
| 528 | 49 | Gilligan | Hand | 165 | 425 | 7 |
| 529 | 49 | Gilligan | Hand | 165 | 425 | 13 |
| 530 | 50 | Gilligan | Arm | 86 | 261 | 15 |
| 531 | 50 | Gilligan | Arm | 86 | 261 | 23 |
| 532 | 50 | Gilligan | Arm | 86 | 261 | 6 |
| 533 | 50 | Gilligan | Arm | 86 | 261 | 10 |
| 534 | 50 | Gilligan | Arm | 86 | 261 | 28 |
| 535 | 51 | Gilligan | Hand | 98 | 451 | 4 |
| 536 | 51 | Gilligan | Hand | 98 | 451 | 18 |
| 537 | 51 | Gilligan | Hand | 98 | 451 | 30 |
| 538 | 52 | Gilligan | Hand | 181 | 460 | 1 |
| 539 | 52 | Gilligan | Hand | 181 | 460 | 9 |
| 540 | 52 | Gilligan | Hand | 181 | 460 | 15 |
| 541 | 52 | Gilligan | Hand | 181 | 460 | 25 |
| 542 | 52 | Gilligan | Hand | 181 | 460 | 8 |
| 543 | 52 | Gilligan | Hand | 181 | 460 | 12 |
| 544 | 52 | Gilligan | Hand | 181 | 460 | 16 |
| 545 | 52 | Gilligan | Hand | 181 | 460 | 26 |
| 546 | 53 | Gilligan | Arm | 321 | 326 | 1 |
| 547 | 53 | Gilligan | Arm | 321 | 326 | 5 |
| 548 | 53 | Gilligan | Arm | 321 | 326 | 21 |
| 549 | 53 | Gilligan | Arm | 321 | 326 | 25 |
| 550 | 53 | Gilligan | Arm | 321 | 326 | 6 |
| 553 | 53 | Gilligan | Arm | 321 | 326 | 30 |
| 554 | 54 | Gilligan | Arm | 231 | 108 | 13 |
| 555 | 54 | Gilligan | Arm | 231 | 108 | 21 |
| 556 | 54 | Gilligan | Arm | 231 | 108 | 25 |
| 557 | 54 | Gilligan | Arm | 231 | 108 | 29 |
| 558 | 54 | Gilligan | Arm | 231 | 108 | 2 |
| 559 | 54 | Gilligan | Arm | 231 | 108 | 6 |
| 560 | 54 | Gilligan | Arm | 231 | 108 | 8 |
| 561 | 54 | Gilligan | Arm | 231 | 108 | 16 |

| UnitID | SiteID | Monkey | Somatotopy | X | Y | Channel |
|---|---|---|---|---|---|---|
| 562 | 54 | Gilligan | Arm | 231 | 108 | 20 |
| 563 | 54 | Gilligan | Arm | 231 | 108 | 24 |
| 564 | 54 | Gilligan | Arm | 231 | 108 | 28 |
| 565 | 55 | Gilligan | Arm | 221 | 68 | 3 |
| 566 | 55 | Gilligan | Arm | 221 | 68 | 11 |
| 567 | 55 | Gilligan | Arm | 221 | 68 | 13 |
| 568 | 55 | Gilligan | Arm | 221 | 68 | 15 |
| 569 | 55 | Gilligan | Arm | 221 | 68 | 19 |
| 570 | 55 | Gilligan | Arm | 221 | 68 | 19 |
| 571 | 55 | Gilligan | Arm | 221 | 68 | 27 |
| 572 | 55 | Gilligan | Arm | 221 | 68 | 27 |
| 573 | 55 | Gilligan | Arm | 221 | 68 | 31 |
| 574 | 55 | Gilligan | Arm | 221 | 68 | 4 |
| 575 | 55 | Gilligan | Arm | 221 | 68 | 32 |
| 576 | 56 | Gilligan | Trunk | 140 | 9 | 3 |
| 577 | 56 | Gilligan | Trunk | 140 | 9 | 7 |
| 578 | 56 | Gilligan | Trunk | 140 | 9 | 17 |
| 579 | 56 | Gilligan | Trunk | 140 | 9 | 23 |
| 580 | 56 | Gilligan | Trunk | 140 | 9 | 27 |
| 581 | 56 | Gilligan | Trunk | 140 | 9 | 31 |
| 582 | 56 | Gilligan | Trunk | 140 | 9 | 2 |
| 583 | 56 | Gilligan | Trunk | 140 | 9 | 4 |
| 584 | 56 | Gilligan | Trunk | 140 | 9 | 12 |
| 585 | 56 | Gilligan | Trunk | 140 | 9 | 18 |
| 586 | 56 | Gilligan | Trunk | 140 | 9 | 22 |
| 587 | 57 | Gilligan | Hand | 117 | 355 | 17 |
| 588 | 57 | Gilligan | Hand | 117 | 355 | 31 |
| 589 | 57 | Gilligan | Hand | 117 | 355 | 22 |
| 590 | 57 | Gilligan | Hand | 117 | 355 | 26 |
| 591 | 58 | Gilligan | Hand | 100 | 277 | 15 |
| 592 | 58 | Gilligan | Hand | 100 | 277 | 19 |
| 594 | 58 | Gilligan | Hand | 100 | 277 | 4 |

| UnitID | SiteID | Monkey | Somatotopy | X | Y | Channel |
|---|---|---|---|---|---|---|
| 595 | 58 | Gilligan | Hand | 100 | 277 | 6 |
| 596 | 58 | Gilligan | Hand | 100 | 277 | 12 |
| 598 | 58 | Gilligan | Hand | 100 | 277 | 30 |
| 599 | 59 | Gilligan | Hand | 157 | 513 | 15 |
| 600 | 59 | Gilligan | Hand | 157 | 513 | 31 |
| 601 | 59 | Gilligan | Hand | 157 | 513 | 2 |
| 602 | 59 | Gilligan | Hand | 157 | 513 | 6 |
| 603 | 59 | Gilligan | Hand | 157 | 513 | 8 |
| 604 | 59 | Gilligan | Hand | 157 | 513 | 8 |
| 605 | 59 | Gilligan | Hand | 157 | 513 | 12 |
| 606 | 60 | Gilligan | Arm | 75 | 207 | 15 |
| 607 | 60 | Gilligan | Arm | 75 | 207 | 17 |
| 608 | 60 | Gilligan | Arm | 75 | 207 | 29 |
| 609 | 60 | Gilligan | Arm | 75 | 207 | 31 |
| 610 | 60 | Gilligan | Arm | 75 | 207 | 6 |
| 611 | 60 | Gilligan | Arm | 75 | 207 | 12 |
| 612 | 60 | Gilligan | Arm | 75 | 207 | 20 |
| 613 | 60 | Gilligan | Arm | 75 | 207 | 22 |
| 614 | 60 | Gilligan | Arm | 75 | 207 | 32 |
| 615 | 61 | Gilligan | Arm | 143 | 134 | 11 |
| 618 | 61 | Gilligan | Arm | 143 | 134 | 4 |
| 619 | 61 | Gilligan | Arm | 143 | 134 | 12 |
| 620 | 61 | Gilligan | Arm | 143 | 134 | 26 |
| 622 | 62 | Gilligan | Trunk | 134 | 87 | 5 |
| 623 | 62 | Gilligan | Trunk | 134 | 87 | 9 |
| 624 | 62 | Gilligan | Trunk | 134 | 87 | 15 |
| 625 | 62 | Gilligan | Trunk | 134 | 87 | 17 |
| 626 | 62 | Gilligan | Trunk | 134 | 87 | 21 |
| 627 | 62 | Gilligan | Trunk | 134 | 87 | 25 |
| 628 | 62 | Gilligan | Trunk | 134 | 87 | 27 |
| 630 | 62 | Gilligan | Trunk | 134 | 87 | 4 |
| 631 | 62 | Gilligan | Trunk | 134 | 87 | 6 |

| UnitID | SiteID | Monkey | Somatotopy | X | Y | Channel |
|---|---|---|---|---|---|---|
| 632 | 62 | Gilligan | Trunk | 134 | 87 | 8 |
| 633 | 62 | Gilligan | Trunk | 134 | 87 | 12 |
| 634 | 62 | Gilligan | Trunk | 134 | 87 | 14 |
| 635 | 62 | Gilligan | Trunk | 134 | 87 | 16 |
| 636 | 62 | Gilligan | Trunk | 134 | 87 | 18 |
| 637 | 62 | Gilligan | Trunk | 134 | 87 | 18 |
| 638 | 62 | Gilligan | Trunk | 134 | 87 | 20 |
| 639 | 62 | Gilligan | Trunk | 134 | 87 | 24 |
| 640 | 62 | Gilligan | Trunk | 134 | 87 | 30 |
| 641 | 62 | Gilligan | Trunk | 134 | 87 | 32 |
| 642 | 63 | Gilligan | Arm | 294 | 147 | 9 |
| 643 | 63 | Gilligan | Arm | 294 | 147 | 11 |
| 644 | 63 | Gilligan | Arm | 294 | 147 | 14 |
| 645 | 63 | Gilligan | Arm | 294 | 147 | 16 |
| 646 | 63 | Gilligan | Arm | 294 | 147 | 16 |
| 647 | 63 | Gilligan | Arm | 294 | 147 | 18 |
| 648 | 63 | Gilligan | Arm | 294 | 147 | 24 |
| 649 | 63 | Gilligan | Arm | 294 | 147 | 26 |
| 650 | 63 | Gilligan | Arm | 294 | 147 | 28 |
| 651 | 63 | Gilligan | Arm | 294 | 147 | 28 |
| 653 | 64 | Gilligan | Hand | 70 | 365 | 3 |
| 655 | 64 | Gilligan | Hand | 70 | 365 | 31 |
| 656 | 64 | Gilligan | Hand | 70 | 365 | 6 |
| 657 | 64 | Gilligan | Hand | 70 | 365 | 8 |
| 658 | 64 | Gilligan | Hand | 70 | 365 | 14 |
| 659 | 64 | Gilligan | Hand | 70 | 365 | 20 |
| 660 | 64 | Gilligan | Hand | 70 | 365 | 28 |
| 661 | 64 | Gilligan | Hand | 70 | 365 | 32 |
| 662 | 65 | Gilligan | Arm | 286 | 165 | 1 |
| 663 | 65 | Gilligan | Arm | 286 | 165 | 3 |
| 664 | 65 | Gilligan | Arm | 286 | 165 | 15 |
| 665 | 65 | Gilligan | Arm | 286 | 165 | 19 |

| UnitID | SiteID | Monkey | Somatotopy | X | Y | Channel |
|---|---|---|---|---|---|---|
| 666 | 65 | Gilligan | Arm | 286 | 165 | 23 |
| 667 | 65 | Gilligan | Arm | 286 | 165 | 29 |
| 668 | 65 | Gilligan | Arm | 286 | 165 | 2 |
| 669 | 65 | Gilligan | Arm | 286 | 165 | 4 |
| 670 | 65 | Gilligan | Arm | 286 | 165 | 4 |
| 671 | 65 | Gilligan | Arm | 286 | 165 | 8 |
| 672 | 65 | Gilligan | Arm | 286 | 165 | 12 |
| 673 | 65 | Gilligan | Arm | 286 | 165 | 12 |
| 674 | 65 | Gilligan | Arm | 286 | 165 | 18 |
| 675 | 65 | Gilligan | Arm | 286 | 165 | 30 |
| 676 | 66 | Gilligan | Arm | 236 | 260 | 1 |
| 677 | 66 | Gilligan | Arm | 236 | 260 | 3 |
| 678 | 66 | Gilligan | Arm | 236 | 260 | 5 |
| 679 | 66 | Gilligan | Arm | 236 | 260 | 15 |
| 681 | 66 | Gilligan | Arm | 236 | 260 | 27 |
| 682 | 66 | Gilligan | Arm | 236 | 260 | 4 |
| 683 | 67 | Gilligan | Trunk | 214 | 523 | 27 |
| 684 | 67 | Gilligan | Trunk | 214 | 523 | 4 |
| 685 | 68 | Gilligan | Arm | 291 | 84 | 7 |
| 686 | 68 | Gilligan | Arm | 291 | 84 | 7 |
| 687 | 68 | Gilligan | Arm | 291 | 84 | 11 |
| 689 | 68 | Gilligan | Arm | 291 | 84 | 19 |
| 690 | 68 | Gilligan | Arm | 291 | 84 | 27 |
| 691 | 68 | Gilligan | Arm | 291 | 84 | 31 |
| 692 | 68 | Gilligan | Arm | 291 | 84 | 2 |
| 693 | 68 | Gilligan | Arm | 291 | 84 | 4 |
| 694 | 68 | Gilligan | Arm | 291 | 84 | 14 |
| 695 | 69 | Gilligan | Hand | 97 | 308 | 7 |
| 696 | 69 | Gilligan | Hand | 97 | 308 | 7 |
| 697 | 69 | Gilligan | Hand | 97 | 308 | 11 |
| 698 | 69 | Gilligan | Hand | 97 | 308 | 15 |
| 699 | 69 | Gilligan | Hand | 97 | 308 | 23 |

| UnitID | SiteID | Monkey | Somatotopy | X | Y | Channel |
|---|---|---|---|---|---|---|
| 700 | 69 | Gilligan | Hand | 97 | 308 | 23 |
| 701 | 69 | Gilligan | Hand | 97 | 308 | 23 |
| 702 | 69 | Gilligan | Hand | 97 | 308 | 27 |
| 703 | 69 | Gilligan | Hand | 97 | 308 | 4 |
| 704 | 69 | Gilligan | Hand | 97 | 308 | 4 |
| 705 | 69 | Gilligan | Hand | 97 | 308 | 10 |
| 706 | 69 | Gilligan | Hand | 97 | 308 | 10 |
| 707 | 69 | Gilligan | Hand | 97 | 308 | 14 |
| 708 | 69 | Gilligan | Hand | 97 | 308 | 14 |
| 709 | 69 | Gilligan | Hand | 97 | 308 | 18 |
| 710 | 69 | Gilligan | Hand | 97 | 308 | 22 |
| 711 | 69 | Gilligan | Hand | 97 | 308 | 22 |
| 712 | 69 | Gilligan | Hand | 97 | 308 | 24 |
| 713 | 69 | Gilligan | Hand | 97 | 308 | 26 |
| 714 | 69 | Gilligan | Hand | 97 | 308 | 26 |
| 715 | 69 | Gilligan | Hand | 97 | 308 | 28 |
| 716 | 69 | Gilligan | Hand | 97 | 308 | 28 |
| 717 | 69 | Gilligan | Hand | 97 | 308 | 30 |
| 718 | 69 | Gilligan | Hand | 97 | 308 | 32 |
| 719 | 70 | Gilligan | Hand | 277 | 333 | 15 |
| 720 | 70 | Gilligan | Hand | 277 | 333 | 31 |
| 721 | 70 | Gilligan | Hand | 277 | 333 | 2 |
| 722 | 70 | Gilligan | Hand | 277 | 333 | 14 |
| 723 | 71 | Gilligan | Arm | 274 | 121 | 1 |
| 724 | 71 | Gilligan | Arm | 274 | 121 | 1 |
| 725 | 71 | Gilligan | Arm | 274 | 121 | 5 |
| 726 | 71 | Gilligan | Arm | 274 | 121 | 7 |
| 727 | 71 | Gilligan | Arm | 274 | 121 | 9 |
| 728 | 71 | Gilligan | Arm | 274 | 121 | 11 |
| 729 | 71 | Gilligan | Arm | 274 | 121 | 13 |
| 730 | 71 | Gilligan | Arm | 274 | 121 | 17 |
| 731 | 71 | Gilligan | Arm | 274 | 121 | 17 |

| UnitID | SiteID | Monkey | Somatotopy | X | Y | Channel |
|---|---|---|---|---|---|---|
| 732 | 71 | Gilligan | Arm | 274 | 121 | 19 |
| 733 | 71 | Gilligan | Arm | 274 | 121 | 25 |
| 734 | 71 | Gilligan | Arm | 274 | 121 | 8 |
| 735 | 71 | Gilligan | Arm | 274 | 121 | 14 |
| 736 | 71 | Gilligan | Arm | 274 | 121 | 20 |
| 737 | 71 | Gilligan | Arm | 274 | 121 | 22 |
| 738 | 71 | Gilligan | Arm | 274 | 121 | 24 |
| 739 | 71 | Gilligan | Arm | 274 | 121 | 28 |
| 740 | 71 | Gilligan | Arm | 274 | 121 | 28 |
| 741 | 71 | Gilligan | Arm | 274 | 121 | 32 |
| 742 | 72 | Gilligan | Arm | 251 | 365 | 11 |
| 743 | 72 | Gilligan | Arm | 251 | 365 | 26 |
| 744 | 72 | Gilligan | Arm | 251 | 365 | 28 |
| 745 | 72 | Gilligan | Arm | 251 | 365 | 32 |
| 746 | 73 | Gilligan | Arm | 284 | 182 | 7 |
| 747 | 73 | Gilligan | Arm | 284 | 182 | 7 |
| 748 | 73 | Gilligan | Arm | 284 | 182 | 11 |
| 749 | 73 | Gilligan | Arm | 284 | 182 | 11 |
| 750 | 73 | Gilligan | Arm | 284 | 182 | 15 |
| 751 | 73 | Gilligan | Arm | 284 | 182 | 19 |
| 752 | 73 | Gilligan | Arm | 284 | 182 | 31 |
| 753 | 73 | Gilligan | Arm | 284 | 182 | 4 |
| 754 | 73 | Gilligan | Arm | 284 | 182 | 16 |
| 755 | 73 | Gilligan | Arm | 284 | 182 | 18 |
| 756 | 73 | Gilligan | Arm | 284 | 182 | 22 |
| 757 | 73 | Gilligan | Arm | 284 | 182 | 22 |
| 758 | 73 | Gilligan | Arm | 284 | 182 | 26 |
| 759 | 73 | Gilligan | Arm | 284 | 182 | 30 |
| 760 | 74 | Gilligan | Trunk | 264 | 202 | 3 |
| 762 | 74 | Gilligan | Trunk | 264 | 202 | 7 |
| 763 | 74 | Gilligan | Trunk | 264 | 202 | 9 |
| 764 | 74 | Gilligan | Trunk | 264 | 202 | 11 |

| UnitID | SiteID | Monkey | Somatotopy | X | Y | Channel |
|---|---|---|---|---|---|---|
| 765 | 74 | Gilligan | Trunk | 264 | 202 | 11 |
| 766 | 74 | Gilligan | Trunk | 264 | 202 | 15 |
| 767 | 74 | Gilligan | Trunk | 264 | 202 | 17 |
| 768 | 74 | Gilligan | Trunk | 264 | 202 | 19 |
| 769 | 74 | Gilligan | Trunk | 264 | 202 | 21 |
| 770 | 74 | Gilligan | Trunk | 264 | 202 | 23 |
| 771 | 74 | Gilligan | Trunk | 264 | 202 | 25 |
| 772 | 74 | Gilligan | Trunk | 264 | 202 | 27 |
| 773 | 74 | Gilligan | Trunk | 264 | 202 | 27 |
| 774 | 74 | Gilligan | Trunk | 264 | 202 | 31 |
| 775 | 74 | Gilligan | Trunk | 264 | 202 | 4 |
| 776 | 74 | Gilligan | Trunk | 264 | 202 | 6 |
| 777 | 74 | Gilligan | Trunk | 264 | 202 | 10 |
| 778 | 74 | Gilligan | Trunk | 264 | 202 | 14 |
| 779 | 74 | Gilligan | Trunk | 264 | 202 | 18 |
| 780 | 74 | Gilligan | Trunk | 264 | 202 | 20 |
| 781 | 74 | Gilligan | Trunk | 264 | 202 | 24 |
| 782 | 74 | Gilligan | Trunk | 264 | 202 | 28 |
| 783 | 74 | Gilligan | Trunk | 264 | 202 | 30 |
| 784 | 74 | Gilligan | Trunk | 264 | 202 | 32 |
| 785 | 75 | Gilligan | Arm | 131 | 198 | 1 |
| 786 | 75 | Gilligan | Arm | 131 | 198 | 5 |
| 788 | 75 | Gilligan | Arm | 131 | 198 | 12 |
| 789 | 75 | Gilligan | Arm | 131 | 198 | 14 |
| 790 | 75 | Gilligan | Arm | 131 | 198 | 16 |
| 791 | 75 | Gilligan | Arm | 131 | 198 | 24 |
| 792 | 75 | Gilligan | Arm | 131 | 198 | 26 |
| 793 | 75 | Gilligan | Arm | 131 | 198 | 28 |
| 794 | 76 | Gilligan | Trunk | 251 | 444 | 9 |
| 796 | 76 | Gilligan | Trunk | 251 | 444 | 16 |
| 797 | 77 | Gilligan | Hand | 123 | 400 | 3 |
| 798 | 77 | Gilligan | Hand | 123 | 400 | 5 |

| UnitID | SiteID | Monkey | Somatotopy | X | Y | Channel |
|---|---|---|---|---|---|---|
| 799 | 77 | Gilligan | Hand | 123 | 400 | 19 |
| 800 | 77 | Gilligan | Hand | 123 | 400 | 29 |
| 801 | 77 | Gilligan | Hand | 123 | 400 | 4 |
| 802 | 77 | Gilligan | Hand | 123 | 400 | 8 |
| 803 | 77 | Gilligan | Hand | 123 | 400 | 12 |
| 804 | 77 | Gilligan | Hand | 123 | 400 | 14 |
| 805 | 77 | Gilligan | Hand | 123 | 400 | 22 |
| 806 | 77 | Gilligan | Hand | 123 | 400 | 26 |
| 807 | 78 | Gilligan | Hand | 184 | 366 | 1 |
| 808 | 78 | Gilligan | Hand | 184 | 366 | 11 |
| 809 | 78 | Gilligan | Hand | 184 | 366 | 27 |
| 810 | 78 | Gilligan | Hand | 184 | 366 | 31 |
| 811 | 78 | Gilligan | Hand | 184 | 366 | 10 |
| 812 | 78 | Gilligan | Hand | 184 | 366 | 14 |
| 813 | 78 | Gilligan | Hand | 184 | 366 | 16 |
| 814 | 78 | Gilligan | Hand | 184 | 366 | 20 |
| 815 | 78 | Gilligan | Hand | 184 | 366 | 22 |
| 816 | 78 | Gilligan | Hand | 184 | 366 | 26 |
| 817 | 78 | Gilligan | Hand | 184 | 366 | 28 |
| 818 | 78 | Gilligan | Hand | 184 | 366 | 32 |
| 819 | 79 | Gilligan | Trunk | 185 | 545 | 1 |
| 820 | 79 | Gilligan | Trunk | 185 | 545 | 3 |
| 821 | 79 | Gilligan | Trunk | 185 | 545 | 5 |
| 823 | 79 | Gilligan | Trunk | 185 | 545 | 7 |
| 824 | 79 | Gilligan | Trunk | 185 | 545 | 9 |
| 825 | 79 | Gilligan | Trunk | 185 | 545 | 11 |
| 826 | 79 | Gilligan | Trunk | 185 | 545 | 13 |
| 827 | 79 | Gilligan | Trunk | 185 | 545 | 13 |
| 828 | 79 | Gilligan | Trunk | 185 | 545 | 15 |
| 829 | 79 | Gilligan | Trunk | 185 | 545 | 17 |
| 830 | 79 | Gilligan | Trunk | 185 | 545 | 19 |
| 831 | 79 | Gilligan | Trunk | 185 | 545 | 21 |

| UnitID | SiteID | Monkey | Somatotopy | X | Y | Channel |
|---|---|---|---|---|---|---|
| 832 | 79 | Gilligan | Trunk | 185 | 545 | 23 |
| 834 | 79 | Gilligan | Trunk | 185 | 545 | 25 |
| 835 | 79 | Gilligan | Trunk | 185 | 545 | 29 |
| 836 | 79 | Gilligan | Trunk | 185 | 545 | 31 |
| 838 | 79 | Gilligan | Trunk | 185 | 545 | 4 |
| 839 | 79 | Gilligan | Trunk | 185 | 545 | 6 |
| 840 | 79 | Gilligan | Trunk | 185 | 545 | 6 |
| 841 | 79 | Gilligan | Trunk | 185 | 545 | 8 |
| 842 | 79 | Gilligan | Trunk | 185 | 545 | 12 |
| 843 | 79 | Gilligan | Trunk | 185 | 545 | 14 |
| 844 | 79 | Gilligan | Trunk | 185 | 545 | 14 |
| 845 | 79 | Gilligan | Trunk | 185 | 545 | 16 |
| 847 | 79 | Gilligan | Trunk | 185 | 545 | 20 |
| 848 | 79 | Gilligan | Trunk | 185 | 545 | 20 |
| 850 | 79 | Gilligan | Trunk | 185 | 545 | 24 |
| 851 | 79 | Gilligan | Trunk | 185 | 545 | 26 |
| 852 | 79 | Gilligan | Trunk | 185 | 545 | 28 |
| 853 | 79 | Gilligan | Trunk | 185 | 545 | 30 |
| 854 | 79 | Gilligan | Trunk | 185 | 545 | 32 |
| 855 | 80 | Gilligan | Hand | 215 | 349 | 32 |
| 856 | 81 | Gilligan | Arm | 159 | 281 | 7 |
| 857 | 81 | Gilligan | Arm | 159 | 281 | 19 |
| 858 | 81 | Gilligan | Arm | 159 | 281 | 21 |
| 859 | 81 | Gilligan | Arm | 159 | 281 | 4 |
| 860 | 81 | Gilligan | Arm | 159 | 281 | 32 |
| 862 | 82 | Gilligan | Arm | 153 | 87 | 7 |
| 863 | 82 | Gilligan | Arm | 153 | 87 | 19 |
| 864 | 82 | Gilligan | Arm | 153 | 87 | 4 |
| 866 | 82 | Gilligan | Arm | 153 | 87 | 22 |
| 867 | 82 | Gilligan | Arm | 153 | 87 | 26 |
| 868 | 82 | Gilligan | Arm | 153 | 87 | 28 |
| 869 | 82 | Gilligan | Arm | 153 | 87 | 30 |

| UnitID | SiteID | Monkey | Somatotopy | X | Y | Channel |
|---|---|---|---|---|---|---|
| 870 | 82 | Gilligan | Arm | 153 | 87 | 32 |
| 871 | 82 | Gilligan | Arm | 153 | 87 | 32 |
| 872 | 83 | Gilligan | Trunk | 153 | 598 | 3 |
| 873 | 83 | Gilligan | Trunk | 153 | 598 | 15 |
| 874 | 83 | Gilligan | Trunk | 153 | 598 | 15 |
| 875 | 83 | Gilligan | Trunk | 153 | 598 | 19 |
| 877 | 83 | Gilligan | Trunk | 153 | 598 | 27 |
| 878 | 83 | Gilligan | Trunk | 153 | 598 | 27 |
| 879 | 83 | Gilligan | Trunk | 153 | 598 | 29 |
| 880 | 83 | Gilligan | Trunk | 153 | 598 | 31 |
| 881 | 83 | Gilligan | Trunk | 153 | 598 | 31 |
| 882 | 83 | Gilligan | Trunk | 153 | 598 | 4 |
| 883 | 83 | Gilligan | Trunk | 153 | 598 | 6 |
| 884 | 83 | Gilligan | Trunk | 153 | 598 | 8 |
| 885 | 83 | Gilligan | Trunk | 153 | 598 | 10 |
| 886 | 83 | Gilligan | Trunk | 153 | 598 | 16 |
| 887 | 83 | Gilligan | Trunk | 153 | 598 | 24 |
| 888 | 83 | Gilligan | Trunk | 153 | 598 | 30 |
| 889 | 84 | Gilligan | Arm | 223 | 128 | 9 |
| 890 | 84 | Gilligan | Arm | 223 | 128 | 19 |
| 891 | 84 | Gilligan | Arm | 223 | 128 | 21 |
| 892 | 84 | Gilligan | Arm | 223 | 128 | 12 |
| 893 | 84 | Gilligan | Arm | 223 | 128 | 26 |
| 894 | 84 | Gilligan | Arm | 223 | 128 | 28 |
| 895 | 84 | Gilligan | Arm | 223 | 128 | 30 |
| 896 | 85 | Skipper | Arm | 115 | 537 | 29 |
| 897 | 85 | Skipper | Arm | 115 | 537 | 28 |
| 898 | 85 | Skipper | Arm | 115 | 537 | 27 |
| 899 | 85 | Skipper | Arm | 115 | 537 | 26 |
| 900 | 85 | Skipper | Arm | 115 | 537 | 24 |
| 901 | 85 | Skipper | Arm | 115 | 537 | 18 |
| 902 | 85 | Skipper | Arm | 115 | 537 | 17 |

| UnitID | SiteID | Monkey | Somatotopy | X | Y | Channel |
|---|---|---|---|---|---|---|
| 903 | 85 | Skipper | Arm | 115 | 537 | 14 |
| 904 | 85 | Skipper | Arm | 115 | 537 | 13 |
| 905 | 86 | Skipper | Hand | 142 | 390 | 30 |
| 906 | 86 | Skipper | Hand | 142 | 390 | 30 |
| 907 | 86 | Skipper | Hand | 142 | 390 | 30 |
| 908 | 86 | Skipper | Hand | 142 | 390 | 22 |
| 909 | 86 | Skipper | Hand | 142 | 390 | 20 |
| 910 | 86 | Skipper | Hand | 142 | 390 | 14 |
| 911 | 86 | Skipper | Hand | 142 | 390 | 13 |
| 912 | 86 | Skipper | Hand | 142 | 390 | 11 |
| 913 | 86 | Skipper | Hand | 142 | 390 | 10 |
| 914 | 87 | Skipper | Arm | 246 | 148 | 31 |
| 915 | 87 | Skipper | Arm | 246 | 148 | 25 |
| 916 | 87 | Skipper | Arm | 246 | 148 | 22 |
| 917 | 87 | Skipper | Arm | 246 | 148 | 7 |
| 918 | 87 | Skipper | Arm | 246 | 148 | 6 |
| 919 | 87 | Skipper | Arm | 246 | 148 | 5 |
| 920 | 88 | Skipper | Trunk | 309 | 136 | 32 |
| 921 | 88 | Skipper | Trunk | 309 | 136 | 32 |
| 922 | 88 | Skipper | Trunk | 309 | 136 | 31 |
| 923 | 88 | Skipper | Trunk | 309 | 136 | 29 |
| 924 | 88 | Skipper | Trunk | 309 | 136 | 27 |
| 925 | 88 | Skipper | Trunk | 309 | 136 | 25 |
| 926 | 88 | Skipper | Trunk | 309 | 136 | 24 |
| 927 | 88 | Skipper | Trunk | 309 | 136 | 23 |
| 928 | 88 | Skipper | Trunk | 309 | 136 | 22 |
| 929 | 88 | Skipper | Trunk | 309 | 136 | 21 |
| 930 | 88 | Skipper | Trunk | 309 | 136 | 11 |
| 931 | 88 | Skipper | Trunk | 309 | 136 | 11 |
| 932 | 88 | Skipper | Trunk | 309 | 136 | 10 |
| 933 | 88 | Skipper | Trunk | 309 | 136 | 10 |
| 934 | 89 | Skipper | Hand | 142 | 290 | 29 |

| UnitID | SiteID | Monkey | Somatotopy | X | Y | Channel |
|---|---|---|---|---|---|---|
| 935 | 89 | Skipper | Hand | 142 | 290 | 7 |
| 936 | 89 | Skipper | Hand | 142 | 290 | 7 |
| 937 | 90 | Skipper | Arm | 71 | 11 | 21 |
| 938 | 90 | Skipper | Arm | 71 | 11 | 20 |
| 940 | 90 | Skipper | Arm | 71 | 11 | 17 |
| 941 | 90 | Skipper | Arm | 71 | 11 | 12 |
| 942 | 90 | Skipper | Arm | 71 | 11 | 12 |
| 943 | 90 | Skipper | Arm | 71 | 11 | 11 |
| 944 | 90 | Skipper | Arm | 71 | 11 | 6 |
| 945 | 90 | Skipper | Arm | 71 | 11 | 5 |
| 946 | 90 | Skipper | Arm | 71 | 11 | 2 |
| 947 | 91 | Skipper | Hand | 219 | 308 | 30 |
| 948 | 91 | Skipper | Hand | 219 | 308 | 25 |
| 950 | 91 | Skipper | Hand | 219 | 308 | 22 |
| 951 | 91 | Skipper | Hand | 219 | 308 | 21 |
| 952 | 91 | Skipper | Hand | 219 | 308 | 15 |
| 953 | 91 | Skipper | Hand | 219 | 308 | 14 |
| 954 | 91 | Skipper | Hand | 219 | 308 | 13 |
| 955 | 91 | Skipper | Hand | 219 | 308 | 12 |
| 956 | 91 | Skipper | Hand | 219 | 308 | 11 |
| 957 | 91 | Skipper | Hand | 219 | 308 | 10 |
| 958 | 91 | Skipper | Hand | 219 | 308 | 2 |
| 959 | 91 | Skipper | Hand | 219 | 308 | 1 |
| 960 | 92 | Skipper | Arm | 265 | 205 | 25 |
| 961 | 92 | Skipper | Arm | 265 | 205 | 8 |
| 962 | 92 | Skipper | Arm | 265 | 205 | 8 |
| 963 | 92 | Skipper | Arm | 265 | 205 | 2 |
| 964 | 93 | Skipper | Arm | 264 | 468 | 31 |
| 965 | 93 | Skipper | Arm | 264 | 468 | 30 |
| 966 | 93 | Skipper | Arm | 264 | 468 | 30 |
| 967 | 93 | Skipper | Arm | 264 | 468 | 28 |
| 969 | 93 | Skipper | Arm | 264 | 468 | 24 |

| UnitID | SiteID | Monkey | Somatotopy | X | Y | Channel |
|---|---|---|---|---|---|---|
| 970 | 93 | Skipper | Arm | 264 | 468 | 15 |
| 971 | 93 | Skipper | Arm | 264 | 468 | 7 |
| 972 | 93 | Skipper | Arm | 264 | 468 | 6 |
| 973 | 94 | Skipper | Arm | 191 | 53 | 27 |
| 974 | 94 | Skipper | Arm | 191 | 53 | 25 |
| 975 | 94 | Skipper | Arm | 191 | 53 | 17 |
| 976 | 94 | Skipper | Arm | 191 | 53 | 15 |
| 977 | 94 | Skipper | Arm | 191 | 53 | 10 |
| 978 | 95 | Skipper | Trunk | 330 | 227 | 26 |
| 979 | 95 | Skipper | Trunk | 330 | 227 | 25 |
| 980 | 95 | Skipper | Trunk | 330 | 227 | 10 |
| 981 | 96 | Skipper | Arm | 316 | 426 | 26 |
| 982 | 96 | Skipper | Arm | 316 | 426 | 21 |
| 983 | 96 | Skipper | Arm | 316 | 426 | 10 |
| 984 | 97 | Skipper | Arm | 20 | 99 | 30 |
| 985 | 97 | Skipper | Arm | 20 | 99 | 24 |
| 986 | 97 | Skipper | Arm | 20 | 99 | 16 |
| 987 | 97 | Skipper | Arm | 20 | 99 | 10 |
| 988 | 97 | Skipper | Arm | 20 | 99 | 9 |
| 989 | 97 | Skipper | Arm | 20 | 99 | 6 |
| 990 | 98 | Skipper | Trunk | 280 | -10 | 31 |
| 991 | 98 | Skipper | Trunk | 280 | -10 | 30 |
| 992 | 98 | Skipper | Trunk | 280 | -10 | 29 |
| 993 | 98 | Skipper | Trunk | 280 | -10 | 24 |
| 994 | 98 | Skipper | Trunk | 280 | -10 | 22 |
| 995 | 98 | Skipper | Trunk | 280 | -10 | 17 |
| 996 | 98 | Skipper | Trunk | 280 | -10 | 16 |
| 997 | 98 | Skipper | Trunk | 280 | -10 | 14 |
| 998 | 98 | Skipper | Trunk | 280 | -10 | 13 |
| 999 | 98 | Skipper | Trunk | 280 | -10 | 12 |
| 1000 | 98 | Skipper | Trunk | 280 | -10 | 5 |
| 1001 | 99 | Skipper | Trunk | 75 | -49 | 31 |

| UnitID | SiteID | Monkey | Somatotopy | X | Y | Channel |
|---|---|---|---|---|---|---|
| 1002 | 99 | Skipper | Trunk | 75 | -49 | 29 |
| 1003 | 99 | Skipper | Trunk | 75 | -49 | 28 |
| 1004 | 99 | Skipper | Trunk | 75 | -49 | 27 |
| 1005 | 99 | Skipper | Trunk | 75 | -49 | 25 |
| 1006 | 99 | Skipper | Trunk | 75 | -49 | 24 |
| 1007 | 99 | Skipper | Trunk | 75 | -49 | 23 |
| 1008 | 99 | Skipper | Trunk | 75 | -49 | 22 |
| 1009 | 99 | Skipper | Trunk | 75 | -49 | 21 |
| 1010 | 99 | Skipper | Trunk | 75 | -49 | 20 |
| 1011 | 99 | Skipper | Trunk | 75 | -49 | 19 |
| 1012 | 99 | Skipper | Trunk | 75 | -49 | 12 |
| 1013 | 99 | Skipper | Trunk | 75 | -49 | 11 |
| 1014 | 99 | Skipper | Trunk | 75 | -49 | 6 |
| 1015 | 99 | Skipper | Trunk | 75 | -49 | 5 |
| 1016 | 99 | Skipper | Trunk | 75 | -49 | 4 |
| 1017 | 99 | Skipper | Trunk | 75 | -49 | 3 |
| 1018 | 99 | Skipper | Trunk | 75 | -49 | 2 |
| 1019 | 100 | Skipper | Arm | 204 | 141 | 31 |
| 1020 | 100 | Skipper | Arm | 204 | 141 | 30 |
| 1021 | 100 | Skipper | Arm | 204 | 141 | 27 |
| 1022 | 100 | Skipper | Arm | 204 | 141 | 26 |
| 1023 | 100 | Skipper | Arm | 204 | 141 | 24 |
| 1024 | 100 | Skipper | Arm | 204 | 141 | 19 |
| 1025 | 100 | Skipper | Arm | 204 | 141 | 18 |
| 1026 | 100 | Skipper | Arm | 204 | 141 | 14 |
| 1027 | 100 | Skipper | Arm | 204 | 141 | 12 |
| 1028 | 100 | Skipper | Arm | 204 | 141 | 11 |
| 1029 | 100 | Skipper | Arm | 204 | 141 | 7 |
| 1030 | 100 | Skipper | Arm | 204 | 141 | 6 |
| 1031 | 100 | Skipper | Arm | 204 | 141 | 5 |
| 1032 | 100 | Skipper | Arm | 204 | 141 | 3 |
| 1033 | 101 | Skipper | Hand | 152 | 111 | 28 |

| UnitID | SiteID | Monkey | Somatotopy | X | Y | Channel |
|---|---|---|---|---|---|---|
| 1034 | 101 | Skipper | Hand | 152 | 111 | 27 |
| 1035 | 101 | Skipper | Hand | 152 | 111 | 26 |
| 1036 | 101 | Skipper | Hand | 152 | 111 | 21 |
| 1037 | 101 | Skipper | Hand | 152 | 111 | 19 |
| 1038 | 101 | Skipper | Hand | 152 | 111 | 19 |
| 1039 | 101 | Skipper | Hand | 152 | 111 | 17 |
| 1040 | 101 | Skipper | Hand | 152 | 111 | 16 |
| 1041 | 101 | Skipper | Hand | 152 | 111 | 14 |
| 1042 | 101 | Skipper | Hand | 152 | 111 | 14 |
| 1043 | 101 | Skipper | Hand | 152 | 111 | 7 |
| 1044 | 101 | Skipper | Hand | 152 | 111 | 4 |
| 1045 | 101 | Skipper | Hand | 152 | 111 | 3 |
| 1046 | 101 | Skipper | Hand | 152 | 111 | 2 |
| 1047 | 101 | Skipper | Hand | 152 | 111 | 2 |
| 1048 | 101 | Skipper | Hand | 152 | 111 | 1 |
| 1049 | 102 | Skipper | Arm | 188 | 215 | 28 |
| 1050 | 102 | Skipper | Arm | 188 | 215 | 27 |
| 1051 | 102 | Skipper | Arm | 188 | 215 | 26 |
| 1052 | 102 | Skipper | Arm | 188 | 215 | 26 |
| 1053 | 102 | Skipper | Arm | 188 | 215 | 25 |
| 1054 | 102 | Skipper | Arm | 188 | 215 | 24 |
| 1055 | 102 | Skipper | Arm | 188 | 215 | 24 |
| 1056 | 102 | Skipper | Arm | 188 | 215 | 23 |
| 1057 | 102 | Skipper | Arm | 188 | 215 | 23 |
| 1058 | 102 | Skipper | Arm | 188 | 215 | 22 |
| 1059 | 102 | Skipper | Arm | 188 | 215 | 14 |
| 1060 | 102 | Skipper | Arm | 188 | 215 | 13 |
| 1061 | 102 | Skipper | Arm | 188 | 215 | 11 |
| 1062 | 102 | Skipper | Arm | 188 | 215 | 10 |
| 1063 | 102 | Skipper | Arm | 188 | 215 | 8 |
| 1064 | 102 | Skipper | Arm | 188 | 215 | 7 |
| 1065 | 103 | Skipper | Hand | 272 | 245 | 31 |

| UnitID | SiteID | Monkey | Somatotopy | X | Y | Channel |
|---|---|---|---|---|---|---|
| 1066 | 103 | Skipper | Hand | 272 | 245 | 30 |
| 1067 | 103 | Skipper | Hand | 272 | 245 | 29 |
| 1068 | 103 | Skipper | Hand | 272 | 245 | 28 |
| 1069 | 103 | Skipper | Hand | 272 | 245 | 26 |
| 1070 | 103 | Skipper | Hand | 272 | 245 | 25 |
| 1071 | 103 | Skipper | Hand | 272 | 245 | 20 |
| 1072 | 103 | Skipper | Hand | 272 | 245 | 19 |
| 1073 | 103 | Skipper | Hand | 272 | 245 | 18 |
| 1074 | 103 | Skipper | Hand | 272 | 245 | 17 |
| 1075 | 103 | Skipper | Hand | 272 | 245 | 16 |
| 1076 | 103 | Skipper | Hand | 272 | 245 | 15 |
| 1077 | 103 | Skipper | Hand | 272 | 245 | 15 |
| 1078 | 103 | Skipper | Hand | 272 | 245 | 14 |
| 1079 | 103 | Skipper | Hand | 272 | 245 | 14 |
| 1080 | 103 | Skipper | Hand | 272 | 245 | 8 |
| 1081 | 103 | Skipper | Hand | 272 | 245 | 7 |
| 1082 | 103 | Skipper | Hand | 272 | 245 | 6 |
| 1083 | 103 | Skipper | Hand | 272 | 245 | 5 |
| 1084 | 103 | Skipper | Hand | 272 | 245 | 4 |
| 1085 | 103 | Skipper | Hand | 272 | 245 | 4 |
| 1086 | 103 | Skipper | Hand | 272 | 245 | 3 |
| 1087 | 103 | Skipper | Hand | 272 | 245 | 2 |
| 1088 | 104 | Skipper | Hand | 178 | 395 | 30 |
| 1089 | 104 | Skipper | Hand | 178 | 395 | 29 |
| 1090 | 104 | Skipper | Hand | 178 | 395 | 28 |
| 1091 | 104 | Skipper | Hand | 178 | 395 | 27 |
| 1092 | 104 | Skipper | Hand | 178 | 395 | 24 |
| 1093 | 104 | Skipper | Hand | 178 | 395 | 23 |
| 1094 | 104 | Skipper | Hand | 178 | 395 | 22 |
| 1095 | 104 | Skipper | Hand | 178 | 395 | 19 |
| 1096 | 104 | Skipper | Hand | 178 | 395 | 16 |
| 1097 | 104 | Skipper | Hand | 178 | 395 | 16 |

| UnitID | SiteID | Monkey | Somatotopy | X | Y | Channel |
|---|---|---|---|---|---|---|
| 1098 | 104 | Skipper | Hand | 178 | 395 | 13 |
| 1099 | 104 | Skipper | Hand | 178 | 395 | 9 |
| 1100 | 104 | Skipper | Hand | 178 | 395 | 9 |
| 1101 | 104 | Skipper | Hand | 178 | 395 | 8 |
| 1102 | 104 | Skipper | Hand | 178 | 395 | 5 |
| 1103 | 104 | Skipper | Hand | 178 | 395 | 5 |
| 1104 | 104 | Skipper | Hand | 178 | 395 | 4 |
| 1105 | 105 | Skipper | Hand | 211 | 412 | 31 |
| 1106 | 105 | Skipper | Hand | 211 | 412 | 30 |
| 1108 | 105 | Skipper | Hand | 211 | 412 | 28 |
| 1109 | 105 | Skipper | Hand | 211 | 412 | 25 |
| 1110 | 105 | Skipper | Hand | 211 | 412 | 24 |
| 1111 | 105 | Skipper | Hand | 211 | 412 | 22 |
| 1112 | 105 | Skipper | Hand | 211 | 412 | 21 |
| 1113 | 105 | Skipper | Hand | 211 | 412 | 16 |
| 1114 | 105 | Skipper | Hand | 211 | 412 | 11 |
| 1115 | 105 | Skipper | Hand | 211 | 412 | 10 |
| 1116 | 105 | Skipper | Hand | 211 | 412 | 6 |
| 1117 | 105 | Skipper | Hand | 211 | 412 | 5 |
| 1118 | 106 | Skipper | Arm | 118 | 366 | 27 |
| 1119 | 106 | Skipper | Arm | 118 | 366 | 26 |
| 1120 | 106 | Skipper | Arm | 118 | 366 | 24 |
| 1121 | 106 | Skipper | Arm | 118 | 366 | 23 |
| 1122 | 106 | Skipper | Arm | 118 | 366 | 22 |
| 1123 | 106 | Skipper | Arm | 118 | 366 | 21 |
| 1124 | 106 | Skipper | Arm | 118 | 366 | 13 |
| 1125 | 106 | Skipper | Arm | 118 | 366 | 12 |
| 1126 | 106 | Skipper | Arm | 118 | 366 | 11 |
| 1127 | 106 | Skipper | Arm | 118 | 366 | 10 |
| 1128 | 106 | Skipper | Arm | 118 | 366 | 1 |
| 1129 | 107 | Skipper | Hand | 203 | 468 | 31 |
| 1130 | 107 | Skipper | Hand | 203 | 468 | 30 |

| UnitID | SiteID | Monkey | Somatotopy | X | Y | Channel |
|---|---|---|---|---|---|---|
| 1131 | 107 | Skipper | Hand | 203 | 468 | 29 |
| 1132 | 107 | Skipper | Hand | 203 | 468 | 29 |
| 1133 | 107 | Skipper | Hand | 203 | 468 | 28 |
| 1134 | 107 | Skipper | Hand | 203 | 468 | 27 |
| 1135 | 107 | Skipper | Hand | 203 | 468 | 26 |
| 1137 | 107 | Skipper | Hand | 203 | 468 | 17 |
| 1138 | 107 | Skipper | Hand | 203 | 468 | 12 |
| 1139 | 107 | Skipper | Hand | 203 | 468 | 11 |
| 1140 | 107 | Skipper | Hand | 203 | 468 | 10 |
| 1141 | 107 | Skipper | Hand | 203 | 468 | 6 |
| 1142 | 107 | Skipper | Hand | 203 | 468 | 5 |
| 1144 | 107 | Skipper | Hand | 203 | 468 | 2 |
| 1145 | 107 | Skipper | Hand | 203 | 468 | 2 |
| 1146 | 108 | Skipper | Trunk | 289 | 168 | 29 |
| 1147 | 108 | Skipper | Trunk | 289 | 168 | 29 |
| 1148 | 108 | Skipper | Trunk | 289 | 168 | 28 |
| 1149 | 108 | Skipper | Trunk | 289 | 168 | 27 |
| 1150 | 108 | Skipper | Trunk | 289 | 168 | 26 |
| 1151 | 108 | Skipper | Trunk | 289 | 168 | 26 |
| 1153 | 108 | Skipper | Trunk | 289 | 168 | 22 |
| 1154 | 108 | Skipper | Trunk | 289 | 168 | 20 |
| 1155 | 108 | Skipper | Trunk | 289 | 168 | 14 |
| 1156 | 108 | Skipper | Trunk | 289 | 168 | 13 |
| 1157 | 108 | Skipper | Trunk | 289 | 168 | 13 |
| 1158 | 108 | Skipper | Trunk | 289 | 168 | 11 |
| 1159 | 108 | Skipper | Trunk | 289 | 168 | 8 |
| 1160 | 108 | Skipper | Trunk | 289 | 168 | 7 |
| 1161 | 108 | Skipper | Trunk | 289 | 168 | 1 |
| 1162 | 109 | Skipper | Hand | 268 | 226 | 30 |
| 1163 | 109 | Skipper | Hand | 268 | 226 | 29 |
| 1164 | 109 | Skipper | Hand | 268 | 226 | 27 |
| 1165 | 109 | Skipper | Hand | 268 | 226 | 26 |

| UnitID | SiteID | Monkey | Somatotopy | X | Y | Channel |
|---|---|---|---|---|---|---|
| 1166 | 109 | Skipper | Hand | 268 | 226 | 15 |
| 1167 | 109 | Skipper | Hand | 268 | 226 | 15 |
| 1168 | 109 | Skipper | Hand | 268 | 226 | 8 |
| 1169 | 109 | Skipper | Hand | 268 | 226 | 8 |
| 1170 | 109 | Skipper | Hand | 268 | 226 | 5 |
| 1173 | 109 | Skipper | Hand | 268 | 226 | 4 |
| 1174 | 109 | Skipper | Hand | 268 | 226 | 3 |
| 1175 | 109 | Skipper | Hand | 268 | 226 | 3 |
| 1176 | 109 | Skipper | Hand | 268 | 226 | 2 |
| 1177 | 109 | Skipper | Hand | 268 | 226 | 2 |
| 1178 | 110 | Skipper | Trunk | 76 | 179 | 31 |
| 1179 | 110 | Skipper | Trunk | 76 | 179 | 30 |
| 1180 | 110 | Skipper | Trunk | 76 | 179 | 29 |
| 1181 | 110 | Skipper | Trunk | 76 | 179 | 28 |
| 1182 | 110 | Skipper | Trunk | 76 | 179 | 28 |
| 1183 | 110 | Skipper | Trunk | 76 | 179 | 19 |
| 1184 | 110 | Skipper | Trunk | 76 | 179 | 18 |
| 1185 | 110 | Skipper | Trunk | 76 | 179 | 18 |
| 1186 | 110 | Skipper | Trunk | 76 | 179 | 17 |
| 1187 | 110 | Skipper | Trunk | 76 | 179 | 17 |
| 1188 | 110 | Skipper | Trunk | 76 | 179 | 16 |
| 1189 | 110 | Skipper | Trunk | 76 | 179 | 15 |
| 1190 | 110 | Skipper | Trunk | 76 | 179 | 15 |
| 1191 | 110 | Skipper | Trunk | 76 | 179 | 15 |
| 1192 | 110 | Skipper | Trunk | 76 | 179 | 14 |
| 1193 | 110 | Skipper | Trunk | 76 | 179 | 13 |
| 1194 | 110 | Skipper | Trunk | 76 | 179 | 7 |
| 1195 | 110 | Skipper | Trunk | 76 | 179 | 7 |
| 1196 | 110 | Skipper | Trunk | 76 | 179 | 6 |
| 1197 | 110 | Skipper | Trunk | 76 | 179 | 6 |
| 1198 | 110 | Skipper | Trunk | 76 | 179 | 5 |
| 1199 | 110 | Skipper | Trunk | 76 | 179 | 5 |

| UnitID | SiteID | Monkey | Somatotopy | X | Y | Channel |
|---|---|---|---|---|---|---|
| 1200 | 110 | Skipper | Trunk | 76 | 179 | 4 |
| 1201 | 111 | Skipper | Arm | 108 | 107 | 23 |
| 1202 | 111 | Skipper | Arm | 108 | 107 | 21 |
| 1203 | 111 | Skipper | Arm | 108 | 107 | 17 |
| 1204 | 111 | Skipper | Arm | 108 | 107 | 14 |
| 1205 | 111 | Skipper | Arm | 108 | 107 | 6 |
| 1206 | 111 | Skipper | Arm | 108 | 107 | 6 |
| 1207 | 111 | Skipper | Arm | 108 | 107 | 4 |
| 1208 | 111 | Skipper | Arm | 108 | 107 | 3 |
| 1209 | 111 | Skipper | Arm | 108 | 107 | 2 |
| 1210 | 112 | Skipper | Trunk | 158 | 241 | 31 |
| 1211 | 112 | Skipper | Trunk | 158 | 241 | 30 |
| 1212 | 112 | Skipper | Trunk | 158 | 241 | 30 |
| 1213 | 112 | Skipper | Trunk | 158 | 241 | 29 |
| 1214 | 112 | Skipper | Trunk | 158 | 241 | 28 |
| 1215 | 112 | Skipper | Trunk | 158 | 241 | 27 |
| 1216 | 112 | Skipper | Trunk | 158 | 241 | 13 |
| 1217 | 112 | Skipper | Trunk | 158 | 241 | 12 |
| 1218 | 112 | Skipper | Trunk | 158 | 241 | 11 |
| 1219 | 112 | Skipper | Trunk | 158 | 241 | 10 |
| 1220 | 112 | Skipper | Trunk | 158 | 241 | 9 |
| 1221 | 112 | Skipper | Trunk | 158 | 241 | 8 |
| 1222 | 113 | Skipper | Hand | 183 | 261 | 32 |
| 1223 | 113 | Skipper | Hand | 183 | 261 | 32 |
| 1224 | 113 | Skipper | Hand | 183 | 261 | 32 |
| 1225 | 113 | Skipper | Hand | 183 | 261 | 27 |
| 1226 | 113 | Skipper | Hand | 183 | 261 | 26 |
| 1227 | 113 | Skipper | Hand | 183 | 261 | 25 |
| 1228 | 113 | Skipper | Hand | 183 | 261 | 25 |
| 1229 | 113 | Skipper | Hand | 183 | 261 | 24 |
| 1230 | 113 | Skipper | Hand | 183 | 261 | 24 |
| 1231 | 113 | Skipper | Hand | 183 | 261 | 23 |

| UnitID | SiteID | Monkey | Somatotopy | X | Y | Channel |
|---|---|---|---|---|---|---|
| 1232 | 113 | Skipper | Hand | 183 | 261 | 22 |
| 1233 | 113 | Skipper | Hand | 183 | 261 | 22 |
| 1234 | 113 | Skipper | Hand | 183 | 261 | 21 |
| 1235 | 113 | Skipper | Hand | 183 | 261 | 20 |
| 1236 | 113 | Skipper | Hand | 183 | 261 | 19 |
| 1238 | 113 | Skipper | Hand | 183 | 261 | 5 |
| 1239 | 113 | Skipper | Hand | 183 | 261 | 4 |
| 1240 | 113 | Skipper | Hand | 183 | 261 | 3 |
| 1241 | 114 | Skipper | Arm | 226 | 182 | 31 |
| 1242 | 114 | Skipper | Arm | 226 | 182 | 31 |
| 1243 | 114 | Skipper | Arm | 226 | 182 | 30 |
| 1244 | 114 | Skipper | Arm | 226 | 182 | 29 |
| 1245 | 114 | Skipper | Arm | 226 | 182 | 29 |
| 1246 | 114 | Skipper | Arm | 226 | 182 | 29 |
| 1247 | 114 | Skipper | Arm | 226 | 182 | 28 |
| 1248 | 114 | Skipper | Arm | 226 | 182 | 27 |
| 1249 | 114 | Skipper | Arm | 226 | 182 | 26 |
| 1250 | 114 | Skipper | Arm | 226 | 182 | 25 |
| 1251 | 114 | Skipper | Arm | 226 | 182 | 25 |
| 1252 | 114 | Skipper | Arm | 226 | 182 | 24 |
| 1253 | 114 | Skipper | Arm | 226 | 182 | 13 |
| 1254 | 114 | Skipper | Arm | 226 | 182 | 13 |
| 1255 | 114 | Skipper | Arm | 226 | 182 | 8 |
| 1256 | 114 | Skipper | Arm | 226 | 182 | 8 |
| 1257 | 114 | Skipper | Arm | 226 | 182 | 7 |
| 1258 | 114 | Skipper | Arm | 226 | 182 | 7 |
| 1259 | 114 | Skipper | Arm | 226 | 182 | 6 |
| 1260 | 114 | Skipper | Arm | 226 | 182 | 6 |
| 1261 | 114 | Skipper | Arm | 226 | 182 | 3 |
| 1262 | 115 | Skipper | Trunk | 328 | 189 | 30 |
| 1263 | 115 | Skipper | Trunk | 328 | 189 | 30 |
| 1264 | 115 | Skipper | Trunk | 328 | 189 | 25 |

| UnitID | SiteID | Monkey | Somatotopy | X | Y | Channel |
|---|---|---|---|---|---|---|
| 1265 | 115 | Skipper | Trunk | 328 | 189 | 24 |
| 1266 | 115 | Skipper | Trunk | 328 | 189 | 20 |
| 1267 | 115 | Skipper | Trunk | 328 | 189 | 19 |
| 1268 | 115 | Skipper | Trunk | 328 | 189 | 18 |
| 1269 | 115 | Skipper | Trunk | 328 | 189 | 17 |
| 1270 | 115 | Skipper | Trunk | 328 | 189 | 11 |
| 1271 | 115 | Skipper | Trunk | 328 | 189 | 10 |
| 1272 | 115 | Skipper | Trunk | 328 | 189 | 8 |
| 1273 | 115 | Skipper | Trunk | 328 | 189 | 6 |
| 1274 | 115 | Skipper | Trunk | 328 | 189 | 5 |
| 1275 | 115 | Skipper | Trunk | 328 | 189 | 5 |
| 1276 | 115 | Skipper | Trunk | 328 | 189 | 4 |
| 1277 | 116 | Skipper | Arm | 236 | 438 | 31 |
| 1279 | 116 | Skipper | Arm | 236 | 438 | 30 |
| 1280 | 116 | Skipper | Arm | 236 | 438 | 30 |
| 1281 | 116 | Skipper | Arm | 236 | 438 | 29 |
| 1282 | 116 | Skipper | Arm | 236 | 438 | 29 |
| 1283 | 116 | Skipper | Arm | 236 | 438 | 28 |
| 1284 | 116 | Skipper | Arm | 236 | 438 | 28 |
| 1285 | 116 | Skipper | Arm | 236 | 438 | 28 |
| 1286 | 116 | Skipper | Arm | 236 | 438 | 27 |
| 1287 | 116 | Skipper | Arm | 236 | 438 | 27 |
| 1288 | 116 | Skipper | Arm | 236 | 438 | 26 |
| 1289 | 116 | Skipper | Arm | 236 | 438 | 26 |
| 1290 | 116 | Skipper | Arm | 236 | 438 | 25 |
| 1291 | 116 | Skipper | Arm | 236 | 438 | 25 |
| 1292 | 116 | Skipper | Arm | 236 | 438 | 25 |
| 1293 | 116 | Skipper | Arm | 236 | 438 | 24 |
| 1294 | 116 | Skipper | Arm | 236 | 438 | 24 |
| 1295 | 116 | Skipper | Arm | 236 | 438 | 24 |
| 1297 | 116 | Skipper | Arm | 236 | 438 | 21 |
| 1298 | 116 | Skipper | Arm | 236 | 438 | 18 |

| UnitID | SiteID | Monkey | Somatotopy | X | Y | Channel |
|---|---|---|---|---|---|---|
| 1299 | 116 | Skipper | Arm | 236 | 438 | 18 |
| 1300 | 116 | Skipper | Arm | 236 | 438 | 17 |
| 1301 | 116 | Skipper | Arm | 236 | 438 | 17 |
| 1302 | 116 | Skipper | Arm | 236 | 438 | 16 |
| 1303 | 116 | Skipper | Arm | 236 | 438 | 11 |
| 1304 | 116 | Skipper | Arm | 236 | 438 | 11 |
| 1305 | 116 | Skipper | Arm | 236 | 438 | 11 |
| 1306 | 116 | Skipper | Arm | 236 | 438 | 6 |
| 1307 | 116 | Skipper | Arm | 236 | 438 | 5 |
| 1308 | 116 | Skipper | Arm | 236 | 438 | 4 |
| 1309 | 116 | Skipper | Arm | 236 | 438 | 3 |
| 1310 | 117 | Skipper | Trunk | 143 | 218 | 29 |
| 1311 | 117 | Skipper | Trunk | 143 | 218 | 26 |
| 1312 | 117 | Skipper | Trunk | 143 | 218 | 25 |
| 1313 | 117 | Skipper | Trunk | 143 | 218 | 23 |
| 1314 | 117 | Skipper | Trunk | 143 | 218 | 23 |
| 1315 | 117 | Skipper | Trunk | 143 | 218 | 23 |
| 1316 | 117 | Skipper | Trunk | 143 | 218 | 22 |
| 1317 | 117 | Skipper | Trunk | 143 | 218 | 13 |
| 1318 | 117 | Skipper | Trunk | 143 | 218 | 12 |
| 1319 | 117 | Skipper | Trunk | 143 | 218 | 9 |
| 1320 | 117 | Skipper | Trunk | 143 | 218 | 8 |
| 1321 | 117 | Skipper | Trunk | 143 | 218 | 7 |
| 1322 | 117 | Skipper | Trunk | 143 | 218 | 6 |
| 1323 | 117 | Skipper | Trunk | 143 | 218 | 1 |
| 1324 | 117 | Skipper | Trunk | 143 | 218 | 1 |
| 1325 | 118 | Skipper | Arm | 86 | 382 | 30 |
| 1326 | 118 | Skipper | Arm | 86 | 382 | 30 |
| 1327 | 118 | Skipper | Arm | 86 | 382 | 29 |
| 1328 | 118 | Skipper | Arm | 86 | 382 | 28 |
| 1329 | 118 | Skipper | Arm | 86 | 382 | 27 |
| 1330 | 118 | Skipper | Arm | 86 | 382 | 25 |

| UnitID | SiteID | Monkey | Somatotopy | X | Y | Channel |
|---|---|---|---|---|---|---|
| 1331 | 118 | Skipper | Arm | 86 | 382 | 24 |
| 1332 | 118 | Skipper | Arm | 86 | 382 | 22 |
| 1333 | 118 | Skipper | Arm | 86 | 382 | 22 |
| 1334 | 118 | Skipper | Arm | 86 | 382 | 21 |
| 1335 | 118 | Skipper | Arm | 86 | 382 | 21 |
| 1336 | 118 | Skipper | Arm | 86 | 382 | 13 |
| 1337 | 118 | Skipper | Arm | 86 | 382 | 13 |
| 1338 | 118 | Skipper | Arm | 86 | 382 | 11 |
| 1339 | 118 | Skipper | Arm | 86 | 382 | 10 |
| 1340 | 118 | Skipper | Arm | 86 | 382 | 9 |
| 1341 | 118 | Skipper | Arm | 86 | 382 | 8 |
| 1342 | 118 | Skipper | Arm | 86 | 382 | 7 |
| 1343 | 118 | Skipper | Arm | 86 | 382 | 6 |
| 1344 | 119 | Skipper | Arm | 295 | 389 | 31 |
| 1345 | 119 | Skipper | Arm | 295 | 389 | 29 |
| 1346 | 119 | Skipper | Arm | 295 | 389 | 25 |
| 1347 | 119 | Skipper | Arm | 295 | 389 | 17 |
| 1349 | 120 | Skipper | Arm | 350 | 130 | 31 |
| 1350 | 120 | Skipper | Arm | 350 | 130 | 30 |
| 1351 | 120 | Skipper | Arm | 350 | 130 | 26 |
| 1352 | 120 | Skipper | Arm | 350 | 130 | 26 |
| 1353 | 120 | Skipper | Arm | 350 | 130 | 24 |
| 1354 | 120 | Skipper | Arm | 350 | 130 | 17 |
| 1355 | 120 | Skipper | Arm | 350 | 130 | 16 |
| 1356 | 120 | Skipper | Arm | 350 | 130 | 16 |
| 1357 | 120 | Skipper | Arm | 350 | 130 | 14 |
| 1358 | 120 | Skipper | Arm | 350 | 130 | 7 |
| 1359 | 120 | Skipper | Arm | 350 | 130 | 1 |
| 1360 | 121 | Skipper | Hand | 268 | 359 | 31 |
| 1361 | 121 | Skipper | Hand | 268 | 359 | 31 |
| 1362 | 121 | Skipper | Hand | 268 | 359 | 29 |
| 1363 | 121 | Skipper | Hand | 268 | 359 | 29 |

| UnitID | SiteID | Monkey | Somatotopy | X | Y | Channel |
|---|---|---|---|---|---|---|
| 1364 | 121 | Skipper | Hand | 268 | 359 | 28 |
| 1365 | 121 | Skipper | Hand | 268 | 359 | 28 |
| 1366 | 121 | Skipper | Hand | 268 | 359 | 27 |
| 1367 | 121 | Skipper | Hand | 268 | 359 | 27 |
| 1368 | 121 | Skipper | Hand | 268 | 359 | 24 |
| 1369 | 121 | Skipper | Hand | 268 | 359 | 24 |
| 1370 | 121 | Skipper | Hand | 268 | 359 | 22 |
| 1371 | 121 | Skipper | Hand | 268 | 359 | 20 |
| 1372 | 121 | Skipper | Hand | 268 | 359 | 19 |
| 1373 | 121 | Skipper | Hand | 268 | 359 | 19 |
| 1374 | 121 | Skipper | Hand | 268 | 359 | 18 |
| 1375 | 121 | Skipper | Hand | 268 | 359 | 12 |
| 1376 | 121 | Skipper | Hand | 268 | 359 | 11 |
| 1377 | 121 | Skipper | Hand | 268 | 359 | 10 |
| 1378 | 121 | Skipper | Hand | 268 | 359 | 9 |
| 1379 | 121 | Skipper | Hand | 268 | 359 | 6 |
| 1380 | 121 | Skipper | Hand | 268 | 359 | 4 |
| 1381 | 122 | Skipper | Hand | 132 | 269 | 17 |
| 1382 | 122 | Skipper | Hand | 132 | 269 | 9 |
| 1383 | 123 | Skipper | Hand | 226 | 47 | 31 |
| 1384 | 123 | Skipper | Hand | 226 | 47 | 27 |
| 1385 | 123 | Skipper | Hand | 226 | 47 | 23 |
| 1386 | 123 | Skipper | Hand | 226 | 47 | 23 |
| 1387 | 123 | Skipper | Hand | 226 | 47 | 22 |
| 1388 | 123 | Skipper | Hand | 226 | 47 | 22 |
| 1389 | 123 | Skipper | Hand | 226 | 47 | 15 |
| 1390 | 123 | Skipper | Hand | 226 | 47 | 13 |
| 1391 | 123 | Skipper | Hand | 226 | 47 | 13 |
| 1392 | 123 | Skipper | Hand | 226 | 47 | 8 |
| 1393 | 123 | Skipper | Hand | 226 | 47 | 7 |
| 1394 | 123 | Skipper | Hand | 226 | 47 | 6 |
| 1395 | 123 | Skipper | Hand | 226 | 47 | 6 |

| UnitID | SiteID | Monkey | Somatotopy | X | Y | Channel |
|---|---|---|---|---|---|---|
| 1396 | 123 | Skipper | Hand | 226 | 47 | 4 |
| 1397 | 123 | Skipper | Hand | 226 | 47 | 4 |
| 1398 | 124 | Skipper | Trunk | 310 | 38 | 31 |
| 1399 | 124 | Skipper | Trunk | 310 | 38 | 29 |
| 1400 | 124 | Skipper | Trunk | 310 | 38 | 28 |
| 1402 | 124 | Skipper | Trunk | 310 | 38 | 23 |
| 1403 | 124 | Skipper | Trunk | 310 | 38 | 21 |
| 1404 | 124 | Skipper | Trunk | 310 | 38 | 20 |
| 1405 | 124 | Skipper | Trunk | 310 | 38 | 19 |
| 1406 | 124 | Skipper | Trunk | 310 | 38 | 14 |
| 1407 | 124 | Skipper | Trunk | 310 | 38 | 13 |
| 1408 | 124 | Skipper | Trunk | 310 | 38 | 12 |
| 1409 | 124 | Skipper | Trunk | 310 | 38 | 11 |
| 1410 | 124 | Skipper | Trunk | 310 | 38 | 6 |
| 1411 | 124 | Skipper | Trunk | 310 | 38 | 5 |
| 1412 | 125 | Skipper | Arm | 60 | 356 | 18 |
| 1413 | 125 | Skipper | Arm | 60 | 356 | 17 |
| 1414 | 125 | Skipper | Arm | 60 | 356 | 16 |
| 1415 | 125 | Skipper | Arm | 60 | 356 | 14 |
| 1416 | 126 | Skipper | Hand | 308 | 291 | 31 |
| 1417 | 126 | Skipper | Hand | 308 | 291 | 31 |
| 1418 | 126 | Skipper | Hand | 308 | 291 | 27 |
| 1419 | 126 | Skipper | Hand | 308 | 291 | 23 |
| 1420 | 126 | Skipper | Hand | 308 | 291 | 23 |
| 1421 | 126 | Skipper | Hand | 308 | 291 | 20 |
| 1422 | 126 | Skipper | Hand | 308 | 291 | 16 |
| 1423 | 126 | Skipper | Hand | 308 | 291 | 13 |
| 1424 | 126 | Skipper | Hand | 308 | 291 | 6 |
| 1425 | 127 | Skipper | Trunk | 349 | 279 | 32 |
| 1426 | 127 | Skipper | Trunk | 349 | 279 | 25 |
| 1427 | 127 | Skipper | Trunk | 349 | 279 | 24 |
| 1428 | 127 | Skipper | Trunk | 349 | 279 | 22 |

| UnitID | SiteID | Monkey | Somatotopy | X | Y | Channel |
|---|---|---|---|---|---|---|
| 1429 | 127 | Skipper | Trunk | 349 | 279 | 20 |
| 1430 | 127 | Skipper | Trunk | 349 | 279 | 17 |
| 1431 | 127 | Skipper | Trunk | 349 | 279 | 16 |
| 1432 | 127 | Skipper | Trunk | 349 | 279 | 14 |
| 1433 | 127 | Skipper | Trunk | 349 | 279 | 12 |
| 1434 | 127 | Skipper | Trunk | 349 | 279 | 9 |
| 1435 | 127 | Skipper | Trunk | 349 | 279 | 7 |
| 1436 | 127 | Skipper | Trunk | 349 | 279 | 1 |
| 1437 | 128 | Skipper | Hand | 216 | 352 | 32 |
| 1438 | 128 | Skipper | Hand | 216 | 352 | 27 |
| 1439 | 128 | Skipper | Hand | 216 | 352 | 23 |
| 1440 | 128 | Skipper | Hand | 216 | 352 | 23 |
| 1441 | 128 | Skipper | Hand | 216 | 352 | 20 |
| 1445 | 128 | Skipper | Hand | 216 | 352 | 16 |
| 1446 | 128 | Skipper | Hand | 216 | 352 | 13 |
| 1447 | 128 | Skipper | Hand | 216 | 352 | 12 |
| 1448 | 128 | Skipper | Hand | 216 | 352 | 7 |
| 1449 | 129 | Skipper | Arm | 287 | 96 | 32 |
| 1450 | 129 | Skipper | Arm | 287 | 96 | 27 |
| 1451 | 129 | Skipper | Arm | 287 | 96 | 19 |
| 1452 | 129 | Skipper | Arm | 287 | 96 | 18 |
| 1453 | 130 | Skipper | Trunk | 29 | 195 | 32 |
| 1454 | 130 | Skipper | Trunk | 29 | 195 | 32 |
| 1455 | 130 | Skipper | Trunk | 29 | 195 | 31 |
| 1456 | 130 | Skipper | Trunk | 29 | 195 | 27 |
| 1457 | 130 | Skipper | Trunk | 29 | 195 | 21 |
| 1458 | 130 | Skipper | Trunk | 29 | 195 | 18 |
| 1459 | 130 | Skipper | Trunk | 29 | 195 | 17 |
| 1460 | 130 | Skipper | Trunk | 29 | 195 | 16 |
| 1461 | 130 | Skipper | Trunk | 29 | 195 | 14 |
| 1462 | 130 | Skipper | Trunk | 29 | 195 | 7 |
| 1463 | 131 | Skipper | Arm | 340 | 180 | 32 |

| UnitID | SiteID | Monkey | Somatotopy | X | Y | Channel |
|---|---|---|---|---|---|---|
| 1464 | 131 | Skipper | Arm | 340 | 180 | 31 |
| 1467 | 131 | Skipper | Arm | 340 | 180 | 28 |
| 1468 | 131 | Skipper | Arm | 340 | 180 | 27 |
| 1469 | 131 | Skipper | Arm | 340 | 180 | 23 |
| 1470 | 131 | Skipper | Arm | 340 | 180 | 22 |
| 1471 | 131 | Skipper | Arm | 340 | 180 | 21 |
| 1472 | 131 | Skipper | Arm | 340 | 180 | 19 |
| 1473 | 131 | Skipper | Arm | 340 | 180 | 18 |
| 1474 | 131 | Skipper | Arm | 340 | 180 | 18 |
| 1475 | 131 | Skipper | Arm | 340 | 180 | 16 |
| 1476 | 131 | Skipper | Arm | 340 | 180 | 15 |
| 1477 | 131 | Skipper | Arm | 340 | 180 | 14 |
| 1478 | 131 | Skipper | Arm | 340 | 180 | 12 |
| 1479 | 131 | Skipper | Arm | 340 | 180 | 11 |
| 1480 | 131 | Skipper | Arm | 340 | 180 | 9 |
| 1481 | 131 | Skipper | Arm | 340 | 180 | 8 |
| 1482 | 131 | Skipper | Arm | 340 | 180 | 6 |
| 1483 | 131 | Skipper | Arm | 340 | 180 | 5 |
| 1484 | 131 | Skipper | Arm | 340 | 180 | 4 |
| 1485 | 132 | Skipper | Hand | 50 | 275 | 32 |
| 1486 | 132 | Skipper | Hand | 50 | 275 | 28 |
| 1487 | 132 | Skipper | Hand | 50 | 275 | 25 |
| 1488 | 132 | Skipper | Hand | 50 | 275 | 23 |
| 1489 | 132 | Skipper | Hand | 50 | 275 | 22 |
| 1490 | 132 | Skipper | Hand | 50 | 275 | 14 |
| 1491 | 133 | Skipper | Hand | 249 | 382 | 32 |
| 1492 | 133 | Skipper | Hand | 249 | 382 | 32 |
| 1493 | 133 | Skipper | Hand | 249 | 382 | 31 |
| 1494 | 133 | Skipper | Hand | 249 | 382 | 29 |
| 1495 | 133 | Skipper | Hand | 249 | 382 | 25 |
| 1496 | 133 | Skipper | Hand | 249 | 382 | 25 |
| 1497 | 133 | Skipper | Hand | 249 | 382 | 23 |

| UnitID | SiteID | Monkey | Somatotopy | X | Y | Channel |
|---|---|---|---|---|---|---|
| 1498 | 133 | Skipper | Hand | 249 | 382 | 22 |
| 1499 | 133 | Skipper | Hand | 249 | 382 | 22 |
| 1500 | 133 | Skipper | Hand | 249 | 382 | 21 |
| 1501 | 133 | Skipper | Hand | 249 | 382 | 18 |
| 1502 | 133 | Skipper | Hand | 249 | 382 | 16 |
| 1503 | 133 | Skipper | Hand | 249 | 382 | 15 |
| 1504 | 133 | Skipper | Hand | 249 | 382 | 14 |
| 1505 | 133 | Skipper | Hand | 249 | 382 | 14 |
| 1506 | 133 | Skipper | Hand | 249 | 382 | 11 |
| 1507 | 133 | Skipper | Hand | 249 | 382 | 9 |
| 1508 | 133 | Skipper | Hand | 249 | 382 | 6 |
| 1509 | 134 | Skipper | Hand | 142 | 456 | 32 |
| 1510 | 134 | Skipper | Hand | 142 | 456 | 32 |
| 1511 | 134 | Skipper | Hand | 142 | 456 | 31 |
| 1512 | 134 | Skipper | Hand | 142 | 456 | 31 |
| 1513 | 134 | Skipper | Hand | 142 | 456 | 30 |
| 1514 | 134 | Skipper | Hand | 142 | 456 | 27 |
| 1515 | 134 | Skipper | Hand | 142 | 456 | 23 |
| 1516 | 134 | Skipper | Hand | 142 | 456 | 23 |
| 1517 | 134 | Skipper | Hand | 142 | 456 | 21 |
| 1518 | 134 | Skipper | Hand | 142 | 456 | 20 |
| 1519 | 134 | Skipper | Hand | 142 | 456 | 19 |
| 1520 | 134 | Skipper | Hand | 142 | 456 | 18 |
| 1521 | 134 | Skipper | Hand | 142 | 456 | 17 |
| 1522 | 134 | Skipper | Hand | 142 | 456 | 15 |
| 1523 | 134 | Skipper | Hand | 142 | 456 | 13 |
| 1524 | 134 | Skipper | Hand | 142 | 456 | 11 |
| 1525 | 134 | Skipper | Hand | 142 | 456 | 11 |
| 1526 | 134 | Skipper | Hand | 142 | 456 | 10 |
| 1527 | 134 | Skipper | Hand | 142 | 456 | 8 |
| 1528 | 134 | Skipper | Hand | 142 | 456 | 7 |
| 1529 | 134 | Skipper | Hand | 142 | 456 | 4 |

| UnitID | SiteID | Monkey | Somatotopy | X | Y | Channel |
|---|---|---|---|---|---|---|
| 1531 | 135 | Skipper | Arm | 170 | 516 | 32 |
| 1532 | 135 | Skipper | Arm | 170 | 516 | 31 |
| 1533 | 135 | Skipper | Arm | 170 | 516 | 31 |
| 1534 | 135 | Skipper | Arm | 170 | 516 | 28 |
| 1535 | 135 | Skipper | Arm | 170 | 516 | 23 |
| 1536 | 135 | Skipper | Arm | 170 | 516 | 22 |
| 1537 | 135 | Skipper | Arm | 170 | 516 | 22 |
| 1538 | 135 | Skipper | Arm | 170 | 516 | 21 |
| 1539 | 135 | Skipper | Arm | 170 | 516 | 21 |
| 1540 | 135 | Skipper | Arm | 170 | 516 | 19 |
| 1541 | 135 | Skipper | Arm | 170 | 516 | 19 |
| 1542 | 135 | Skipper | Arm | 170 | 516 | 19 |
| 1543 | 135 | Skipper | Arm | 170 | 516 | 18 |
| 1544 | 135 | Skipper | Arm | 170 | 516 | 18 |
| 1545 | 135 | Skipper | Arm | 170 | 516 | 18 |
| 1546 | 135 | Skipper | Arm | 170 | 516 | 17 |
| 1547 | 135 | Skipper | Arm | 170 | 516 | 16 |
| 1548 | 135 | Skipper | Arm | 170 | 516 | 14 |
| 1549 | 135 | Skipper | Arm | 170 | 516 | 12 |
| 1550 | 135 | Skipper | Arm | 170 | 516 | 11 |
| 1551 | 135 | Skipper | Arm | 170 | 516 | 5 |
| 1552 | 136 | Skipper | Arm | 148 | 31 | 32 |
| 1553 | 136 | Skipper | Arm | 148 | 31 | 32 |
| 1554 | 136 | Skipper | Arm | 148 | 31 | 31 |
| 1555 | 136 | Skipper | Arm | 148 | 31 | 30 |
| 1556 | 136 | Skipper | Arm | 148 | 31 | 30 |
| 1558 | 136 | Skipper | Arm | 148 | 31 | 22 |
| 1560 | 136 | Skipper | Arm | 148 | 31 | 21 |
| 1561 | 136 | Skipper | Arm | 148 | 31 | 19 |
| 1563 | 136 | Skipper | Arm | 148 | 31 | 18 |
| 1565 | 136 | Skipper | Arm | 148 | 31 | 16 |
| 1566 | 136 | Skipper | Arm | 148 | 31 | 15 |

| UnitID | SiteID | Monkey | Somatotopy | X | Y | Channel |
|---|---|---|---|---|---|---|
| 1567 | 136 | Skipper | Arm | 148 | 31 | 14 |
| 1568 | 136 | Skipper | Arm | 148 | 31 | 12 |
| 1569 | 136 | Skipper | Arm | 148 | 31 | 12 |
| 1571 | 136 | Skipper | Arm | 148 | 31 | 10 |
| 1572 | 136 | Skipper | Arm | 148 | 31 | 9 |
| 1573 | 136 | Skipper | Arm | 148 | 31 | 7 |

## Appendix C Supplemental figures

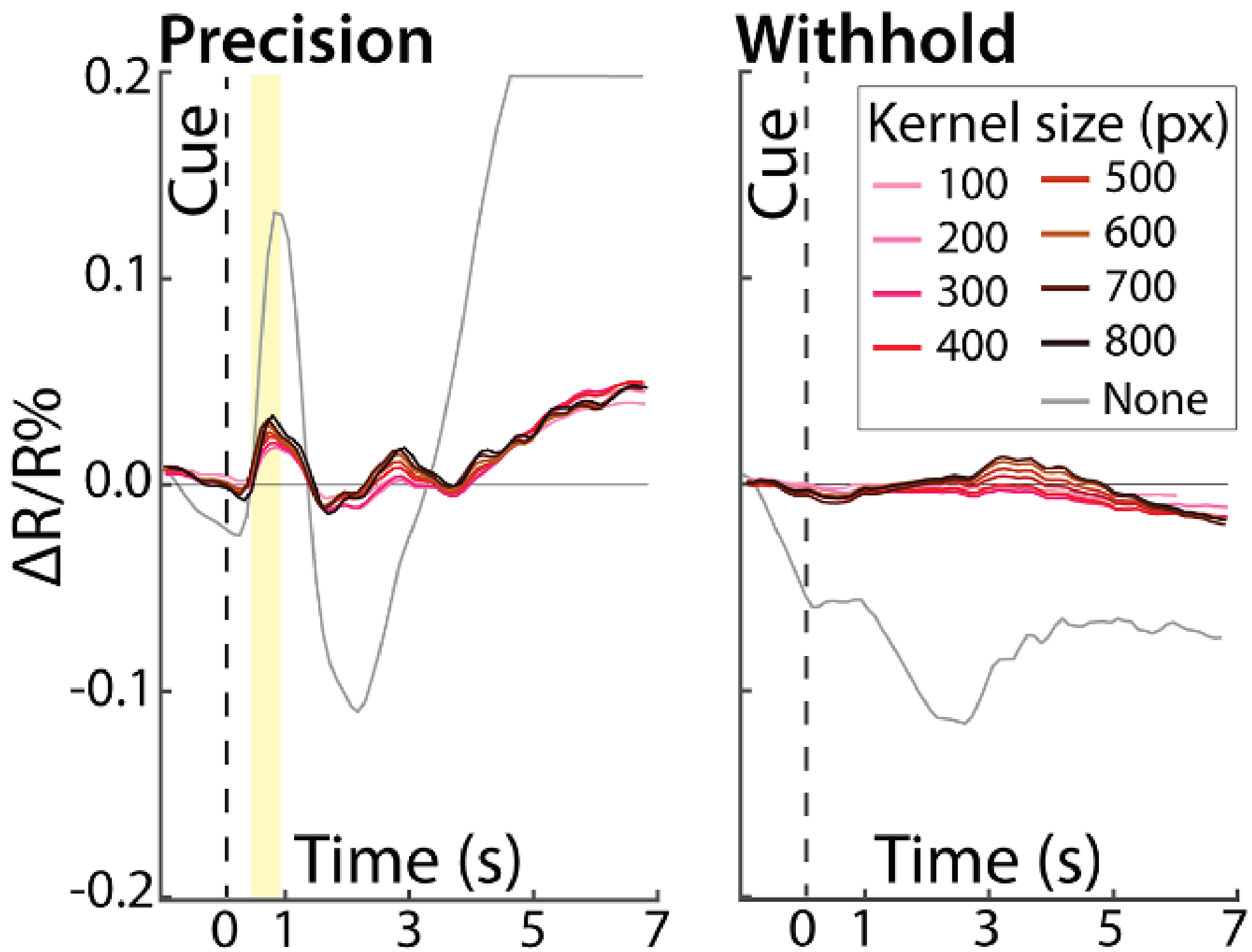

We tested the impact of the high-pass filter on the reflectance change time courses from a region-of-interest that covers the entire field of view. Thus, we systematically varied the size of the filtering kernel (100-800 pixels in steps of 100 pixels) then examined the time courses. The results of those tests are plotted on the left for the precision grip (movement condition) and on the right for the withhold conditions (no movement). The yellow highlighted region mark the mean movement start and end time. We note that in both conditions, most of the colored lines overlapped. This indicates that kernel size had little/no impact on time courses measured from the large ROIs. Nevertheless, the necessity for high pass filtering is evident from the gray time course, which was generated without high pass filtering. That line plot shows the dominance of global signals over the intrinsic signals of interest.

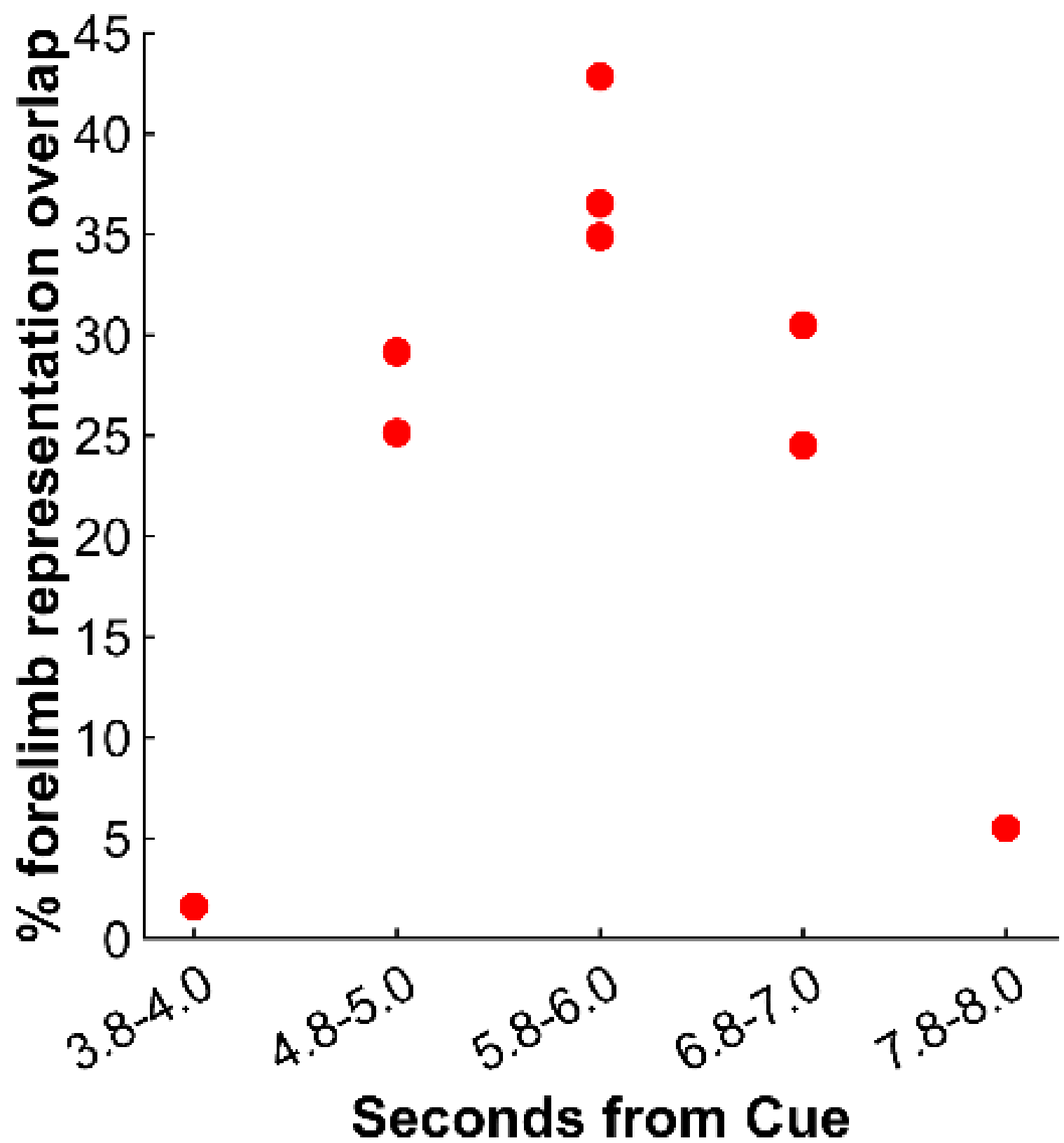

Activity maps reported pixels that were significantly darker in the movement frame range than in the baseline frame range. In Figure 5, we systematically delayed the Cue to demonstrate that increased reflectance change was time-locked to movement. Figure 5C shows nine binarized maps generated from different time points in relation to Cue. We measured the size of each map (i.e., overlap with the M1/PMd forelimb representations) and plotted the results above. The largest maps came from an average frame captured +5.8-6.0s from Cue. Those maps are on the diagonal in Figure 5E (top left to bottom right). The movement frame range for subsequent activity maps was chosen by calculating the mean peak reflectance time of all cortical (non-vessel) pixels (~4.0 s from Cue). This result validates that the chosen movement frame range does not overestimate the amount of active cortex resulting from the signal point spread (~5.0 to 6.0 s from Cue). See Figure 6I, 6K, 7, and 8 for support that the movement frame range does not underestimate the amount of active cortex.